\documentclass[12pt]{report}         
\usepackage{utthesis2}               
\usepackage{graphicx}
\usepackage{tikz}
\usepackage{cite}
\usepackage{epsfig}
\usepackage{amssymb}
\usepackage{bm}
\DeclareMathAlphabet{\mathscr}{OT1}{pzc}%
                                 {m}{it}
\usepackage{amsfonts, amsmath, amssymb}
\usepackage{subfigure}
\usepackage{amsthm}
\usepackage[T1]{fontenc}
\usepackage[ansinew]{inputenc}
\usepackage{amsmath}
\usepackage{amssymb}
\usepackage{graphicx}
\usepackage{color}
\usepackage[colorlinks]{hyperref}
\usepackage{rotating}
\usepackage[all]{xy}
\usepackage{float}
\newcommand{\be}{\begin{equation}}
\newcommand{\ee}{\,. \end{equation}}
\newcommand{\ba}{\begin{eqnarray}}
\newcommand{\ea}{\,. \end{eqnarray}}

\allowdisplaybreaks[1]

\renewcommand{\b}{\bar}

\phdthesis                           

 \doublespace                       

\renewcommand{\thesisauthor}{Jerry Michael Schirmer}    

\renewcommand{\thesismonth}{December}     

\renewcommand{\thesisyear}{2010}      

\renewcommand{\thesistitle}{Boundary conditions for Black Holes using the Ashtekar Isolated and Dynamical Horizons formalism}     

\renewcommand{\thesisauthorpreviousdegrees}{B.A.}

\renewcommand{\thesissupervisor}{Richard Matzner}

\renewcommand{\thesiscommitteemembera}{Duane Dicus}
\renewcommand{\thesiscommitteememberb}{Sonia Paban}
\renewcommand{\thesiscommitteememberc}{Dan Knopf}
\renewcommand{\thesiscommitteememberd}{Philip Morrison}

\renewcommand{\thesisauthoraddress}{2736 Stonewall Sta. \,   \\ St. Charles, MO 63303}

\renewcommand{\thesisdedication}{Dedicated to the generosity that I have received over the course of the past year, while my life has been turned upside down: \\
  
  From Casey Clough, who kept the household running while I was moving back and forth and never in one place, and for always being ready with some delicious noodles or a pot of tea.  Even after the most rough of work days. \\
  
  From Jennifer Kreft-Pierce and Austin Pearce, who opened their home to me when I was under duress, and whose help I am incapable of understating. \\
  
  From Allyson Whipple, during the afternoons at Mozarts and Bouldin Creek working on this, and who always had an ear to lend to me. \\
  
  And from my parents, Anne and Michael Schirmer, who have always been there for me, no matter what.\thispagestyle{empty}}

\begin{document}
\numberwithin{equation}{section}

\thesiscopyrightpage                 

\thesiscertificationpage             

\thesistitlepage                     


\thesisdedication                



\begin{thesisabstract}               
Isolated and Dynamical horizons are used to generate boundary conditions upon the lapse and shift vectors.  Numerous results involving the Hamiltonian of General relativity are derived, including a self-contained derivation of the Hamiltonian equations of general relativity using both a direct 'brute force' method of directly computing Lie derivatives, as well as the standard Hamiltonian approach.  Conclusions are compared to numerous examples, including the Kerr, Schwarzschild-De Sitter, McVittie, and Vaiyda spacetimes.  

\end{thesisabstract}                 

\tableofcontents                     


\listoffigures                     

\setcounter{chapter}{0}

\chapter{Overview}
\section{Conventions}

First, a basic note on convention.  For what follows, we will be taking the signature $(-,+,+,+)$ for the metric tensor and the convention $(\nabla_{a}\nabla_{b}-\nabla_{b}\nabla_{a})v_{c} = R_{abc}{}^{d}v_{d}$ for the Riemann tensor.  We will also take $\epsilon_{0123} = +1$.  If a mathematical quantity (i.e., $R_{ab}$,$ \nabla_{a}$, etc.) is written unmodified, it will be considered to be a 4-dimensional object.  If the same object has a tilde placed over it (i.e., $\tilde \nabla_{i}$), then that object will be considered to be an object intrinsic to a null hypersurface embedded in a Minkowskian 4-space.  If that same object has an overbar ($\bar R_{ij}$), then that object will be considered to be an object intrinsic to a 3-dimensional spacelike submanifold.  If the object in question is denoted with an overhat ($\hat R_{AB}$), then that object will be considered to be intrinsic to a two-dimensional submanifold to the 4-dimensional space.  We will also take the convention that indices that run over 4-dimensional labels will be labeled with lower case Latin letters from the beginning of the alphabet (a,b,c,...).  3-dimensional indices labeled in the intrinsic coordinates of the 3-dimensional space will be labeled with lower case Latin letters running from the middle of the alphabet (i,j,k,...).  Finally, 2-dimensional indices running over the coordinates intrinsic to a 2-dimensional space will be denoted by capital Latin letters running from the beginning of the alphabet (A,B,C,...).  Note that, with the use of the inclusion operators to be defined below, it $\mathbf{is}$ possible to have (for instance) a 2-dimensional quantity whose indices are 4-dimensional.  
\section{Scope of Dissertation}
Over the past three and a half decades, there has been a broad array of work on formalizing and redefining the notion of the black hole horizons, starting with the work of Regge and Tietelboim \cite{RTboundary}.  The primary thrust of this work has been to take the heavy mathematical machinery of immersions and manifolds and to use them to obtain a broad swath of powerful results in Relativity.  However, as this has progressed, there has only recently been a focus upon how to compute and analyze these quantities.  And while there are numerous attempts to use these quantities to set boundary conditions for numerical simulations \cite{CookPaper}\cite{Schnetter}\cite{Jaramillo:2007p1092}\cite{Jaramillo:2006p1066}\cite{Jaramillo:2007p1083}
and for quantum gravitational computations\cite{QuantumGrav}, there has not been a clear method to use 3+1 conditions in order to track the motion of a dynamical horizon across a 3+1 slice.  In section \ref{sec: DH stuff}, this is precisely what is done.  It is also explicitly shown, using the Vaiyda solution, how this term can be used to track the area growth of a black hole apparent horizon without assuming any prior knowledge about the dynamics of the black hole.  

This work, therefore, will have two goals.  The first one will be to re-derive existing results in a way that makes it transparent how one can do concrete calculations using the listed techniques.  Theorems from differential topology and geometry will be used with a focus towards indicating how one can choose a particular coordinate system and calculate quantities.  In particular, with null geometry, this question is treated with great care, as a null surface will have a tangent space and a cotangent space that are not metrically related to each other.  Thus, this work has many explicit examples and concrete calculations, both in the body of the paper and in the appendicesf.  The explicit null decomposition of Minkowski spacetime and of the Kerr spacetime done in Appendix \ref{sec: Kerr} is, to the author's knowledge, not currently published in the literature.  The formalism used in doing this decomposition actually improves existing literature, as it resolves an ambiguity in the definition of the degenerate metric tensor and the metric compatible null connection noted by Ashtekar et al.\cite{Ashtekar:2000p814}

Additional original work provided here includes the analysis of the McVittie spacetime\footnote{The McVittie spacetime fuses properties of the Schwarzschild and Robertson-Walker spacetimes, and thus can be considered to describe a black hole sitting in a Robertson-Walker background.}.  Discussion of the effects of the expansion of the universe on geodesics is described, as well as actual numerical computation of the geodesics.  There is a discussion regarding the singularity of this spacetime at the surface $r=2\,M$, where curvature invariants $C_{abcd}C^{abcd}$, $R_{ab}R^{ab}$ and $R$ are computed in order to establish the still-disputed claim that the McVittie ``horizon'' represents a spacetime singularity, and not a coordinate singularity \' a la Schwarzschild.  Then the spacetime is used to generate results related to the Hamiltonian dynamics of Relativity, in particular, as an example of how one may make sense of an ADM mass even in a non-asymptotically flat spacetime.  

This work also treats the elegant classical counterterm treatment of the ADM quantities, as originally computed by Hawking et al.  
rather than the more standard treatment which rejects boundary terms in a variational principle, and then inserts them at the last step in order to ensure consistency of the Hamiltonian equations of motion with the Hamiltonian \cite{RTboundary}.  While the treatment of Hamiltonian Relativity using counterterms is not original, the discussion to follow does provide a new spin upon the discussion, focused on the interaction between boundary degrees of freedom and bulk degrees of freedom.  

Finally, this work derives boundary conditions upon the lapse function and a component of the shift vector, and does so both in the intrinsic language of the Isolated and Dynamical horizons of Ashtekar and in the 3+1 Hamiltonian language that is best suited for numerical simulations of Einstein's equation.  And this is all done with an aim of developing a clear technique to show how the relevant objects can actually be evaluated and computed in a 3+1 context.

\chapter{Basic Definitions and Decomposition Process} 
\label{sec: Definitions}
\section{Choice of a 3+1 slice and normal vector}
In the literature, there has been a dearth of clarity on the topic of projecting tensors from higher dimensional spaces onto lower dimensional spaces.  In this section, a consistent procedure for doing this will be described, and then a direct application to the derivation of the (3+1) ADM split will be given.  The key point to remember is that the projection operation has different behavior depending on whether one is using the projection map or the inclusion map and whether or not these maps are acting on vectors or one-forms.

Now, consider a decomposition of a general manifold $\mathbb{M}$ with metric tensor $g_{ab}$.  Let $\mathbb{M}$ have the local topology $\mathbb{R}\times \mathbf{m}$, where $\mathbf{m}$ is a submanifold to $\mathbb{M}$.  Now, let $\tau$ be some function defined on $\mathbb{M}$ which takes a constant value on each fibre $\mathbf{m}$ of $\mathbb{R}$.  Then, $\nabla_{a}\tau$ will be a normal vector to the fiber.  For now, assume that $\nabla_{a}\tau\nabla^{a}\tau \neq 0$, (a case that will be dealt with on page \pageref{NullVectorSection}), and define $\xi$ as the sign of $\nabla_{a}\tau\nabla^{a}\tau$.  After making the definition $\alpha \equiv \frac{1}{\sqrt{\xi\nabla^{a}\tau\nabla_{a}\tau}} \label{DefAlpha}$, it is simple enough to define the unit normal to $\mathbf{m}$ by the formula \label{sec: normals}

\begin{equation}
n_{a} \equiv \alpha \nabla_{a}\tau \label{DefN} 
\,. \end{equation}

Typically, the next step would be to go and define the normal projection operator $\gamma_{ab} \equiv g_{ab}-\xi n_{a}n_{b}$, which we will soon do, but some care should be taken with this definition, as there are two different senses in which we can take the phrase ``3-metric''.  First, we wish to choose a special coordinate system for our manifold.  In particular, we wish to choose $\tau$ as one of the coordinates in the neighborhood of one of the fibers.  Clearly, this should be a valid coordinate choice, as $\tau$ is constant on each fiber, so advancing in $\tau$ merely advances from one fibre to the next, just as stepping forward in $t$ in standard Minkowski spacetime merely advances one from one spacelike copy of $\mathbb{R}^{3}$ to the next.  So, after replacing one of the original manifold coordinates with $\tau$ we have a coordinate system of the form $(\tau, x^{i})$, where the index i ranges over the remaining coordinates in the spacetime.\footnote{While the language of this example is geared for the most common case, a timelike $\nabla_{a}\tau$ in a 4-dimensional Minkowskian spacetime, and the argument will advance as such, note that no such assumption is being made here.  The above methodology will work for any non-null $\nabla_{a}\tau$.}  Therefore, it is now clear that, by assumption, we have $\nabla_{a}\tau = (\xi,0,0,...,0)$\footnote{The factor of $\xi$ is inserted so as to make $g^{ab}\nabla_{a}\tau$ future pointing}, and $\alpha = \frac{1}{\sqrt{\xi g^{\tau \tau}}}$.  Therefore, this gives a relatively simple coordinate expression for the raised form of $n^{a}$:

\begin{equation}
n^{a} = \frac{\xi}{\sqrt{\xi g^{\tau \tau}}}g^{\tau a}
\,. \end{equation}

We therefore have an expression for the raised operator $\gamma^{ab}$ in terms of the original metric components:

\begin{equation}
\gamma^{ab}= g^{ab} -\xi n^{a}n^{b} = g^{ab} - \xi \frac{1}{\xi g^{\tau \tau}}\xi g^{\tau a}\xi g^{\tau b}=g^{ab} - \frac{1}{g^{\tau \tau}}g^{\tau a}g^{\tau b}\,. \label{DefGamma}
\,. \end{equation}

Where we use the fact that $\xi = ±1$ to set $\xi^{2} = 1$.  It can now clearly be seen that all of the components of $\gamma^{ab}$ where either index takes on the value $\tau$ vanishes.  It is therefore natural to consider $\gamma^{ab}$ to be an operator that takes four-component one-forms on $\mathbb{M}$ and projects them onto three-component vectors on $\mathbf{m}$.  Therefore, we can consider the operator $\gamma^{ab}$ in two senses:  First, we can think of it as a projection operator living in a 4-dimensional space whose output is vectors with three nonzero entries.  Secondly, we can think of it as an intrinsic 3-dimensional object that serves to define the inner product over the covector space to $\mathbf{m}$.  Now, we will show that the natural inverse of the three by three matrix $\gamma^{ij}$ is in fact obtained simply by restricting the 4-metric to the appropriate 3-dimensional indices, $g_{ij}$:

\begin{align}
\gamma^{ij}g_{jk}&=(g^{ij}-\frac{1}{g^{\tau \tau}}g^{\tau i}g^{\tau j})g_{jk} \, ,  \\
&= g^{ia}g_{ak} - g^{i\tau}g_{\tau k} - \frac{1}{g^{\tau \tau}}g^{\tau i}(g^{\tau a}g_{ak} - g^{\tau \tau}g_{\tau k})\, ,  \\
&= \delta^{i}{}_{k}-g^{i\tau}g_{\tau k}-\frac{g^{\tau i}}{g^{\tau \tau}}\delta^{\tau}{}_{k} + g^{\tau i}g_{\tau k}\, ,  \\
&= \delta^{i}{}_{k}
\,. \end{align}

Here, we change the summation over 3-dimensional indices to a summation over 4-dimensional indices by the identity $v^{i} v_{i} = v^{a}v_{a}-v^{\tau}v_{\tau}$, and, in the transition to the last line, we realize that k ranges over only spatial indices, so therefore $\delta^{\tau}{}_{k}$ is identically zero, and we use the symmetry of the metric tensor to cancel the first and third terms.  Therefore, we can see that the projection operator on covariant indices merely requires that we drop the $\tau$ component of the one-form, and this is where care should be taken, as if we use $g^{ab}$ to raise and lower the indices of $\gamma^{ab}$, we could na\''{i}vely believe that the proper three projection of a one-form $\omega_{a}$ might be $\gamma^{a}{}_{b}\omega_{a}$\footnote{And, as we will see on page \pageref{VectorList}, in a certain sense it is}.  A simple computation, however, will show that this form does, in fact, have a $\tau$ component, and therefore, is not properly a 3-dimensional one-form, in the sense that it is three quantities along the three coordinate basis one-forms on the cotangent space to $\mathbf{m}$.  Instead, as the above example involving the metric indicates, the proper thing to do is, instead, simply drop the $d\tau$ component of the covector $\omega_{a}$.

Finally, we might want to project a 4-dimensional vector onto our 3-dimensional space.  This action, however, has a more complicated form than simply dropping the $\tau$ component.  In order to accomplish this, we first lower the index of the vector, follow the above projection procedure, and then raise the vector with $\gamma^{ij}$:

\begin{align}
v^{a} \rightarrow& g_{ia}v^{a}\, ,  \\
g_{ia}v^{a} =& g_{i\tau}v^{\tau} + g_{ij}v^{j}\, ,  \\
\bar v^{k}=& \gamma^{ki}g_{i\tau}v^{\tau} +\gamma^{ki}g_{ij}v^{j}\, ,  \\
=&(g^{ki}+n^{k}n^{i})g_{it}v^{t} +\gamma_{j}{}^{k}v^{j}\, ,  \\
=&(\delta_{\tau}{}^{k}+n^{k}n_{\tau})v^{\tau} + \gamma_{j}{}^{k}v^{j}\, ,  \\
=&\gamma_{a}{}^{k}v^{a}
\,. \end{align}

So, while the operator $\gamma_{a}{}^{b}$ does not take four component one forms on $\mathbb{M}$ and spit out three component one forms on $\mathbf{m}$, it $\mathit{does}$ do this to vectors.  

On the other hand, it should be clear that we might have a problem going in the other direction, from 3-dimensional space to 4-dimensional space.   In particular, if we were to map from the 3-space back into the 4-space, we would want our target vector/one-form to be normal to the appropriate version of $n_{a}$, since we want to consider $n_{a}$ to be the normal to the 3-space.  But, if we were to merely map a general 3-dimensional one form in the trivial manner:

\begin{equation}
s_{i} = (x_{1}, x_{2}...,x_{n})\rightarrow s_{a}=(0,x_{1}, x_{2}...,x_{n}) \label{TrivialInclusionOneFormMap}
\,. \end{equation}

We can see right away that we will get a result that is inconsistent with our desire to be normal to $n^{a}$, since $n^{a} = \frac{-\xi}{\sqrt{\xi g^{tt}}}g^{ta}$ and, in general, the $g^{ti} \neq 0$.  So, what are we to do?  The answer is that, for one-forms, we use the operator $\gamma^{a}{}_{b}$ not as a projection operator from $\mathbb{M}$ onto $\mathbf{m}$, but rather as an $\mathit{inclusion}$ $\mathit{operator}$ of $\mathbf{m}$ into $\mathbb{M}$.  If we use the above mapping, and then follow that with having the one-form be acted upon by $\gamma^{a}{}_{b}$, we get a vector that is normal to $n^{a}$, since 

\begin{equation}
n^{a}\gamma_{a}{}^{b} = n^{a}(\delta_{a}{}^{b} + n_{a}n^{b}) = n^{b} - n^{b} = 0 \label{Ntozero}
\,. \end{equation}

Furthermore, we have

\begin{equation}
\gamma^{ab}\gamma_{bc} = (g^{ab}+n^{a}n^{b})(\gamma_{bc}+n_{b}n_{c}) = \delta^{a}{}_{c} +n^{a}n_{c} +n^{a}n_{c} -n^{a}n_{c}=\delta^{a}{}_{c} +n^{a}n_{c} = \gamma^{a}{}_{c}
\,. \end{equation}

Now, we will show that use of $\gamma^{ab}$ enables us to find a vector in $\mathbb{M}$ that is, in a sense, equivalent to a one form living in the cotangent space of $\mathbf{m}$, thereby avoiding this problem.  Then, once we know this duality, we can then use the appropriate metric tensor to raise and lower indices as we will.  First, consider a one-form $s_{i}$ living in $\mathbf{m}$.  Define $\bar s_{a}$ according to the trivial inclusion map seen in \eqref {TrivialInclusionOneFormMap}.  Since we know that the components of $g_{ij}$ are equivalent to the components of $\gamma_{ij}$, and that the matrix $\gamma^{ij}$ is the inverse of the matrix $\gamma_{ij}$, while, at the same time, we know that $\gamma_{ab}$ satisfies the identity

\begin{align}
\gamma^{ab}\gamma_{bc} =& \left(g^{ab} -\xi n^{a}n^{b}\right)\left(g_{bc} - \xi n_{b}n_{c}\right) \nonumber \, ,  \\
=& \delta^{a}{}_{c} - \xi n^{a}n_{c} - \xi n^{a}n_{c} + \xi^{3}n^{a}n_{c} \nonumber \, ,  \\
=& \delta^{a}{}_{c} - \xi n^{a}n_{c} \nonumber \, ,  \\
=& \gamma^{a}{}_{c}
\,. \end{align}

Then, we can show that if we define $s^{a} \equiv \gamma^{ab}\bar s_{b}$, we can show that the norm of $\bar s_{i}$ under $\gamma^{ij}$ is equivalent to the norm of $s^{a}$ under $g_{ab}$:

\begin{align}
\bar s_{i} \bar s^{i} =& \gamma^{ij}\bar s_{i}\bar s_{j}\,. \, ,  \\
=&\gamma^{ab}\bar s_{a} \bar s_{b} \, ,  \\
=& \gamma_{cd} \gamma^{ca}\gamma^{db} \bar s_{a}\bar s_{b}\, ,  \\
=& (g_{cd} -\xi n_{c}n_{d})\gamma^{ca}\bar s_{a}\gamma^{db}\bar s_{b}\, ,  \\
=& g_{cd}s^{c}s^{d}\, ,  \\
=& s^{a}s_{a}
\,. \end{align}

So, then, we can lower the left hand index of $\gamma^{ab}$ using $g_{ab}$, and then, we can use the result, $\gamma_{a}^{b}$, as an inclusion map whose target has the same norm in $\mathbb{M}$ as the source had in $\mathbf{m}$.  Also, note that since $n_{a}$ has only a $\tau$ component, we can simply drop the $\tau$ component of $s_{a}$ to recover $\bar s_{i}$.  Henceforward, this paper will drop the bar notation on one forms, and let the type of index denote whether we are discussing the 3-dimensional one-form or its inclusion into the four-manifold.

Meanwhile, using the trivial inclusion map for vectors creates no problem, since $\nabla_{a}\tau$ and $n_{a}$ have no components along any other direction than the $\tau$ coordinate.  Therefore, if we take:

\begin{equation}
s^{i} = (\bar x^{1}, \bar x^{2}... \bar x^{n}) \rightarrow s^{a}=(0, \bar x^{1}, \bar x^{2}... \bar x^{n})
\,. \end{equation}

We are guaranteed to not run into any trouble, since the right hand of the above equation is normal to all of the one-forms with no components that project along the space, and it is guaranteed to have the same norm, since $g_{ij} = \gamma_{ij}$.  So, we have the following rules for the projection and inclusion maps:

\begin{enumerate}
\item{If we are using the projection map:} \label{VectorList}
\begin{enumerate}
\item{One forms project trivially: just drop the appropriate component, and the projection is done}
\item{Vectors, however, do NOT project trivially, and need to be projected using the $\gamma^{a}{}_{b}$ operator, which is equivalent to lowering with the full metric, operating with the trivial projection map, and then raising with the three-metric}
\end{enumerate}
\item{If we are acting with the inclusion map (we are taking 3-component objects intrinsic to a 3-surface, and seeing how they look in the full spacetime)}
\begin{enumerate}
\item{in this case, it is the vectors who have the trivial behaviour.  Just add a zero in the appropriate component, and then you have a vector in the full tangent space} \label{inclusionmap}
\item{One forms, however, need to have the inclusion matrix $\gamma_{a}{}^{b}$ operate on them.  This is ultimately equivalent to raising them using the induced metric $\gamma^{ij}$, operating with the trivial inclusion map for vectors and then lowering with the enveloping four-dimensional metric.}
\end{enumerate}
\end{enumerate}

The above procedure is justified by a look at any differential topology textbook (see, e.g., \cite{GP}), which will define the action of the pullback operator (that maps from a manifold with higher dimension to one with lower dimension) in terms of the differential of the map from the higher dimensional space to the lower dimensional space.  It will therefore have a trivial action on forms.  Meanwhile, the book will define the inclusion operator (which acts in the opposite direction) in terms of the vectors in the target space that are being acted on by the above forms, and therefore, the inclusion operator will act trivially on vectors, but not forms.  The topology book will typically stop short of saying what to do with the non-trivial actions, as a metric tensor is typically not assumed.  In our case, however, we can use the metric and induced metric to move the tangent space into the cotangent space, and then operate with the inclusion map and the projection map, and then move back between the tangent space and cotangent space to get the correct operator.  

So, in summary, there are two different senses in which we can talk about 3-dimensional tensor indices.  The first is to consider the tensors as arising from three component indices intrinsic to the 3-dimensional submanifold.  The other is to consider the tensors as being the targets of the inclusion map described above.  These objects have four indices, and live in the 4-dimensional spacetime, but are ``3-dimensional'' in the sense that they are normal to the vector $n^{a}$ and also to the one-form $n_{a}$.  Above, we have shown that the two methods are equivalent, and have shown the manner in which we can map from one version of a 'three-vector' to another.  From here on out, we will work primarily with the version of a three vector that lives in the 4-dimensional space, and then project down when necessary.

Finally, one might object to our above procedure, since we took the ten metric tensor components, and projected them down to get the lapse $\alpha$, as defined in \ref{DefAlpha} and the three-metric $\gamma_{ab}$ as defined in \eqref{DefGamma}, which has six independent components.  What of the other three 4-metric components?  These three components are encoded in the shift vector, defined according to:

\begin{equation}
\beta_{i} \equiv g_{\tau i} \label{DefBeta}
\,. \end{equation}

It should be clear that this vector does, in fact, have three components that correspond to the metric tensor components that are dropped when the projection is done.  And with that, we have encoded all of the ten independent metric tensor components in terms of equivalent 3-dimensional quantities.  At times, it is convenient to speak of the time evolution vector, which is the vector that generates the evolution in the three-geometry as one advances along surfaces with increasing values of $\tau$:

\begin{equation}
t^{a} \equiv  \alpha n^{a} + \gamma^{ab}\beta_{b} \label{DefT}
\,. \end{equation}

In fact, it is easy to show that, in our coordinate system, this vector has only a $\tau$ component:

\begin{align}
t^{a} =& \alpha n^{a} + \gamma^{ab}\beta_{b} \, ,  \\
=& (\frac{1}{\sqrt{\xi g^{\tau \tau}}})\frac{\xi g^{\tau a}}{\sqrt{\xi g^{\tau \tau}}} + \gamma^{ab}g_{b\tau}\, ,  \\
=&\frac{\xi g^{\tau a}}{\xi g^{\tau \tau}} + g^{ab}g_{b\tau} -\xi n^{a}n^{b}g_{b \tau}\, ,  \\
=&\frac{g^{\tau a}}{g^{\tau \tau}} +\delta^{a}{}_{\tau} - \xi n^{a}n_{\tau}\, ,  \\
=&\frac{g^{\tau a}}{g^{\tau \tau}} + \delta^{a}{}_{\tau} - \xi (\frac{\xi g^{\tau a}}{\sqrt{\xi g^{\tau \tau}}})\frac{\xi}{\sqrt{ \xi g^{\tau \tau}}}\, ,  \\
=&\delta^{a}{}_{\tau} +\frac{g^{\tau a}}{g^{\tau \tau}} - \frac{\xi^{3}g^{\tau a}}{\xi g^{\tau \tau}}\, ,  \\
=& \delta^{a}{}_{\tau}
\,. \end{align}

Now, let us consider the appropriate way to define a 3-dimensional derivative operator.  We remember from basic differential geometry that a derivative operator is unique if it (i) has domain and range within the manifold, (ii) maps the metric tensor to zero, and is (iii) torsion free.  Therefore, consider the following definition, for any vector $v_{a}$ such that $v_{a}n^{a}=0$:

\begin{equation}
\bar \nabla_{a}v_{b} = \gamma_{a}^{c}\gamma_{b}^{d}\nabla_{c}v_{d} \label{defD}
\,. \end{equation}

Condition (iii) is automatically satisfied if the original connection is torsion free, since you can't introduce antisymmetry in the Christoffel symbols by multiplying them by matrices.  Condition (i) is satisfied due to a combination of  the restriction on $v_{a}$ and equation \eqref {Ntozero} above.  Appealing to the metric compatibility of the original connection, we can explicitly show that condition (ii) is satisfied:

\begin{align}
\bar \nabla_{a}\gamma_{bc} =& \gamma^{m}{}_{a}\gamma^{n}{}_{b}\gamma^{k}{}_{c}\nabla_{m}(g_{nk}+n_{n}n_{k})\, ,  \\
=&0+\gamma^{m}{}_{a}\gamma^{n}{}_{b}\gamma^{k}{}_{c}(n_{n}\nabla_{m}n_{k} + n_{k}\nabla_{m}n_{n})\, ,  \\
=&0
\,. \end{align}

Since $\gamma^{a}{}_{b}$ annihilates $n_{a}$.  Therefore, $\bar \nabla_{a}$ is the unique derivative operator on $\mathbf{m}$.  Furthermore, we can note that this is, in fact, just equivalent to the standard definition one would make of a covariant derivative, if we first note that the operator $\gamma^{a}{}_{b}$  merely serves to project out any derivatives along the $t$ direction.  We can therefore take $\gamma_{i}{}^{a}\gamma_{j}{}^{b}\partial_{a}v_{b} = \partial_{i}v_{j}$ after having dropped the $t$ indices according to the procedure outlined above.  Now, we can check the equivalence of the definition \eqref {defD} with the standard definition of the covariant derivative:

\begin{align*}
\gamma_{i}{}^{a}\gamma_{j}{}^{b}\nabla_{a}v_{b} =& \gamma_{i}{}^{a}\gamma_{j}{}^{b}(\partial_{a}v_{b} - \Gamma_{ab}{}^{c}v_{c})\, ,  \\
=&\partial_{i}v_{j} -\frac{1}{2}\gamma_{i}{}^{a}\gamma_{j}{}^{b}v_{k}\gamma^{k}{}_{c}g^{cm}(g_{am,b}+g_{bm,a}-g_{ab,m})\, ,  \\
=& \partial_{i}v_{j} -\frac{1}{2}\gamma_{i}{}^{a}\gamma_{j}{}^{b}v_{k}\gamma^{km}[\partial_{a}(\gamma_{bm}-n_{b}n_{m})+\partial_{b}(\gamma_{am}-n_{a}n_{m})  \nonumber \\
&- \partial_{k}(\gamma_{ab}-n_{a}n_{b})]\, ,  \\
=&\partial_{i}v_{j}-\frac{1}{2}\gamma_{i}{}^{a}\gamma_{j}{}^{b}v_{k}\gamma^{km}[\gamma_{am,b}+\gamma_{bm,a}- \gamma_{ab,m}]\, ,  \\
=&\partial_{i}v_{j} - \bar \Gamma_{ij}{}^{k}v_{k}
\end{align*}

Where, in the transition from the third to fourth lines, we saw, that when the partial derivative operator acted on the $n_{m}n_{n}$ terms and we expanded with the product rule, we saw that it was impossible to not contract one of the $n_{m}$ on a corresponding $\gamma^{m}{}_{n}$, yielding zero.  Similarly, when going to the last line, we realized that, if all of the quantities involved are 3-dimensional, then the $\gamma_{a}{}^{b}$ operator is merely the identiy operator on the 3-dimensional space, and we defined the $\bar \Gamma_{ab}{}^{c}$ operator to be the obvious Christoffel symbol of the 3-dimensional space.  

Now, beyond just the basic metric and connection of the three manifold, we should consider the information about the embedding of $\mathbf{m}$ in $\mathbf{M}$.  In order to do this, we introduce the extrinsic curvature tensor according to the definition:

\begin{equation}
K_{ab} \equiv -\gamma_{a}^{c}\gamma_{b}^{d}\nabla_{c}n_{d}
\,. \end{equation}

Where the second factor of $\gamma_{b}^{d}$ can be dropped in favor of $\delta_{b}^{d}$ since those two tensors differ only by a factor of $n_{b}n^{d}$, which vanishes when contracted on the covariant derivative term since $n_{a}$ has a fixed norm.  This concludes our basic discussion of how to encode 4-geometry in terms of 3-geometry.  In the next chapter, we will investigate how to translate the curvature and Einstein's equation into three-dimensional language, using a more explicit (albeit less elegant) methodology than is typically used in the literature.

\chapter{ADM decomposition of Einstein's Equation} \label{sec: Brute Force}

In this section, we will now proceed to take the Einstein Equation, and express it in terms of the 3-dimensional quantities described above.  This can be (and in the literature, usually is) done by writing the Hilbert action $\frac{1}{16\pi G}\int R d^{4}x$ in terms of the 3-dimensional quantities, defining the conjugate momentum to $\gamma_{ab}$, taking a Legendre transformation of the Lagrangian in order to find the Hamiltonian, and then writing down Hamilton's equations of motion.  In this chapter, we take a different approach, explicitly calculating Lie derivatives of appropriate quantities using a 3+1 split.  While this approach is much less elegant than the Hamiltonian technique, it serves a few purposes.  First, it gives desired results much more directly--there is no need to Legendre transform, define conjugate momenta, and the like.   In the Hamiltonian approach, the equations you naturally get involve the canonical momentum $\Pi^{ab} \equiv \frac{\sqrt{\gamma}}{16 \pi G}\left(\gamma^{ab} K - K^{ab}\right)$, and the equation for $\dot \Pi^{ab}$ must be used to derive an equation for $\dot K_{ab}$, which involves significant amounts of algebra involving factors of $\dot \gamma_{ab}$.  In the brute force method, with little embellishment, one can directly calculate $\dot \gamma_{ab}$ and $\dot K_{ab}$.    Second, while both techniques are available for splitting a Lorentzian 4-manifold of local topology $\mathbb{R} \times \mathbf{m}$, only the brute force technique used here is generalizable to the splitting a neighborhood of the boundary of a Riemannian 3-surface into its 2-dimensional boundary and a 1-dimensional normal space.  

For completeness, and for the sake of investigating the boundary terms that the Hilbert action necessarily gives rise to, we will compute the ADM equations using the Hamiltonian picture in Chapter \eqref {sec:Hamiltonian General Relativity}

\section{Constraint Equations}
First, given an arbitrary four vector $v_{a}$, let us recall the definition of the Riemann Curvature tensor:
\begin{equation}
R_{abc}{} ^{d}v_{d} \equiv 2\nabla_{[a}\nabla_{b]}v_{c}
\,. \end{equation}

Since we defined the 3-dimensional connection above in equation \eqref {defD}, we also have a simple enough expression for the 3-dimensional curvature tensor, of an arbitrary 3-dimensional one-form $\omega_{c}$ (and using the metric compatibility of the 4-connection, combined with $n^{a}\gamma_{a}^{b}=0$):

\begin{align}
\bar R_{abc}{}^{d}\omega_{d} =& 2 \bar \nabla_{[a}\bar \nabla_{b]}\omega_{c}\, ,  \\
=& 2\gamma^{m}{}_{[a}\gamma^{n}{}_{b]}\gamma^{k}{}_{c}\nabla_{m}(\gamma^{r}{}_{n}\gamma^{s}{}_{k}\nabla_{r}\omega_{s})\, ,  \\
=&2\gamma^{m}{}_{[a}\gamma^{r}{}_{b]}\gamma^{s}{}_{c}\nabla_{m}\nabla_{r}\omega_{s}+2\gamma^{m}{}_{[a}\gamma^{r}{}_{b]}\gamma^{k}{}_{c}(\nabla_{m}\gamma_{k}{}^{s})\nabla_{r}\omega_{s}\nonumber   \\
&+2\gamma^{m}{}_{[a}\gamma^{n}{}_{b]}\gamma^{s}{}_{c}(\nabla_{m}\gamma_{n}{}^{r})\nabla_{r}\omega_{s} \, ,  \\
=& \gamma_{a}{}^{m}\gamma_{b}{}^{n}\gamma_{c}{}^{k}R_{mnk}{}^{d}\omega_{d}-2\xi\gamma^{m}{}_{[a}\gamma^{r}{}_{b]}\gamma^{k}{}_{c}n^{s}(\nabla_{m}n_{k})\nabla_{r}\omega_{s}\nonumber  \\
& -2\xi\gamma^{m}{}_{[a}\gamma^{n}{}_{b]}\gamma^{s}{}_{c}n^{r}(\nabla_{m}n_{n})\nabla_{r}\omega_{s}\label{threeconnectionstep}
\,. \end{align}

Before we proceed farther, we should derive some properties of the gradient of $n_{a}$.  First, as asserted above, we know that $n^{b}\nabla_{a}n_{b} = 0$, since it is easy to show that it is equal to minus itself.  Now, let us evaluate $n^{a}\nabla_{a}n_{b}$.  Using Equation \eqref {DefN}:

\begin{align}
n^{a}\nabla_{a}n_{b} =& n^{a}\nabla_{a}(\xi \alpha \nabla_{b}\tau)\, ,  \\
=& \xi n^{a}\nabla_{b}\tau\nabla_{a}\alpha + \xi\alpha n^{a}\nabla_{a}\nabla_{b}\tau\, ,  \\
=& n^{a}\frac{n_{b}}{\alpha}\nabla_{a} \alpha + \xi\alpha n^{a} \nabla_{b}\nabla_{a}\tau\, ,  \\
=& n^{a} n_{b}\nabla_{a}ln(\alpha) + \alpha n^{a} \nabla_{b} (\frac{n_{a}}{\alpha})\, ,  \\
=& n^{a} n_{b}\nabla_{a}ln(\alpha) -  \alpha n^{a}n_{a}\frac{\nabla_{b}\alpha}{\alpha^{2}}+ n^{a} \nabla_{b}n_{a}\, ,  \\
=& n^{a}n_{b} \nabla_{a} ln(\alpha) - \xi \nabla_{b} ln(\alpha)\, ,  \\
=& -\xi \nabla_{b}ln(\alpha)\left(\delta_{b}{}^{a} - \xi n^{a}n_{b}\right)\, ,  \\
=& -\xi \gamma_{bc}\gamma^{ca}\nabla_{a}ln(\alpha)\, ,  \\
=& -\xi \bar \nabla_{b} ln(\alpha)
\,. \end{align}

Finally, we can see that, since $\gamma^{ab}$ annihilates all vectors parallel to $n_{a}$, we have 

\begin{equation}
\gamma^{am}\gamma^{bn}\nabla_{m}n_{n} = \gamma^{am}\gamma^{bn}\nabla_{m}(\alpha \nabla_{n}\tau) =\alpha \gamma^{am}\gamma^{bn}\nabla_{m}\nabla_{n}\tau
\,. \end{equation}

By the vanishing of the torsion tensor, $K^{ab}$ is symmetric\footnote{Note that although $K^{ab}$ is symmetric, $\nabla_{a}n_{b}$ clearly is not, as should be explicitly clear in equation \eqref {4GradientN}.  It is only the projection of $\nabla_{a}n_{b}$ onto the 3-surface that is symmetric.}, and we have:

\begin{equation}
\nabla_{a}n_{b} =  \xi n_{a}\bar \nabla_{b}\ln(\alpha) - K_{ab} \label{4GradientN}
\,. \end{equation}.

Using these results, we can see that the second line of equation \eqref {threeconnectionstep} vanishes, since it is a 3-dimensional antisymmetrization of \eqref{4GradientN}.  Using the fact that $n^{a}\omega_{a} = 0$, we use the product rule to replace \eqref {threeconnectionstep} with the following.  We then proceed by factoring out the factor of $\omega_{s}$ and multiplying by $\gamma_{sd}$:

\begin{align}
\bar R_{abc}{}^{d}\omega_{d} =& \gamma_{a}^{m}\gamma_{b}^{n}\gamma_{c}^{k}R_{mnk}{}^{d}\omega_{d} +\xi \gamma_{a}^{m}\gamma_{b}^{r}\gamma_{c}^{k}\omega_{s}(\nabla_{m}n_{k})\nabla_{r}n^{s}\nonumber  \\
&-\xi \gamma_{b}^{m}\gamma_{a}^{r}\gamma_{c}^{k}\omega_{s}(\nabla_{m}n_{k})\nabla_{r}n^{s}\, ,  \\
=&\gamma_{a}^{m}\gamma_{b}^{n}\gamma_{c}^{k}R_{mnk}{}^{s}\omega_{s}+\xi K_{ac}K_{b}^{s}\omega_{s} -\xi K_{bc}K_{a}^{s}\omega_{s}\, ,  \\
\bar R_{abcd} =& \gamma_{a}{}^{m}\gamma_{b}{}^{n}\gamma_{c}{}^{k}\gamma^{l}{}_{d}R_{mnkl} -\xi K_{ad}K_{bc}+ \xi K_{ac}K_{bd} \label{DecR}\, ,  \\
\bar R =& \gamma^{ac}\gamma^{bd}R_{abcd} -\xi K^{ab}K_{ab}+\xi K^{2}\, ,  \\
=& R - 2\xi R_{ab}n^{a}n^{b} -\xi K^{ab}K_{ab} +\xi K^{2} \label{Hamiltonianequationtwo}
\,. \end{align}

Having come this far, now we have to assume that we are making a timelike slicing of our spacetime.  For the rest of this section, let us therefore assume that $\xi = -1$, meaning that $n^{a}$ is a timelike vector.  

Now, after making the definition $\rho \equiv n^{a}n^{b}T_{ab}$ and recalling Einstein's equation, we can derive the following identity:

\begin{align}
8\pi T_{ab}n^{a}n^{b} =& R_{ab}n^{a}n^{b} - \frac{1}{2} R g_{ab}n^{a}n^{b} \nonumber \, ,  \\
8 \pi \rho =& \frac{1}{2}(2R_{ab}n^{a}n^{b}-\xi R)\nonumber \, ,  \\ 
16 \pi \rho =& 2 R_{ab}n^{a}n^{b} -\xi R \nonumber \, ,  \\
2 R_{ab}n^{a}n^{b}=& 16 \pi \rho +\xi R \label{Hamiltonianequationone}
\,. \end{align}

Combining equations \eqref {Hamiltonianequationtwo} and \eqref {Hamiltonianequationone}, we obtain the Hamiltonian constraint\footnote{we set $\xi =-1$ in the last step to get the explicit form of the Hamiltonian constraint for a 3+1 split.  Also, in the second to last line, both sides are multiplied through with a factor of $\xi$, and all factors of $\xi^{2}$ are set equal to 1.}:

\begin{align}
\bar R =& R - \xi \left(16 \pi \rho +\xi R\right) +\xi \left(K^{2} - K^{ab}K_{ab}\right) \nonumber \, ,  \\
16 \xi \pi \rho =& - \bar R +\xi \left(-K_{ab}K^{ab} + K^{2}\right) \nonumber \, ,  \\
16 \pi \rho =& - \xi \bar R -K^{ab}K_{ab} + K^{2} \nonumber \, ,  \\
16 \pi \rho =&  \bar R - K^{ab}K_{ab} + K^{2} \label{HamiltonianConstraint}
\,. \end{align}

Now, using a slightly different technique, we can derive the Momentum Constraint.  First, project the Einstein Equation onto $n^{a}\gamma^{b}{}_{c}$, and define $j_{a} \equiv -T_{bc}n^{c}\gamma^{b}{}_{a}$:

\begin{align}
8\pi T_{ab}n^{a}\gamma^{b}{}_{c}=&R_{ab}n^{a}\gamma^{b}{}_{c}-\frac{1}{2}g_{ab}n^{a}\gamma^{b}{}_{c}R\nonumber \, ,  \\
-8\pi j_{c} =& n^{a}\gamma^{b}{}_{c}g^{de}R_{dbea} - \frac{1}{2}n_{b}\gamma^{b}{}_{c}R\nonumber \, ,  \\
=&2\gamma^{b}{}_{c}g^{de}\nabla_{[d}\nabla_{b]}n_{e}\nonumber \, ,  \\
=&2\gamma_{c}{}^{b}\nabla_{[d}\nabla_{b]}n^{d}=\gamma_{c}^{b}\nabla_{d}\nabla_{b}n^{d} - \gamma_{c}{}^{b}\nabla_{b}\nabla_{d}n^{d}\nonumber \, ,  \\
=&\nabla_{d}(\gamma^{b}{}_{c}\nabla_{b}n^{d})-(\nabla_{d}\gamma_{c}^{b})\nabla_{b}n^{d}+\bar \nabla_{c}K\nonumber \, ,  \\
=&-\nabla_{d}K_{c}{}^{d}+\xi n^{b}(\nabla_{d}n_{c})\nabla_{b}n^{d}+\xi n_{c}(\nabla_{d}n^{b})\nabla_{b}n^{d}+\bar \nabla_{c}K\nonumber \, ,  \\
8\pi j_{c} =& \nabla_{d}K^{d}{}_{c} +\xi K_{cd}n^{b}\nabla_{b}n^{d}+\xi n_{c}K_{d}{}^{b}\nabla_{b}n^{d} -\bar \nabla_{c}K\nonumber \, ,  \\
=&g^{bd}\nabla_{d}K_{bc} -\xi  n^{b}n^{d}\nabla_{b}K_{cd}-\xi n_{c}n^{d}\nabla_{b}K^{b}{}_{d}-\bar \nabla_{c}K \label{Momentumstepone}\, ,  \\
=&\left(g^{bd}-\xi n^{b}n^{d}\right)\left(g_{c}{}^{e}-\xi n_{c}n^{e}\right)\nabla_{d}K_{be}-\bar \nabla_{c}K \label{Momentumsteptwo}\, ,  \\
=&\bar \nabla_{b}K^{b}{}_{c} - \bar \nabla_{c}K\nonumber \, ,  \\
8\pi j^{a} =& \bar \nabla_{b}(K^{ab} - \gamma^{ab}K) \label{MomentumConstraint}
\,. \end{align}

Here, Equation \eqref {MomentumConstraint} is the momentum constraint, and in moving from equation \eqref {Momentumstepone} to equation \eqref {Momentumsteptwo}, we used the fact that $n^{a}n^{b}n^{c}n_{d}\nabla_{a}K_{bc}=n^{a}n^{c}n_{d}\nabla_{a}(n^{b}K_{bc}) - n^{a}n^{c}K_{bc}n_{d}\nabla_{a}n^{b}=0-0$\footnote{Note that the common factor here is included simply because it also appears in the derivation of Equation \eqref{MomentumConstraint}} .  We have now derived expressions for the Constraint equations for the standard ADM 3+1 split. 

\section{Evolution Equations}
Now, we will derive expressions for the ADM evolution equations.  Once again, rather than going with the more elegant Hamiltonian framework, we are instead going to explicitly evaluate the time derivative (in this case, the Lie derivative along the vector $t^{a}$ defined in equation \eqref {DefT}) of our evolution variables $\gamma_{ab}$ and $K_{ab}$.  First, let us find the evolution equation for $\gamma_{ab}$:

\begin{align}
£_{t}\gamma_{ab} =& £_{\alpha n}\gamma_{ab} + £_{\beta}\gamma_{ab}\, ,  \\
=& \alpha n^{c}\nabla_{c}\gamma_{ab} + \gamma_{cb}\nabla_{a}(\alpha n^{c}) +\gamma_{ac}\nabla_{b}(\alpha n^{c}) + \beta^{c}\bar \nabla_{c}\gamma_{ab} \nonumber  \\
&+ \gamma_{cb}\bar \nabla_{a}\beta^{c} +\gamma_{ac}\bar \nabla_{b} \beta^{c}\, ,  \\
=&\alpha[n_{a}n^{c}\nabla_{c}n_{b}+n_{b}n^{c}\nabla_{c}n_{a}+\nabla_{a}n_{b}+\nabla_{b}n_{a}]+2\bar \nabla_{(b}\beta_{a)}\, ,  \\
=&\alpha[(n_{a}n^{c}+g_{a}{}^{c})\nabla_{c}n_{b} + (n_{b}n^{c}+g_{b}{}^{c})\nabla_{c}n_{a}]+2\bar \nabla_{(b}\beta_{a)}\, ,  \\
=&-2\alpha K_{ab} + 2 \bar \nabla_{(a}\beta_{b)} \label{Gevolution}
\,. \end{align}

Since it is expressed entirely in terms of 3-dimensional quantities (namely, we've eliminated any reference to $n^{a}$), we can consider equation \eqref {Gevolution} to be the evolution equation for $\gamma_{ab}$.  Note that no reference to the Einstein Equation was required in this derivation.  We can consider this evolution equation, therefore, to be more of a definition of the relationship between $t^{a}$, $K_{ab}$ and $\beta_{a}$ than anything else.  

Now, let us derive an expression for the evolution equation for $K_{ab}$:

\begin{align}
£_{t}K_{ab}=&£_{\alpha n}K_{ab} + £_{\beta}K_{ab}\, ,  \\
=&£_{\alpha n}K_{ab} + \beta^{c} \bar \nabla_{c}K_{ab} +K_{cb}\bar \nabla_{a}\beta^{c} + K_{ac}\bar \nabla_{b}\beta^{c}\, ,  \\
(£_{t}-£_{\beta})K_{ab}=&\alpha n^{c}\nabla_{c}K_{ab}+K_{cb}\nabla_{a}(\alpha n^{c}) + K_{ac}\nabla_{b}(\alpha n^{c})\, ,  \\
=&\alpha[-n^{c}\nabla_{c}(\gamma_{a}^{d}\gamma_{b}^{e}\nabla_{d}n_{e})+K_{cb}\nabla_{a}n^{c}+K_{ac}\nabla_{b}n^{c}]\, ,  \\
=&\alpha[-n^{c}\gamma_{a}^{d}\gamma_{b}^{e}\nabla_{c}\nabla_{d}n_{e}-n^{c}(\nabla_{c}\gamma_{b}^{e})\gamma_{a}^{d}\nabla_{d}n_{e}\nonumber  \\
&-n^{c}(\nabla_{c}\gamma_{a}^{d})\nabla_{d}n_{b}-K_{cb}K_{a}^{c}-K_{cb}n_{a} \bar \nabla^{c}ln(\alpha)\nonumber  \\
&-K_{ac}K_{b}^{c}-K_{ac}n_{b}\bar \nabla^{c}ln(\alpha)]\, ,  \\
=&\alpha[-n^{c}\gamma_{a}^{d}\gamma_{b}^{e}R_{cde}{}^{f}n_{f}-n^{c}\gamma_{a}^{d}\gamma_{b}^{e}\nabla_{d}\nabla_{c}n_{e}\nonumber  \\
&+n^{c}n_{b}(\nabla_{c}n^{e})K_{ae}-n^{c}n^{d}(\nabla_{c}n_{a})\nabla_{d}n_{b}\nonumber  \\
&-n^{c}n_{a}(\nabla_{c}n^{d})\nabla_{d}n_{b}-2K_{ac}K_{b}^{c}]-n_{a}K_{b}^{c}\bar \nabla_{c}\alpha\nonumber  \\
& -n_{b} K_{a}^{c} \bar \nabla_{c} \alpha\, ,  \\
=&\alpha[(\gamma^{cf}-g^{cf})\gamma_{a}^{d}\gamma_{b}^{e}R_{dcef}-\gamma_{a}^{d}\gamma_{b}^{e}\nabla_{d}(n^{c}\nabla_{c}n_{e})\nonumber  \\
&+\gamma_{a}^{d}\gamma_{b}^{e}(\nabla_{d}n^{c})\nabla_{c}n_{e}+n_{b}K_{ae}\bar \nabla^{e}ln(\alpha)\nonumber  \\
&-(\bar \nabla_{a}ln(\alpha))\bar \nabla_{b}ln(\alpha)+n_{a}K_{db}\bar \nabla^{d}ln(\alpha)-2K_{ac}K^{c}{}_{b}]\nonumber  \\
&-n_{a}K^{c}{}_{b}\bar \nabla_{c}\alpha -n_{b}K_{a}{}^{c}\bar \nabla_{c}\alpha\, ,  \\
=&\alpha[\gamma_{a}{}^{l}\gamma^{nk}\gamma_{b}{}^{m}R_{lnmk}-\gamma_{a}{}^{m}\gamma_{b}{}^{n}R_{mn}\nonumber  \\
&-\gamma_{a}{}^{d}\gamma_{b}{}^{e}\nabla_{d}(\bar \nabla_{e}ln(\alpha))+K_{a}{}^{c}K_{cb}-2K_{ac}K^{c}{}_{b}]\nonumber  \\
&-\frac{1}{\alpha}(\bar \nabla_{a}\alpha)\bar \nabla_{b}\alpha \displaybreak[0]\, ,  \\
=&\alpha[\gamma_{a}{}^{l}\gamma^{nk}\gamma_{b}{}^{m}R_{lnmk}-\gamma_{a}{}^{m}\gamma_{b}{}^{n}(8\pi T_{mn}+\frac{1}{2}Rg_{ab}) \label{DefS} \nonumber  \\
&-K_{ac}K^{c}{}_{b}]-\frac{1}{\alpha}(\bar \nabla_{a}\alpha)\bar \nabla_{b}\alpha-\alpha \bar \nabla_{a}(\frac{1}{\alpha}\bar \nabla_{b}\alpha)\, ,  \\
=&\alpha[\bar R_{ab}-K_{ac}K^{c}{}_{b}+K K_{ab}-8 \pi S_{ab}-\frac{1}{2}R \gamma_{ab} \label{Ridentity}\nonumber  \\
&-K_{ac}K^{c}{}_{b}]-\bar \nabla_{a}\bar \nabla_{b}\alpha\, ,  \\
=&\alpha[\bar R_{ab} - 2K_{ac}K^{c}{}_{b}+KK_{ab} - 8 \pi S_{ab} +4\pi\gamma_{ab}(S-\rho)]\nonumber  \\
&-\bar \nabla_{a}\bar \nabla_{b}\alpha\, ,  \\
£_{t}K_{ab}=&\alpha[\bar R_{ab} - 2K_{ac}K^{c}{}_{b}+KK_{ab} - 8 \pi S_{ab} +4\pi\gamma_{ab}(S-\rho)]\nonumber  \\
&-\bar \nabla_{a}\bar \nabla_{b}\alpha+ \beta^{c} \bar \nabla_{c}K_{ab} +K_{cb}\bar \nabla_{a}\beta^{c} + K_{ac}\bar \nabla_{b}\beta^{c} \label{Kevolution}
\,. \end{align}

In going from lines \eqref {DefS} to the next line, the definition $T_{ab}\gamma^{a}{}_{m}\gamma^{b}{}_{n} \equiv S_{mn}$ was made, and the result from equation \eqref {DecR} was used.  In going forward from \eqref {Ridentity}, we contracted the Einstein Equation on $g^{ab}$ in the following way:

\begin{align}
g^{ab}R_{ab} - \frac{1}{2} g^{ab}g_{ab}R =& 8\pi g^{ab}T_{ab}\, ,  \\
R-2R =&8\pi (\gamma^{ab}-n^{a}n^{b})T_{ab}\, ,  \\
-R=&8 \pi (S- \rho)
\,. \end{align}

Since, once again, equation \eqref {Kevolution} depends only on 3-dimensional quantities, we can consider it to be the proper evolution equation for $K_{ab}$, and we now have the complete set of ADM equations, namely equations \eqref {Kevolution}, \eqref {Gevolution}, \eqref {HamiltonianConstraint} and \eqref {MomentumConstraint}, and this concludes our ``brute force'' derivation of the ADM equations.

\chapter{Hamiltonian General Relativity}
\label{sec:Hamiltonian General Relativity}
Now, we above performed the brute force derivation of the ADM equations for a few reasons.  First, there was the goal of seeing true completeness to our derivation--there was no direct reference to non-geometric quantities above--we were able to show that the ADM equations are a direct consequence of merely the Einstein Equation and the rules of Lorentizan geometry.  Second, the above procedure generalizes quite nicely to a foliation of a spacelike two surface embedded in a spacelike three surface--the relevant situation to many initial value boundary problems in General Relativity.  

This is not to say, however, that the more standard Hamiltonian approach to this derivation is without merit.  It is only through this approach that we can investigate several important issues--in particular, the Hamiltonian formalism enables us to investigate the angular and linear momentum stored in a spacetime, in addition to that spacetime's net energy.  The brute force approach given above necessarily only deals directly with bulk terms--these are things that are best viewed through an action principle.  So, here, we will work through this action principle in general relativity.
\section{From the Lagrangian to the Hamiltonian}
We begin with the well-known (see, for instance, \cite{MTW}) Hilbert action:

\begin{equation}
S = \frac{1}{16\pi G}\int \sqrt{-g}_{ } R_{ } d^{4}x \; .\label{HilbertAction}
\,. \end{equation}

Here, $g_{ab}$ is the four-dimensional Minkowskian metric of the spacetime, $g$ is the determinant of $g_{ab}$, and R is the trace of its Ricci curvature.  We wish to take this Lagrangian density and form a Hamiltonian density with proper equations of motion.  It is a known result that no second time derivatives of the $g_{ta}$ appear in the action \eqref{HilbertAction}, and therefore, if we were to take all of the components $g_{ab}$ to be our canonical variables, we would then define our canonical momenta to be $P^{ab} \equiv \frac{\delta S}{\delta \dot g_{ab}}$, but then we would be stuck, because since the Hessian

\begin{equation}
H^{abcd} \equiv \frac{\delta^{2}S}{\delta \dot g_{ab} \delta \dot g_{cd}}= \frac{\delta P^{ab}}{\delta \dot g_{cd}}
\,. \end{equation}

\noindent would have a vanishing determinant \footnote{To see this, if we set a and c to t, then we know that the Hessian is zero, regardless of the value of c and d (there are no second time derivatives of the $g_{ta}$).  This, in turn, then indicates that an entire row of the Hessian Matrix vanishes, which then implies that the determinant of the Hessian vanishes}, meaning that it would be impossible to solve for all of the $P^{ab}$ in terms of all of the $\dot g_{ab}$.  In particular, in transforming the variables of our system from the $g_{ab}$ and the $\dot g_{ab}$ to the $g_{ab}$ and the $P^{ab}$, there would be a noninvertible transformation operator, meaning that degrees of freedom would be lost in going from the $g_{ab}$ to the $P^{ab}$.  The Lagrangian and Hamiltonian formulations of this theory would therefore not be compatible.  

In order to avoid this problem, we reduce the number of variables in our system to the $g_{ij}$, which, as shown in section \ref{sec: Definitions}, are exactly equal to the 3-components of the 3-dimensional metric $\gamma_{ab}$ if we were to do a 3+1 decomposition using $t$ as our slicing variable \footnote{of course, if we want \textit{another} slicing, we can make this choice simply by coordinate transforming $t$ to some other, more general function $\tau$ with a timelike gradient}.   Then, using Equations \eqref {DefAlpha} and \eqref {DefBeta}
 to define $\alpha$ and $\beta_{i}$, a straightforward calculation relates $g$ to the determinant $\gamma$ of $\gamma_{ab}$:
 
 \begin{equation}
 g = -\alpha^{2} \gamma
 \,. \end{equation}

Then, we use Equation \eqref {Hamiltonianequationone} to replace $R$ with $\bar R$, giving us:
 
 \begin{equation}
 S = \frac{1}{16\pi G}\int \alpha \sqrt{\gamma}\left( \bar R +\xi K_{ab}K^{ab} -\xi K^{2} + 2\xi R_{ab}n^{a}n^{b} \right) _{} d^{4}x  \label{MidAction}
 \,. \end{equation} 
 
 Since $\bar R$ manifestly has no time derivatives in it, and $\frac{\delta K_{ab}}{\delta \dot g_{cd}} = \frac{1}{2}(\delta_{a} {}^{c} \delta_{b} {}^{d} + \delta_{a}{} ^{d} \delta_{b} {}^{c} )(-2\alpha)$, we can see that \eqref {MidAction} brings us closer to the canonical form necessary to Legendre transform the Hilbert action: every term except for the last either has no time derivatives, or is expressible in terms of the $\dot g_{ij}$.  So, what of that fourth term?  Let us investigate it further:
 
 \begin{align}
 n^{a}n^{b}R_{ab} =& n^{a}n^{b}g^{cd}R_{cadb}\, ,  \\
 =& n^{a}g^{cd}\left( \nabla_{c}\nabla_{a}n_{d} - \nabla_{a}\nabla_{c}n_{d} \right)\, ,  \\
 =& n^{a} \left(\nabla_{c}\nabla_{a} n^{c} -\nabla_{a}\nabla_{c}n^{c} \right)\, ,  \\
 =& \nabla_{c}\left(n^{a}\nabla_{a}n^{c}\right) - \left(\nabla_{c}n^{a}\right)\nabla_{a}n^{c} - \nabla_{a}\left(n^{a}\nabla_{c}n^{c}\right) + \left(\nabla_{a}n^{a}\right)\nabla_{c}n^{c} \nonumber \, ,  \\
 =& \nabla_{a}\left(-\xi \bar \nabla^{a} ln(\alpha)\right)+\nabla_{a}\left(n^{a} K\right)-K^{ab}K_{ab} + K^{2}   \label{RicciTrick}
 \,. \end{align}

Before we progress further, we will quickly derive two identities.  The first simplifies the term involving the lapse function:

\begin{align}
\nabla_{a}\bar \nabla^{a} ln(\alpha) =&\nabla_{a} \left(\gamma^{ab}\nabla_{b} ln(\alpha)\right) \nonumber \, ,  \\
=&\nabla_{a}\left( \frac{1}{\alpha}\gamma^{ab}\nabla_{b}\alpha \right)\nonumber \, ,  \\
=& \frac{1}{\alpha}\gamma^{ab}\nabla_{a}\nabla_{b}\alpha + \frac{1}{\alpha}(\nabla_{b}\alpha)\nabla_{a}\gamma^{ab} - \frac{1}{\alpha^{2}}\gamma^{ab}(\nabla_{a}\alpha)(\nabla_{b}\alpha)\nonumber \, ,  \\
=& \frac{1}\alpha \left[\gamma^{ac}\gamma_{c}{}^{b}\nabla_{a}\nabla_{b}\alpha + \nabla_{b}\alpha \nabla_{a}\gamma^{ab} -\frac{1}{\alpha}(\bar \nabla^{a}\alpha)\bar \nabla_{a}\alpha\right]\nonumber\, ,  \\
=& \frac{1}{\alpha}\left[\gamma^{ac}\nabla_{a}\left(\gamma_{c}{}^{b}\nabla_{b}\alpha\right) - \gamma^{ac}\left(\nabla_{a}\gamma_{c}{}^{b}\right)\nabla_{b}\alpha+(\nabla_{b}\alpha)\nabla_{a}\gamma^{ab} \right.\nonumber  \\
&\left.- \frac{1}{\alpha}(\bar \nabla_{a}\alpha)\bar \nabla^{a}\alpha\right] \nonumber \, ,  \\
=& \frac{1}{\alpha}\left[ \bar \nabla^{2}\alpha +\xi \gamma^{ac}\nabla_{b}\alpha(n_{c}\nabla_{a}n^{b} +  n^{b}\nabla_{a}n_{c}) \right.\nonumber  \\
&\left.- \xi(\nabla_{b}\alpha)(n^{a}\nabla_{a}n^{b} + n^{b}\nabla_{a}n^{a})-\frac{1}{\alpha}(\bar \nabla^{a} \alpha)\bar \nabla_{a}\alpha \right] \nonumber \, ,  \\
=& \frac{1}{\alpha}\left[\bar \nabla^{2}\alpha -0- \xi K n^{b}\nabla_{b}\alpha + \xi^{2} \nabla_{b}\alpha \bar \nabla^{b} ln(\alpha) + \xi (\nabla_{b}\alpha)n^{b}K \right.\nonumber   \\*
&\left.-\frac{1}{\alpha}(\bar \nabla^{a}\alpha)\bar \nabla_{a}\alpha \right]\nonumber\, ,  \\
=& \frac{1}{\alpha} \bar \nabla^{2}\alpha  \label{AlphaBoundaryTerm}
\,. \end{align}

Which eliminates the 4-divergences of $\alpha$ and replaces them with 3-divergences.
\newline
The second identity simplifies the term that involves the total divergence of $n^{a}K$\footnote{Here, we consider the boundary to $\mathbb{M}$ to be divided into three parts--the initial and final timeslices of our evolution--namely, the endpoints of the function $\tau$, and a part topologically equivalent to $\mathbb{R}\times \partial \mathbf{m}$.  We will worry about the latter below, and take the former only as an integral over initial and final timeslices, which we shall neglect, as the initial and final states of our variation will be considered fixed.}  In the following derivation double overbarred terms like $\bar {\bar {\gamma}}$ will refer to the metric intrinsic to the boundary to the 4-dimensional manifold $\mathbb{M}$.\footnote{Also note that following the argument by York outlined in Appendix 
 we must make a distinction between the normal to boundary in the 4-manifold and its projection onto a 3-manifold.  The former will be labeled as $s_{a} = B \nabla_{a}R$, and the latter will be labeled as $\bar s_{a} = \bar B \bar \nabla_{a}R$.  Note that while the two are defined by the same function R, they are quite distinct objects--the latter is normal to the unit timelike normal $n^{a}$, and the normalization constants are fixed, respectively, by the 4-metric $g^{ab}$ and the 3-metric $\gamma^{ab}$.}

\begin{align}
\int d^{4}x \sqrt{-g} \nabla_{a}\left(n^{a}K\right) =& \oint d^{3}x \sqrt{\bar {\bar {\gamma}}}s_{a}n^{a}K \nonumber \, ,  \\
=& \oint d^{3}x\left(\frac{\sqrt{-g}}{B}\right)K\,g^{ab}n_{a}\,s_{b} \nonumber \, ,  \\
=&\oint d^{3}x\,\alpha^2 \sqrt{\gamma}K\,g^{ab}\left(\nabla_{a}\tau\right)\nabla_{b}R \nonumber\, ,  \\
=&\oint d^{3}x\,\alpha^{2}\sqrt{\gamma}K\,\left(-\frac{1}{\alpha^{2}}\delta^{b}{}_{\tau}+\frac{1}{\alpha^{2}}\beta^{b}\right)\nabla_{b}R\nonumber \, ,  \\
=&\int dt \oint d^{2}x \bar B \sqrt{q}\,K\,\beta^{a}\bar \nabla_{a}R \nonumber \, ,  \\
=&\int dt \oint d^{2}x \sqrt{q}\, K \bar s_{a}\beta^{a} \nonumber \, ,  \\
=&\int dt \int d^{3}x \sqrt{\gamma} \bar \nabla_{a}\left(\beta^{a}\,K\right) \nonumber \, ,  \\
=&\int d^{4} x \sqrt{-g} \frac{1}{\alpha} \bar \nabla_{a} \left(\beta^{a}\,K\right)\label{KtoBetaTerm}
\,. \end{align}

So, therefore, this term is also equivalent to an appropriate integral over $\partial\mathbf{m}$ integrated over time, or equivalently, a 3-divergence integrated over the 4-manifold.  
\newline
Now, we can combine equations \eqref {AlphaBoundaryTerm} and \eqref {KtoBetaTerm} with their corresponding terms in \eqref {RicciTrick} to get:

\begin{equation}
n^{a}n^{b}R_{ab}=-\xi \frac{1}{\alpha} \bar \nabla^{2}\alpha + \frac{1}{\alpha} \bar \nabla_{a}\left(\beta^{a}\,K\right) -K^{ab}K_{ab} + K^{2}\label{YetMoreEquationeering}
\,. \end{equation}
 \newline
 Now, putting \eqref {YetMoreEquationeering} back into the fourth term of \eqref {MidAction} we get that the term involving $R_{ab}n^{a}n^{b}$ reduces to:
 
 \begin{align}
\frac{1}{16\pi G} &\int \sqrt{|g|} d^{4}{}x{} \left(2R_{ab}n^{a}n^{b}\right) =\nonumber \, ,  \\
&= \frac{2}{8 \pi G}\int d^{4}{}x \alpha \sqrt{\gamma} \left[-\xi \bar \nabla^{2}\alpha + \nabla_{a}\left(n^{a}K\right) - K^{ab}K_{ab} + K^{2}\right]\nonumber   \\
&+\frac{1}{8 \pi G}\oint d^{3}{}x s_{a}n^{a} K \nonumber\, ,  \\
&= \frac{1}{8 \pi G}\oint d^{3}{}x s_{a}n^{a} K + \frac{1}{16 \pi G}\int d^{4}x\,\alpha \sqrt{\gamma}\left[ 2K^{2}-2K^{ab}K_{ab} -\frac{2\xi}{\alpha} \bar \nabla^{2}\alpha \right]\,. \label{ActionJackson}  
\,. \end{align}

Finally, we can substitute \eqref{ActionJackson} into the Hilbert action \eqref {MidAction} to obtain

\begin{equation}
S =\frac{1}{16\pi G}\int d^{4}x{}\sqrt{\gamma}\left[\alpha\left(\bar R - K^{2} + K^{ab}K_{ab}\right) +2 \bar \nabla^{2}\alpha\right]+\frac{1}{8 \pi G}\oint d^{3}{}x s_{a}n^{a} K\,.\label{FinalAction}
\,. \end{equation}

Where we have taken $\xi=-1$ again to reflect the timelike slicing of our spacetime in the bulk.  However, the explicit inclusion of the factor of $\xi$ in going to equation \eqref {ActionJackson} shows us that there is no sign difference in the boundary term, no matter whether we are evaluating the boundary term on the initial timeslice or on the outer boundary of our spacetime.  This is fortunate, as it avoids any contradictions on the intersection of the initial timeslice and the outer boundary.  
 
 Now, we can (due to our assumption of topology $\mathbb{R} \times \mathbf{m}$ for our spacetime) remove the integral over our time coordinate $\tau$ from \eqref {FinalAction}, and infer the Lagrangian density\footnote{Note that the text has replaced the normal $n^{a}$ with a term $r^{a}$.  We do this in order to indicate that $r_{a}$ is explicitly unit spacelike, as we mean for the Lagrangian density to be defined on a single timeslice, which has a timelike outer boundary, and therefore, this boundary has a spacelike normal.  Also note that we have converted the boundary term in \eqref{FinalAction} with a divergence.} :
 
 \begin{equation}
 \mathscr{L} = \frac{1}{16\pi G}  \alpha \sqrt{\gamma}(\bar R + K^{ab}K_{ab} -K^{2}) + \frac{1}{8 \pi G} \sqrt{\gamma} \nabla_{c}(r^{c}K) \label{HilbertLagrangian}
 \,. \end{equation}
 Note that \eqref {HilbertLagrangian} is now written explicitly in terms of $\gamma_{ab}$ and $K_{ab}$, which is directly related to $£_{t}\gamma_{ab}$, so we have successfully separated out the ``good'' canonical variables from the Lagrange multipliers.  Therefore, we can now feel safe defining conjugate momenta and taking a Legendre transform.   Recalling from the ADM equation for $\gamma_{ab}$, we have $£_{t}\gamma_{ab} = \dot \gamma_{ab} = -2 \alpha K_{ab}$, we therefore can now define the momentum conjugate to $\gamma_{ab}$:
 
 \begin{align}
 \Pi^{ab} \equiv& \frac{\delta \mathscr{L}}{\delta\dot \gamma_{ab}} \, ,  \\
 =& \frac{1}{8 \pi G} \alpha \sqrt{\gamma}(\frac{1}{2\alpha})(K \gamma^{ab} - K^{ab}) + \frac{1}{4\pi G}\frac{\delta}{\delta \dot \gamma_{ab}} \sqrt{\gamma} \nabla_{c}(r^{c}\gamma^{ab}K_{ab}) \label{StUfF}\, ,  \\
 =& \frac{1}{16\pi G} \sqrt{\gamma}(K\gamma^{ab} - K^{ab})\,. \label{DefPi}
 \,. \end{align}

Where we dropped the $\nabla_{c}\left(r^{c}K\right)$ term in equation \eqref{StUfF} due to the fact that since $\bar \nabla_{c}\gamma^{ab}=0$, we therefore have, after integrating by parts\footnote{
Note that while it is true that $\frac{\delta K_{ab}}{\delta \dot \gamma_{mn}} = \delta_{ab}^{mn}$, this is not true of $\frac{\delta }{\delta \dot \gamma_{mn}}\nabla_{c} K_{ab}$.  In the latter case, it is appropriate to interchange the functional derivative with the covariant derivative and integrate by parts.  This originates from the fact that a formal definition of the variational derivative is based in taking terms proportional to differentials of functions.  The entire functional derivative should be the multiplier of the differential of the function, which does not work if it is inside of a derivative.}:


\begin{align}
\lefteqn{\frac{\delta}{\delta \dot \gamma_{ab}}\int d^{3}{}x \sqrt{\gamma}\bar \nabla_{c}(r^{c}\gamma^{de}K_{de})=}\, ,  \\
=& \frac{\delta}{\delta \dot \gamma_{ab}}\int d^{3}{}x \sqrt{\gamma}\gamma^{de}\bar \nabla_{c}(r^{c}K_{de})\, ,  \\
=&\frac{\delta}{\delta \dot \gamma_{ab}}\int d^{3}{}x \sqrt{\gamma}\gamma^{de}(K_{de}\bar \nabla_{c}(r^{c})+r^{c}\bar \nabla_{c}K_{de})\, ,  \\
=&\int d^{3}{}x \sqrt{\gamma}(\gamma^{de}(\bar \nabla_{c}r^{c})\frac{\delta K_{de}}{\delta \dot \gamma_{ab}}-\bar \nabla_{c}(\gamma^{de} r^{c})\frac{\delta K_{de}}{\delta \dot \gamma_{ab}})\, ,  \\
=&0
\,. \end{align}

So, now, we have our canonical variables $\gamma_{ab}$, their conjugate momenta $\Pi^{ab}$, and we are left with four Lagrange multipliers in $\alpha$ and the $\beta_{a}$.  Inverting equation \eqref {DefPi} to obtain a expressions for $K$ and $K^{ab}$ in terms of $\Pi^{ab}$ gives us:

\begin{equation}
K= \frac{8\pi G}{\sqrt{\gamma}}\Pi\;\;\;\;\;\;\;\;\;\;\;\;\;\;\;\;K^{ab} = \frac{16\pi G}{\sqrt{\gamma}}(\frac{1}{2}\gamma^{ab}\Pi - \Pi^{ab})    \label{InvertedPi}
\,. \end{equation}

An obvious consequence of this is that:

\begin{align}
K^{2} =& \left(\frac{8\pi G}{\sqrt{\gamma}}\right)^{2}\Pi^{2} \, ,  \\
K^{ab}K_{ab} =& \left(\frac{8\pi G}{\sqrt{\gamma}}\right)^{2}(\gamma^{ab}\Pi - 2\Pi^{ab})(\gamma_{ab}\Pi - 2\Pi_{ab}) \nonumber \, ,  \\
=& \left(\frac{8\pi G}{\sqrt{\gamma}}\right)^{2}(3\Pi^{2} - 4\Pi^{2} + 4 \Pi^{ab}\Pi_{ab}) \nonumber \, ,  \\
=&\left(\frac{8\pi G}{\sqrt{\gamma}}\right)^{2}(4\Pi^{ab}\Pi_{ab} - \Pi^{2})
\,. \end{align}

Giving us the useful result:

\begin{equation}
K^{ab}K_{ab} - K^{2} = \left(\frac{16\pi G}{\sqrt{\gamma}}\right)^{2}(\Pi^{ab}\Pi_{ab} - \frac{1}{2}\Pi^{2})
\,. \end{equation}

Knowing this\footnote{And note, that if we had chosen to try and write a Canonical theory based on the $g_{ab}$ and the $\dot g_{ab}$ as our variables, this program would have failed here, as we would not have been able to invert equation \eqref {DefPi} in order to get equation \eqref{InvertedPi}, and therefore, we would not have been able to write down a Hamiltonian in terms of the $g_{ab}$ and the $\Pi^{ab}$} we are ready to take the Legendre transform of our Lagrangian density $\mathscr{L}$ in order to define a Hamiltonian density $\mathscr{H}$:

\begin{equation}
\mathscr{H} \equiv \Pi^{ab}\dot \gamma_{ab} - \mathscr{L}
\,. \end{equation}

Remembering from equation \eqref {Gevolution} that $£_{t}\gamma_{ab}=\dot \gamma_{ab} = -2\alpha K_{ab} + 2 \bar \nabla_{(a}\beta_{b)}$, we can then explicitly perform this procedure:

\begin{align}
\mathscr{H} =& \Pi^{ab}\dot \gamma_{ab} - \mathscr{L} \nonumber \, ,  \\
=&\Pi^{ab}\left(-2\alpha K_{ab} + 2 \bar \nabla_{(a}\beta_{b)}\right)-\left( \frac{1}{16\pi G}  \alpha \sqrt{\gamma}(\bar R + K^{ab}K_{ab} -K^{2} \right.\nonumber   \\
&\left.+\frac{1}{\alpha}\bar \nabla^{2}\alpha+ \frac{1}{8 \pi G} \sqrt{\gamma} \bar \nabla_{c}(r^{c}K)  \right) \nonumber \, ,  \\
=& -2\alpha \Pi^{ab}\left(\left(\frac{16\pi G}{\sqrt{\gamma}}\right)\left(\frac{1}{2}\gamma_{ab}\Pi - \Pi_{ab}\right)\right) + 2 \Pi^{ab} \bar \nabla_{(a} \beta_{b)} - \frac{\alpha \sqrt{\gamma}}{16 \pi G}\bar R \nonumber    \\
&- \frac{\sqrt{\gamma}}{16\,\pi\,G}\bar \nabla^{2}\alpha- \alpha \left(\frac{16 \pi G}{\sqrt{\gamma}}\right)\left(\Pi^{ab}\Pi_{ab} - \frac{1}{2}\Pi^{2}\right) - \frac{1}{8 \pi G} \sqrt{\gamma}\nabla_{a}\left(n^{a}K\right) \nonumber \, ,  \\
=&\alpha\left(\frac{16 \pi G}{\sqrt{\gamma}}\right)\left(\Pi^{ab}\Pi_{ab} - \frac{1}{2} \Pi^{2}\right)-\frac{\alpha \sqrt{\gamma}}{16 \pi G}\bar R + 2\Pi^{ab}\bar \nabla_{(a}\beta_{b)} \nonumber  \\
&-\frac{\sqrt{\gamma}}{16\,\pi\,G}\bar \nabla^{2}\alpha - \frac{1}{8 \pi G} \sqrt{\gamma}\nabla_{a}\left(n^{a}K\right) \label{ADM Bulk Hamiltonian}
\,. \end{align}

Which is the ADM Hamiltonian density, as found originally by Arnowitt, Deser and Misner \cite{ADMPaper}.  A 3-dimensional integral of this quantity would then yield the ADM Hamiltonian.  Note that it includes boundary terms, and that these boundary terms must be treated properly.  In order to do this, one must be quite careful about the proper treatment of boundary {\it charges}.  In particular, it can be shown that the the above Hamiltonian can lead to inconsistencies if just na\''ively varied.  The reader is referred to the \ref{sec: Gauss's Theorem} for some background on properly treating these boundary terms.
\section{Constraints}
Now, we take the variation of the Hamiltonian density in order to find equations of motion.  Total divergences will be neglected in this section, but will be treated explicitly according to the perspective of Chapter \ref{sec: Gauss's Theorem}\footnote{In this Chapter, we will use what we know about the bulk equations of motion in order to pick the appropriate boundary terms.  Therefore, it is not explicitly necessary to carefully and faithfully preserve the boundary terms dictated by the Hilbert action.  Instead, we faithfully maintain the appropriate behaviour {\it in the bulk}, and then let that behaviour dictate what our boundary terms should be}.  Note that the above Hamiltonian must be modified by the addition of a series of surface charges in order for the overall variation to be zero.  

Having said all of this, it is now easy to derive the constraint equations.  The Hamiltonian constraint is derived by taking the variation of equation \eqref {ADM Bulk Hamiltonian} with respect to $\alpha$.  Neglecting the boundary terms, the result is:

\begin{align}
\frac{\delta \mathscr{H}}{\delta \alpha} =&0=\left(\frac{16 \pi G}{\sqrt{\gamma}}\right)\left(\Pi^{ab}\Pi_{ab} - \frac{1}{2}\Pi^{2}\right)-\frac{\sqrt{\gamma}}{16 \pi G}\bar R \nonumber \, ,  \\
=& - \left(\frac{\sqrt{\gamma}}{16 \pi G}\right)\left(\bar R - K^{ab}K_{ab} + K^{2}\right) \label{partial alpha variation}
\,. \end{align}

If it is desired, we could have added a matter density to the original Lagrangian, which typically would have been a function of $g^{ab}$ and $g_{ab}$, but not their derivatives.  Had we done this, we would have gone through the same (3+1) decomposition and Legendre transform process with these degrees of freedom\footnote{This procedure is traced in detail for a Klein-Gordon field coupled to an external gravitational field in Appendix \eqref {sec: Klein-Gordon 3+1}}.  This would give a well-defined matter Hamiltonian $\mathscr{H}_{m}$, which we could add to the Hamiltonian given in equation \eqref {ADM Bulk Hamiltonian}.  Making the definition $\rho =\left(\frac{1}{\sqrt{\gamma}}\right) \frac{\partial \mathscr{H}_{m}}{\partial \alpha}$ then makes equation \eqref {partial alpha variation} equal to:

\begin{equation}
\frac{\delta \mathscr{H}_{total}}{\delta \alpha} =0= \frac{\delta \mathscr{H}_{Einstein}}{\delta \alpha}+\frac{\delta \mathscr{H}_{m}}{\delta \alpha} = \sqrt{\gamma}\left[\rho - \left(\frac{1}{16 \pi G}\right)\left(\bar R -K^{ab}K_{ab} + K^{2}\right)\right]
\,. \end{equation}

Which is the Hamiltonian constraint, as seen in equation \eqref {HamiltonianConstraint}.  

Similarly, we can derive the Momentum constraint, but taking the variation of equation \eqref {ADM Bulk Hamiltonian} with respect to $\beta_{a}$.  As we do in Chapter \ref{sec: Gauss's Theorem}, we do a little bit of algebra gymnastics, and ultimately, we discard the boundary term, in anticipation of dealing with it in Chapter \ref{sec: Gauss's Theorem}.  Since $\beta_{a}$ only appears in a single term above, we only worry about that single term.  Furthermore, since $\Pi^{ab}$ is a symmetric tensor, the symmetrization over the derivative of $\beta_{a}$ is omitted as irrelevant.

\begin{align}
\frac{\delta \mathscr{H}}{\delta \beta_{b}} =& 0=\frac{\delta \mathscr{H}_{Einstein}}{\delta \beta_{b}}+\frac{\delta \mathscr{H}_{m}}{\delta \beta_{b}} \nonumber \, ,  \\
=&\frac{\delta}{\delta \beta_{b}}\left( 2 \Pi^{ab} \bar \nabla_{a} \beta_{b} \right)+\frac{\delta \mathscr{H}_{m}}{\delta \beta_{b}}\nonumber \, ,  \\ 
=& \frac{\delta}{\delta \beta_{b}}\left(\left(\frac{\sqrt{\gamma}}{\sqrt{\gamma}}\right)2\Pi^{ab}\bar \nabla_{a} \beta_{b}\right)+\frac{\delta \mathscr{H}_{m}}{\delta \beta_{b}}\nonumber \, ,  \\
=& \frac{\delta}{\delta \beta_{b}}\left(2\sqrt{\gamma}\left[\bar \nabla_{a}\left(\left(\frac{1}{\sqrt{\gamma}}\right)\Pi^{ab}\beta_{b}\right) - \beta_{b}\bar \nabla_{a}\left(\left(\frac{1}{\sqrt{\gamma}}\right)\Pi^{ab}\right)\right]\right)+\frac{\delta \mathscr{H}_{m}}{\delta \beta_{b}}\nonumber \, ,  \\
=& -2\sqrt{\gamma}\, \bar \nabla_{a}\left(\left(\frac{1}{\sqrt{\gamma}}\right)\Pi^{ab}\right) +\frac{\delta \mathscr{H}_{m}}{\delta \beta_{b}}\nonumber \, ,  \\
=& \frac{1}{8 \pi G}\bar \nabla_{a} \left(K^{ab} -\gamma^{ab}K\right) - j^{b}
\,. \end{align}

Where we used equation \eqref {DefPi} and made the definition $j^{a} =- \frac{\delta \mathscr{H}_{m}}{\delta \beta_{a}}$ in the last step.  This equation is the Momentum Constraint, as seen in equation \eqref {MomentumConstraint}. 

\section{Evolution equations}

Now, we follow the standard Hamiltonian procedure to derive equations for the canonical field variables $\gamma_{ab}$ and $\Pi^{ab}$.\footnote{For simplicity, the matter fields are set to zero in this section.  We have already derived the correct ADM equations with matter above in Chapter \ref{sec: Brute Force}, and we are soon to be drowned under a large number of terms.  Adding matter back in doesn't massively increase the complexity of what follows, but it does enough that it is not worth our trouble here.}  Namely, after deriving the appropriate Hamiltonian density and solving for the appropriate constraints, we vary the phase space action 

\begin{equation}
S = \int d^{4}x \, \left(\Pi^{ab}\dot \gamma_{ab} - \mathscr{H}\right)
\,. \end{equation}

with respect to the canonical variables.  This gives:

\begin{align}
\delta S =& \int d^{4}x \, \left[\left(\delta \Pi^{ab}\right)\dot \gamma_{ab} + \Pi^{ab}\delta\left(\dot \gamma_{ab}\right) - \frac{\delta \mathscr{H}}{\delta \Pi^{ab}}\delta \Pi^{ab} -\frac{\delta \mathscr{H}}{\delta \gamma_{ab}}\delta \gamma_{ab}\right]\nonumber \, ,  \\
=&\int d\tau \int d^{3}x \, \left[\delta \Pi^{ab}\left(\dot \gamma_{ab} - \frac{\delta \mathscr{H}}{\delta \Pi^{ab}}\right) + \frac{d}{dt}\left(\Pi^{ab}\delta\gamma_{ab}\right)  -\delta \gamma_{ab}\left(\dot \Pi^{ab} + \frac{\delta \mathscr{H}}{\delta \gamma_{ab}}\right)\right] \label{Hamiltonian Variation A}
\,. \end{align}

The first and third terms in equation \eqref{Hamiltonian Variation A} are simply equivalent to the Hamiltonian equations of motion $\dot \gamma_{ab} = \frac{\delta \mathscr{H}}{\delta \Pi^{ab}}$ and $\dot \Pi^{ab} = - \frac{\delta \mathscr{H}}{\delta \gamma_{ab}}$.  The second term is a total time derivative.  We can therefore do the time integral first over this term, giving us $\int d^{3}x \Pi^{ab}\delta \gamma_{ab}$ over the initial and final timeslices.  We, however, assume that our system is set up in such a way that it has a definite set of initial conditions, which therefore means that the value of $\gamma_{ab}$ is held fixed on the initial and final time slices.  Therefore, any variation over these regions must vanish identically.  We can therefore drop the second term in \eqref{Hamiltonian Variation A}.  Therefore, deriving the evolution equations is as simple as varying the Hamiltonian with respect to the 3-metric and its canonical momentum, $\Pi^{ab}$.  This will then give us equations for $\dot \Pi^{ab}$ and $\dot \gamma_{ab}$.  Note, however, that this does \emph{not} directly give us an equation for $\dot K_{ab}$.  However, an inspection of equation \eqref {DefPi} shows that one can solve for $\dot K^{ab}$ if one knows the values of $\dot \Pi^{ab}$ and $\dot \gamma_{ab}$\footnote{In accomplishing this, it is absolutely useful to use the identity derived from the definition of the inverse metric:
\begin{align}
£_{t}\gamma_{ab}& = £_{t}\left(\gamma_{ac}\gamma^{cd}\gamma_{db}\right)\nonumber \, ,  \\
 &=\left(£_{t}\gamma_{ac}\right)\gamma^{cd}\gamma_{db} +\gamma_{ac}\left(£_{t}\gamma^{cd}\right)\gamma_{db} + \gamma_{ac}\gamma^{cd}£_{t}\gamma_{db}\nonumber \, ,  \\
 &=£_{t}\gamma_{ab}+\gamma_{ac}\gamma_{db}£_{t}\gamma^{cd} + £_{t}\gamma_{ba}\nonumber \, ,  \\
 -£_{t}\gamma_{ab} &= \gamma_{ac}\gamma_{db}£_{t}\gamma^{cd}
\,. \end{align}\label{Inverse metric identity} }.  Having derived the equation for $\dot K_{ab}$ in equation \eqref {Kevolution}, we will not concern ourselves with performing this transformation here.  It will just be noted that this transformation can be performed through a relatively direct, if laborious, computation.  

First, we find the equation for $\dot \gamma_{ab}$, neglecting terms not involving $\Pi^{ab}$

\begin{align}
\dot \gamma_{ab} &= \frac{\delta \mathscr{H}}{\delta \Pi^{ab}}\nonumber \, ,  \\
&= \frac{\delta}{\delta \Pi^{ab}}\left[\alpha \left(\frac{16 \pi G}{\sqrt{\gamma}}\right)\left(\Pi^{ab}\Pi_{ab} - \frac{1}{2}\Pi^{2}\right) + 2\Pi^{ab}\bar \nabla_{(a} \beta_{b)}\right] \nonumber \, ,  \\
&=2\alpha \left(\frac{16 \pi G}{\sqrt{\gamma}}\right)\left(\Pi_{ab} - \frac{1}{2}\gamma_{ab}\Pi \right)+ 2\bar \nabla_{(a}\beta_{b)}\nonumber \, ,  \\
&=-2\alpha K_{ab} + 2\bar \nabla_{(a}\beta_{b)} \label{Gamma Equation B}
\,. \end{align}

Where we used equation \eqref {InvertedPi} in order to convert the $\Pi_{ab}$ terms into $K_{ab}$ terms.  Equation \eqref{Gamma Equation B} is clearly the same equation as \eqref {Gevolution}.

Before we derive the equation for $\dot \Pi^{ab}$, we derive a few properties of the variation of quantities with respect to $\delta \gamma_{ab}$.  First, we tackle the determinant of the 3-metric \footnote{This derivation works in any dimension, not just three--in d dimensions, the Levi-Civita symbol has d indices, requiring d copies of the metric tensor, and making the factorial term $\frac{1}{d!}$.  Then, the rest of this proof follows in exactly the same way as this one does.}:

\begin{align}
\frac{\delta}{\delta \gamma_{ab}} \gamma &= \frac{\delta}{\delta \gamma_{ab}}\left(\frac{1}{3!}\gamma_{mr}\gamma_{ns}\gamma_{kt}\epsilon^{mnk}\epsilon^{rst}\right)\nonumber\, ,  \\
&=\frac{1}{2!}\gamma_{ns}\gamma_{kt}\epsilon^{ank}\epsilon^{bst} \nonumber \, ,  \\
&=\gamma \gamma^{ab}\,. \label{Trick with Determinant}
\,. \end{align}

Where the last step in \eqref{Trick with Determinant} involved recognizing two things. First, that the matrix $\gamma_{ns}\gamma_{kl}\epsilon^{ank}\epsilon^{bst}$ is constructed entirely from the metric and the Levi-Civita symbol, and thus, can depend only on these two things.   Second, that if one were to multiply $\gamma_{ns}\gamma_{kl}\epsilon^{ank}\epsilon^{bst}$ by $\gamma_{ab}$, the answer would be three times $\gamma$.  Therefore, one can conclude the answer on the last line of \eqref{Trick with Determinant}.
It is also not too hard to work out the variation of $\gamma^{ab}$ with respect to $\gamma_{ab}$:

\begin{align}
\frac{\delta \gamma^{mn}}{\delta \gamma_{ab}} &= \frac{\delta}{\delta \gamma_{ab}}\left(\gamma^{mr}\gamma_{rs}\gamma^{sn}\right)\nonumber \, ,  \\
&=\left(\frac{\delta \gamma^{mr}}{\delta \gamma_{ab}}\right)\gamma_{rs}\gamma^{sn} + \gamma^{mr}\left(\frac{\delta \gamma_{rs}}{\delta \gamma_{ab}}\right)\gamma^{sn}+ \gamma^{mr}\gamma_{rs}\left(\frac{\delta \gamma^{sn}}{\delta \gamma_{ab}}\right)\nonumber\, ,  \\
&=2\left(\frac{\delta \gamma^{mn}}{\delta \gamma_{ab}}\right) +\gamma^{mr}\gamma^{sn}\delta_{(r}{}^{a}\delta_{s)}{}^{b}\nonumber \, ,  \\
-\left(\frac{\delta \gamma^{mn}}{\delta \gamma_{ab}}\right)&=\gamma^{m(a}\gamma^{b)n} \label{InverseMetricVariationTrick}
\,. \end{align}

And we can use equation \eqref {InverseMetricVariationTrick} to easily change the variable of the variation between the metric and its inverse when it is convenient to do so--in particular, when taking the variation of the Ricci scalar with respect to the 3-metric.

Now, we embark upon taking the variation of the Hamiltonian \eqref{ADM Bulk Hamiltonian} with respect to the 3-metric\footnote{Note that the variation of the $\Pi^{ab}\nabla_{a}\beta_{b}$ term with respect to $\gamma_{ab}$ depends only on the variation of the 3-Christoffel symbol with respect to the metric, due to the fact that the canonical form of the $\Pi^{ab}$ is considered to be the raised one, and the canonical form of $\beta_{a}$ is the one that has the lowered index.  Thus, there is no dependence on $\gamma_{ab}$ in this term.}:

\begin{align}
\left.\delta {\bf H}\right|_{\gamma} =& \left.\delta\right|_{\gamma} \int d^{3}x\left[\frac{16\,\pi\,G}{\sqrt{\gamma}}\alpha\left(\Pi^{ab}\Pi_{ab} - \frac{1}{2}\Pi^{2}\right) -\frac{\alpha \sqrt{\gamma}}{16\,\pi\,G}\bar R \right.\nonumber   \\
&\left.+ 2 \Pi^{ab}\bar \nabla_{a} \beta_{b}-\frac{\sqrt {\gamma}}{16\,\pi\,G} \bar \nabla^{2}\alpha - \frac{1}{8\,\pi \, G}\sqrt{\gamma}\nabla_{a}\left(n^{a}K\right)\right] \nonumber \, ,  \\
=&\int d^{3}x\left[-\frac{8\,\pi \,G}{\sqrt{\gamma}}\gamma^{ab}\alpha\delta \gamma_{ab}\left(\Pi^{cd}\Pi_{cd} - \frac{1}{2}\Pi^{2}\right)-\frac{\sqrt{\gamma}}{32\,\pi\,G}\gamma^{ab}\delta\gamma_{ab}\bar \nabla^{2}\alpha\right.\nonumber   \\
&\left.+\frac{32\,\pi\,G}{\sqrt{\gamma}}\alpha \delta\gamma_{ab}\left(\Pi^{a}{}_{c}\Pi^{cb} - \frac{1}{2}\Pi^{ab}\Pi\right)-\frac{\sqrt{\gamma}}{16\,\pi\,G}\delta \gamma_{ab}\bar \nabla^{a}\bar \nabla^{b}\alpha \right.\nonumber   \\ 
&\left. -\frac{\alpha\,\sqrt{\gamma}}{32\,\pi\,G}\bar R\gamma^{ab}\delta\gamma_{ab} - \frac{\alpha \sqrt{\gamma}}{16\,\pi\,G}\left(\bar R^{ab} \delta \gamma_{ab} + \gamma^{ab} \left.\delta \bar R_{ab}\right|_{\gamma}\right)-2\beta_{c}\Pi^{ab}\delta \bar {\Gamma}_{ab}{}^{c}\right] \label{RicciTermA} 
\,. \end{align}

We will deal with the rest of these terms in their due course, but let us first consider the last term involving the variation of the Ricci tensor in \eqref{RicciTermA}.  

\subsection{The variation of the Ricci tensor}
Tackling this term, we start with:

\begin{equation}
\delta\mathscr{H}_{Ricci} =-\int d^{3}x \frac{1}{16 \pi G}\left(\alpha \sqrt{\gamma}\gamma^{ab}\delta \bar R_{ab}\right)
\,. \end{equation}

Following \cite{ADM Paper}, define $\kappa \equiv \frac{1}{16\pi G}$ for simplicity.  Our goal is then to compute this variation of $\bar R_{ab}$ in terms of the variation of $\gamma_{ab}$.    To do this, we first expand $\bar R_{ab}$ in terms of Christoffel symbols.  We then remember that since the difference of two connections is a tensor, therefore, the difference of two Christoffel symbols is also a tensor, making the variation of a Christoffel symbol a tensor.  We can then turn partial derivatives into covariant derivatives relative to the central metric about which we take the variation \footnote{also, note that variations are defined in terms of partial derivatives in function space.  Therefore, they can be freely commuted through partial derivatives}:

\begin{align}
\delta \mathscr{H}_{Ricci} =& - \int d^{3}x \, \kappa \, \alpha \sqrt{\gamma}\gamma^{ab} \delta \left(\partial_{c} \bar \Gamma_{ab}{}^{c}-\partial_{a} \bar \Gamma_{bc}{}^{c} + \bar \Gamma_{ab}{}^{d}\bar \Gamma_{dc}{}^{c}-\bar \Gamma_{ac}{}^{d}\bar \Gamma_{db}{}^{c}\right)\nonumber \, ,  \\
=& -\int \sqrt{\gamma}\, d^{3}x\, \kappa \alpha \gamma^{ab}\left( \partial_{c} \delta \bar \Gamma_{ab}{}^{c}-\partial_{a} \delta \bar \Gamma_{bc}{}^{c}+\bar \Gamma_{dc}{}^{c} \delta \bar \Gamma_{ab}{}^{d} + \bar \Gamma_{ab}{}^{d}\delta \bar \Gamma_{dc}{}^{c}\right. \nonumber   \\
&\left.-\bar \Gamma_{ac}{}^{d} \delta \bar \Gamma_{db}{}^{c}- \bar \Gamma_{db}{}^{c} \delta \bar \Gamma_{ac}{}^{d}\right)\nonumber \, ,  \\
=&-\int \sqrt{\gamma}\, d^{3}x\, \kappa \alpha \gamma^{ab} \left(\bar \nabla_{c} \delta \bar \Gamma_{ab}{}^{c} + \bar \Gamma_{ca}{}^{d}\delta \bar \Gamma_{db}{}^{c} + \bar \Gamma_{cb}{}^{d}\delta \bar \Gamma_{ad}{}^{c} -\bar \Gamma_{dc}{}^{d} \delta \bar \Gamma_{ab}{}^{c}\right. \nonumber   \\
&\left.- \bar \nabla_{a} \delta \bar \Gamma_{bc}{}^{c} - \bar \Gamma_{ab}{}^{d} \delta \bar \Gamma_{dc}{}^{c}+\bar \Gamma_{dc}{}^{c} \delta \bar \Gamma_{ab}{}^{d} + \bar \Gamma_{ab}{}^{d}\delta \bar \Gamma_{dc}{}^{c} \right. \nonumber   \\
&\left. -\bar \Gamma_{ac}{}^{d} \delta \bar \Gamma_{db}{}^{c}- \bar \Gamma_{db}{}^{c} \delta \bar \Gamma_{ac}{}^{d}\right) \nonumber \, ,  \\
=&-\int \sqrt{\gamma}\,d^{3}x\,\kappa\alpha \gamma^{ab}\left(\bar \nabla_{c} \delta\bar \Gamma_{ab}{}^{c}-\bar \nabla_{a} \delta\bar \Gamma_{bc}{}^{c}\right) \nonumber \, ,  \\
=&-\int \sqrt{\gamma}\,d^{3}x\,\kappa \alpha  \left(\gamma^{ab} \delta^{d}{}_{c} - \gamma^{da}\delta^{b}{}_{c} \right)\bar \nabla_{d}\delta\bar \Gamma_{ab}{}^{c}
\,. \end{align}

So, we have now reduced this problem to evaluating the gradient of the variation of the Christoffel symbol in terms of the variation of the 3-metric.  As with almost all variational problems, we proceed by integrating by parts\footnote{Note that if the factor of $\alpha$ were not present, we could simply pull $\nabla_{d}$ through to the beginning of the equation and the integration by parts would produce nothing but a boundary term}.  

\begin{align}
\delta \mathscr{H}_{Ricci}=& - \int \sqrt{\gamma}\,d^{3}x\left\{\kappa \bar \nabla_{d}\left[\alpha\left(\gamma^{ab}\delta^{d}{}_{c}-\gamma^{ad}\delta^{b}{}_{c}\right)\delta \bar \Gamma_{ab}{}^{c}\right] \right. \nonumber   \\
&\left.-\kappa \bar \nabla_{d}\left[\alpha\left(\gamma^{ab}\delta^{d}{}_{c}-\gamma^{ad}\delta^{b}{}_{c}\right)\right]\delta \Gamma_{ab}{}^{c}\right\}\nonumber \, ,  \\
=& -\oint \sqrt{q}\,d^{2}x \, \kappa \alpha \left(\gamma^{ab} \bar r_{c}- \bar r^{a} \delta^{b}{}_{c}\right)\delta \bar \Gamma_{ab}{}^{c} \nonumber   \\
&+\int\sqrt{\gamma}\,d^{3}x \, \kappa \left( \gamma^{ab} \bar \nabla_{c}\alpha - \delta^{b}{}_{c} \bar \nabla^{a} \alpha\right)\delta \bar \Gamma_{ab}{}^{c} \, ,  \\
=& \oint \sqrt{q}\, d^{2}x \,  B^{ab}{}_{c}\delta \bar \Gamma_{ab}{}^{c} + \int \sqrt{\gamma}\,d^{3}x\, T^{ab}{}_{c}\delta \bar \Gamma_{ab}{}^{c} \label{RicciTermVariationMid}
\,. \end{align}

Where we make the definitions $B^{ab}{}_{c} \equiv \kappa \alpha \left(\frac{1}{2}\bar r^{a} \delta ^{b}{}_{c}+\frac{1}{2}\bar r^{b} \delta^{a}{}_{c} - \gamma^{ab} \bar r_{c}\right)$ and $T^{ab}{}_{c} \equiv \kappa\left(\gamma^{ab} \bar \nabla_{c}\alpha - \frac{1}{2}\delta^{b}{}_{c} \bar \nabla^{a} \alpha-\frac{1}{2}\delta^{a}{}_{c}\bar \nabla^{b} \alpha\right)$\footnote{Since no allowed variation of the metric will induce torsion in the connexion, we know that $\delta \Gamma_{ab}{}^{c}$ is symmetric on a and b.  Therefore, we can symmetrize on these indices of $B^{ab}{}_{c}$ and $T^{ab}{}_{c}$}.  Note that both objects are true tensors whose indices can be freely raised and lowered with $\gamma_{ab}$ and its inverse.  Now, our variational problem has been reduced to finding the variation of a three index tensor multiplied by $\delta \Gamma_{ab}{}^{c}$.  In order to do this, we quickly work out a property relating the variation of the metric to the variation of its inverse:

\begin{align}
\delta \gamma^{ab} =& \delta \left(\gamma^{ac}\gamma_{cd}\gamma^{db}\right) \nonumber \, ,  \\
=& \left(\delta \gamma^{ac}\right)\gamma_{cd}\gamma^{db} + \gamma^{ac}\left(\delta\gamma_{cd}\right)\gamma^{db}+ \gamma^{ac}\gamma_{cd}\left(\delta \gamma^{db}\right)\nonumber \, ,  \\
=& 2 \delta \gamma^{ab} + \gamma^{ac}\gamma^{db}\delta \gamma_{cd} \nonumber \, ,  \\
\delta \gamma^{ab} =& - \gamma^{ac}\gamma^{db}\delta \gamma_{cd} \label{InverseMetricVariation}
\,. \end{align}

Having this identity in hand, we now work out the general expression for a tensor multiplying a variation of a Christoffel symbol.  Using the decomposition the the Christoffel symbol into metric components, and converting partial derivatives into covariant derivatives, we obtain the following:

\begin{align}
M^{ab}{}_{c}\delta \Gamma_{ab}{}^{c} =& \frac{1}{2} M^{ab}{}_{c}\, \delta \left[\gamma^{cd}\left(\partial_{a}\gamma_{bd}+ \partial_{b}\gamma_{ad}-\partial_{d}\gamma_{ab}\right)\right] \nonumber \, ,  \\
=& -\frac{1}{2}M^{ab}{}_{c}\left(\gamma^{ce}\gamma^{df}\delta\gamma_{ef}\right)\left(\partial_{a}\gamma_{bd}+ \partial_{b}\gamma_{ad}-\partial_{d}\gamma_{ab}\right) \nonumber   \\
&+\frac{1}{2}M^{ab}{}_{c}\gamma^{cd}\left(\partial_{a} \delta \gamma_{bd} + \partial_{b} \delta\gamma_{ad} - \partial_{d}\delta \gamma_{ab}\right) \nonumber \, ,  \\
=& - M^{abe}\bar \Gamma_{ab}{}^{f}\delta \gamma_{ef}+ \frac{1}{2}M^{abd}\left(\bar \nabla_{a} \delta \gamma_{bd} + \bar \Gamma_{ab}{}^{c}\delta \gamma_{cd} + \bar \Gamma_{ad}{}^{c} \delta \gamma_{bc} \right. \nonumber   \\
&\left.+ \bar \nabla_{b} \delta \gamma_{ad} + \bar \Gamma_{ba}{}^{c} \delta \gamma_{cd} + \bar \Gamma_{bd}{}^{c} \delta \gamma_{ac} - \bar \nabla_{d} \delta \gamma_{ab} - \bar \Gamma_{da}{}^{c} \delta \gamma_{cb} - \bar \Gamma_{db}{}^{c} \delta \gamma_{ac}\right)\nonumber \, ,  \\
=& \frac{1}{2}M^{abc} \left(\bar \nabla_{a} \delta \gamma_{bc} + \bar \nabla_{b} \delta \gamma_{ac} - \bar \nabla_{c}\delta \gamma_{ab}\right) \nonumber \, ,  \\
=& \frac{1}{2}\left(M^{abc} + M^{bac} - M^{cab}\right)\bar \nabla_{a}\delta \gamma_{bc} \label{ChristoffelVariation}
\,. \end{align}

Using the above definitions and symmetrizing over the indices b and c, since the multiplier is $\bar \nabla_{a} \gamma_{bc}$:

\begin{align}
B^{abc} + B^{bac} - B^{cab} =& \kappa \alpha \left(\frac{1}{2}\bar r^{a} \gamma^{bc} + \frac{1}{2} \bar r^{b} \gamma^{ac}-\gamma^{ab} \bar r^{c}+ \frac{1}{2} \bar r^{b} \gamma^{ac} + \frac{1}{2} \bar r^{a} \gamma^{bc} \right. \nonumber   \\
&\left.  - \gamma^{ab} \bar r^{c}-\frac{1}{2} \bar r^{c} \gamma^{ab} - \frac{1}{2}\bar r^{a} \gamma^{bc} + \gamma^{ca} \bar r^{b}\right)\nonumber \, ,  \\
=&\kappa \alpha \left(\bar r^{a} \gamma^{bc} + 2 \bar r^{b} \gamma^{ac} - \frac{5}{2} \bar r^{c} \gamma^{ab}\right) \nonumber \, ,  \\
\left(B^{abc}+B^{bac} - B^{cab}\right)_{sym, bc} =&\kappa \alpha \left( \bar r^{a} \gamma^{bc} - \gamma^{a(b}\bar r^{c)}\right) \label{B symmetrized} \, ,  \\
T^{abc} + T^{bac} - T^{cab} =& \kappa \left(\gamma^{ab} \bar \nabla^{c} \alpha - \frac{1}{2} \gamma^{bc} \bar \nabla^{a} \alpha - \frac{1}{2} \gamma^{ac} \bar \nabla^{b} \alpha + \gamma^{ba} \bar \nabla^{c} \alpha\right. \nonumber   \\
&\left.  - \frac{1}{2} \gamma^{ac} \bar \nabla^{b} \alpha - \frac{1}{2} \gamma^{bc} \bar \nabla^{a} \alpha-\gamma^{cb} \bar \nabla^{a} \alpha \right. \nonumber   \\
&\left. + \frac{1}{2} \gamma^{ab} \bar \nabla^{c} \alpha + \frac{1}{2} \gamma^{cb} \bar \nabla^{a} \alpha\right) \nonumber \, ,  \\
=& \kappa \left(\frac{5}{2} \gamma^{ab} \bar \nabla^{c} \alpha -  \gamma^{bc} \bar \nabla^{a} \alpha - 2 \gamma^{ac} \bar \nabla^{b} \alpha \right) \nonumber \, ,  \\
\left(T^{abc} + T^{bac} - T^{cab}\right)_{sym, bc} =&\kappa\left(\gamma^{a(b}\bar \nabla^{c)}\alpha - \gamma^{bc}\bar \nabla^{a} \alpha\right) \label{T symmetrized}
\,. \end{align}

Thus, equations \eqref{B symmetrized} and \eqref{T symmetrized} give us the final expressions for the terms multiplying $\bar \nabla_{a} \gamma_{bc}$ in the variation in equation \eqref{RicciTermVariationMid}, since equation \eqref{ChristoffelVariation} tells us how to convert the variation of a Christoffel symbol into the gradient of the variation of the 3-metric.  Putting all of these equations together, and then integrating by parts\footnote{note that we automatically apply Gauss's theorem to the divergence term originating from the bulk term.}, we obtain:

\begin{align}
\delta \mathscr{H}_{Ricci} &= \oint \sqrt{q}\,d^{2}x\,\frac{1}{2}\kappa \alpha\left(\bar r^{a} \gamma^{bc} - \gamma^{a(b}\bar r^{c)}\right) \bar \nabla_{a} \delta \gamma_{bc} \nonumber   \\
&+ \int \sqrt{\gamma}\,d^{3}x\, \frac{1}{2}\kappa\left(\gamma^{a(b}\bar \nabla^{c)}\alpha - \gamma^{bc} \bar \nabla^{a} \alpha\right)\bar \nabla_{a} \delta \gamma_{bc} \nonumber \, ,  \\
&=\oint \sqrt{q}\,d^{2}x\,\left\{\frac{\kappa}{2}\bar \nabla_{a} \left[\alpha\left(\bar r^{a} \gamma^{bc} - \gamma^{a(b}\bar r^{c)}\right)\delta \gamma_{bc}\right] \right. \nonumber   \\
&\left.- \frac{\kappa}{2}\delta\gamma_{bc} \bar \nabla_{a} \left[\alpha\left(\bar r^{a} \gamma^{bc} -\gamma^{a(b}\bar r^{c)}\right)\right]+ \frac{\kappa}{2}\bar r_{a} \left(\gamma^{a(b}\bar \nabla^{c)} \alpha - \gamma^{bc} \bar \nabla^{a} \alpha \right)\delta \gamma_{bc}\right\} \nonumber   \\
&+\int \sqrt{\gamma}\,d^{3}x\, \frac{\kappa}{2}\bar \nabla_{a}\left(\gamma^{bc} \bar \nabla^{a}\alpha - \gamma^{a(b}\bar \nabla^{c)} \alpha\right) \delta \gamma_{bc} \nonumber 
\,. \end{align}
\begin{align}
&=\oint \sqrt{q}\,d^{2}x\,\frac{\kappa}{2}\left\{ \nabla_{a} \left[\alpha\left(\bar r^{a} \gamma^{bc} -\gamma^{a(b}\bar r^{c)}\right)\delta \gamma_{bc}\right]\right. \nonumber   \\
&\left. +\delta \gamma_{bc} \left[\alpha \left(\gamma^{bc} \bar \nabla_{a}\bar r^{a} -\bar \nabla^{(b}\bar r^{c)}\right)+ 2 \bar r^{(b}\bar \nabla^{c)}\alpha - 2 \gamma^{bc} \bar r^{a} \bar \nabla_{a} \alpha \right] \right\} \nonumber   \\
&+\int \sqrt{\gamma}\,d^{3}x\, \frac{\kappa}{2}\left(\gamma^{bc} \bar \nabla^{a}\bar \nabla_{a}\alpha - \bar \nabla^{(b}\bar \nabla^{c)} \alpha\right) \delta \gamma_{bc} \label{Almost a variation}
\,. \end{align}

Now, equation \eqref{Almost a variation} nearly has the form that we want.  Most of the terms are direct multipliers of the variation of the 3-metric $\gamma_{bc}$.  We are left, however, with a 3-divergence to be integrated over the boundary.  The na\''ive thought would be to simply eliminate this term using Gauss's theorem, since it should be convertible to an integral over the boundary of the boundary, and a well-known theorem of topology tells us that the boundary of the boundary is zero for any manifold.  

Finally, remembering our promise to wait until Chapter \ref{sec: Gauss's Theorem} to deal with the boundary terms, we drop everything in \eqref{Almost a variation} except for the bulk term, and our final answer is:

\begin{equation}
\int d^{3}x \gamma^{ab}\left.\delta \bar R_{ab} \right|_{\gamma} = \int d^{3}x\sqrt\gamma \frac{\kappa}{2}\left(\gamma^{ab} \bar \nabla^{2}\alpha - \bar \nabla^{a} \bar \nabla^{b}\alpha\right)\delta \gamma_{ab} \label{Final Ricci Variation}
\,. \end{equation}

\subsection{The equation for $\dot \Pi^{ab}$}
Now, having gone through the work of that extensive aside, we are ready to return to the equation for $\dot \Pi^{ab}$ that we obtained from the ADM Hamiltonian.  Substituting \eqref{Final Ricci Variation} into \eqref{RicciTermA}, and using \eqref{ChristoffelVariation} to deal with the term in \eqref{RicciTermA} involving the shift vector\footnote{Once again, neglecting boundary terms, as they will be treated with in Chatper \ref{sec: Gauss's Theorem}}, we get:

\begin{align}
\left.\delta {\bf H}\right|_{\gamma} =&\int d^{3}x\left[-\frac{8\,\pi \,G}{\sqrt{\gamma}}\gamma^{ab}\alpha\delta \gamma_{ab}\left(\Pi^{cd}\Pi_{cd} - \frac{1}{2}\Pi^{2}\right)-\frac{\sqrt{\gamma}}{32\,\pi\,G}\gamma^{ab}\delta\gamma_{ab}\bar \nabla^{2}\alpha\right.\nonumber   \\
&\left.+\frac{32\,\pi\,G}{\sqrt{\gamma}}\alpha \delta\gamma_{ab}\left(\Pi^{a}{}_{c}\Pi^{cb} - \frac{1}{2}\Pi^{ab}\Pi\right)-\frac{\sqrt{\gamma}}{16\,\pi\,G}\delta \gamma_{ab}\bar \nabla^{a}\bar \nabla^{b}\alpha \right.\nonumber   \\ 
&\left. -\frac{\alpha\sqrt{\gamma}}{32\,\pi\,G}\bar R\gamma^{ab}\delta\gamma_{ab} - \frac{\alpha \sqrt{\gamma}}{16\,\pi\,G}\left(\bar R^{ab} \delta \gamma_{ab} + \frac{1}{2}\left(\gamma^{ab} \bar \nabla^{2}\alpha - \bar \nabla^{a} \bar \nabla^{b}\alpha\right)\delta \gamma_{ab}\right)\right. \nonumber  \\
&\left.-\left(\frac{\sqrt{\gamma}}{\sqrt{\gamma}}\right)\left(\Pi^{ab}\beta^{c}-\Pi^{bc}\beta^{a}-\Pi^{ca}\beta^{b}\right)\bar \nabla_{c}\delta \gamma_{ab} \right]\nonumber \, ,  \\
=&\int d^{3}x\,\delta\gamma_{ab}\left[\frac{8\,\pi\,G\,\alpha}{\sqrt{\gamma}}\left(\frac{1}{2}\gamma^{ab}\Pi^{2}-\gamma^{ab}\Pi^{cd}\Pi_{cd}+ 4 \Pi^{a}{}_{c}\Pi^{cb} - 2 \Pi^{ab}\Pi\right)\right. \nonumber   \\
&\left.+\frac{\sqrt{\gamma}}{16\,\pi\,G}\left(-\alpha \bar R^{ab} -\alpha \frac{1}{2}\gamma^{ab} \bar R- \bar \nabla^{a}\bar \nabla^{b} \alpha \right)\right.\nonumber   \\
&\left. +\sqrt{\gamma}\delta\gamma_{ab}\b\nabla_{c}\left\{\frac{1}{\sqrt{\gamma}}\left(\Pi^{ab}\beta^{c}-\Pi^{ac}\beta^{b}-\Pi^{bc}\beta^{a}\right)\right\}\right]
\,. \end{align}

Finally for all of that effort, we get our final solution, after referring to our original phase space action variation in \eqref{Hamiltonian Variation A}: 

\begin{align}
\dot \Pi^{ab} =& - \frac{\left.\delta\mathscr{H}\right|_{\gamma}}{\delta \gamma_{ab}}\nonumber \, ,  \\
=&\frac{8\,\pi\,G\,\alpha}{\sqrt{\gamma}}\left(-\frac{1}{2}\gamma^{ab}\Pi^{2}+\gamma^{ab}\Pi^{cd}\Pi_{cd}- 4 \Pi^{a}{}_{c}\Pi^{cb} + 2 \Pi^{ab}\Pi\right) \nonumber   \\
&\left.+\frac{\sqrt{\gamma}}{16\,\pi\,G}\left(\alpha \bar R^{ab} +\alpha \frac{1}{2}\gamma^{ab} \bar R+ \bar \nabla^{a}\bar \nabla^{b} \alpha \right)\right.\nonumber   \\
& +\sqrt{\gamma}\delta\gamma_{ab}\b\nabla_{c}\left[ \frac{1}{\sqrt{\gamma}}\left(-\Pi^{ab}\beta^{c}+\Pi^{ac}\beta^{b}+\Pi^{bc}\beta^{a}\right)\right] \label{Pi Equation B}
\,. \end{align}

Equation \eqref{Pi Equation B} can be shown, through a laborious calculation involving \eqref{InvertedPi} and \eqref{Gamma Equation B}, to be equivalent to \eqref{Kevolution}.  We have now completed our derivation of the 3+1 Einstein equations in the bulk using the Hamiltonian formalism.  Now, we move on and work with the boundary terms in the next Chapter.

\chapter{Boundary Charges and Counterterms}\label{sec: Gauss's Theorem}
\section{Gauss's Theorem} 
Now, let us consider the way to extract information about boundary charges out of the ADM formalism.  For our current purposes, we will not concern ourselves with the $\int \sqrt{\gamma} d^{3}{}x K$ terms over the initial and final slices in \eqref{ADM Bulk Hamiltonian}.  Instead, we focus on the term that has the form $\int d\tau\int d^{3} x \sqrt{\gamma} \bar \nabla^{2} \alpha$.  Before we progress, we will restate Gauss's theorem as described in Hawking and Ellis \cite{HawkingEllis}:  

First, consider the integral of the following quantity over an arbitrary region $\mathbf{\mathscr{R}}$ of a n-dimensional manifold $\mathbf{\mathscr{M}}$, endowed with metric tensor $g_{ab}$, and metric-compatible connection $\nabla_{a}$:

\begin{equation}
\int_{\mathscr{R}} d^{n}x \sqrt{|g|} \nabla_{a}V^{a}
\,. \end{equation}

Since the exterior derivative acting on the volume element necessarily yields zero (there are no totally antisymmetric (n+1) dimensional objects over a n dimensional manifold), we can apply integration by parts to the above term.  This reduces the dimension of the volume element by one, and gives us an integral over the boundary of $\mathbf{\mathscr{R}}$: 

\begin{equation}
\int_{\mathscr{R}} d^{n}x \sqrt{|g|} \nabla_{a}V^{a} = \int_{\partial \mathscr{R}}d^{n-1}x \sqrt{|\gamma|} r_{a}V^{a} \label{Gauss's Theorem}
\,. \end{equation}

Where the integral in \eqref {Gauss's Theorem} is over the boundary of $\mathbf{\mathscr{R}}$, and $r_{a}$ is the unit ``outward'' normal to this boundary.  If one prefers to think of $n$ dimensional quantities instead of $n-1$ dimensional quantities, one can define the boundary of $\mathbf{\mathscr{R}}$ as the zero of a scalar function $\rho$\footnote{For most cases, we would take this to be something supremely simple, like $\rho = r - r_{0}$, but it is not absolutely necessary to do so.}.  We can then define $``r_{a}'' = \nabla_{a} \rho$.  

$``r_{a}''$, however, differs from $r_{a}$ merely by a factor of $\frac{1}{\sqrt{\nabla_{a}\rho \nabla^{a}\rho}}$.  A simple computation then shows that $(\sqrt{\nabla_{a}\rho\nabla^{a}\rho})(\sqrt{|\gamma|}) = \sqrt{|g|}$.  Therefore, if one would prefer, \eqref {Gauss's Theorem} could be rewritten as:

\begin{equation}
\int_{\mathscr{R}} d^{n}x \sqrt{|g|} \nabla_{a}V^{a} = \int_{\partial \mathscr{R}}d^{n-1}x \sqrt{|g|}``r_{a}'' V^{a}
\,. \end{equation}  

For a great many applications, this conversion greatly simplifies the calculation of an integral, as the domain of dependence for the problem has been reduced from the entire bulk of the problem to merely its boundary.  Furthermore, since the boundary of the boundary is zero by a famous theorem from topology, we can immediately conclude,  that the right hand side of \eqref {Gauss's Theorem} vanishes if we are able to repeat this procedure.  We should not, however, take this to mean that the integral of a $n$-dimensional Laplacian over the bulk always vanishes.  First, we will see a simple counterexample to show that na\''ive double application of Gauss's Theorem results in nonsense, and then it will be explained why this is the case.  First, consider the differential form of Gauss's Law:

\begin{equation}
\nabla_{a}E^{a} = \frac{1}{\epsilon_{0}} \rho
\,. \end{equation}

It is a simple matter to integrate this equation over space on both sides, and then apply Gauss's theorem in order to obtain the integral form of Gauss's Law (for clarity, we absorb the metric determinant factors into the volume elements):

\begin{equation}
\int_{\mathscr{M}} d^{3}x_{} \nabla_{a}E^{a} = \int_{\partial \mathscr{M}} d^{2}x_{} r_{a} E^{a} = \int d^{3}x_{} \frac{\rho}{\epsilon_{0}} = \frac{Q_{inc}}{\epsilon_{0}} \label{Gauss's Law}
\,. \end{equation}

Now, one might then recall that the Electric field is merely -1 times the gradient of the electrostatic potential $\phi$, and then replace $E^{a}$ with $-\nabla^{a}\phi$ in \eqref {Gauss's Law}:

\begin{align}
\frac{Q_{inc}}{\epsilon_{0}} =& -\int_{\partial \mathscr{M}} d^{2}x_{}r_{a} \nabla^{a}\phi \nonumber\, ,  \\
=& \int_{\partial \mathscr{M}} d^{2}x_{} \phi \nabla_{a}r^{a} - \int_{\partial \mathscr{M}} d^{2}x_{}\nabla_{a}(\phi r^{a}) \label{misleading}\, ,  \\
``=''& \int_{\partial \mathscr{M}} d^{2}x_{} \phi \nabla_{a}r^{a} - \int_{\partial \partial \mathscr{M}} d^{1}x_{}s_{a} \phi r^{a}\, ,  \\
``=''& \int_{\partial \mathscr{M}} d^{2}x_{} \phi \nabla_{a}r^{a} \label{Wrong!}
\,. \end{align}

Where we used the metric compatibility of $\nabla_{a}$ to raise and lower indices both inside and outside the gradient.  A computation of \eqref {Wrong!} for even the simplest examples from Electrostatics will show that \eqref {Wrong!} is nonsense, however.  For instance, if we simply consider the static point charge field from Coloumb's law, we have $\phi = \frac{1}{4\pi \epsilon_{0}} \frac{q}{r}$, while for $r^{a}$ normal to a sphere in Euclidean space, we have $\nabla_{a}r^{a}= \frac{2}{r}$ Therefore, the right hand side of \eqref {Wrong!} comes out to $\int_{0}^{2\pi}\int_{0}^{\pi}d \phi_{} d\theta_{} r^{2}\sin(\theta)\frac{q}{2\pi \epsilon_{0}r^{2}}$, which then causes \eqref {Wrong!} to yield $\frac{Q_{inc}}{\epsilon_{0}}= \frac{2q}{\epsilon_{0}}$, which is clearly incorrect.  

What is the origin of this error, however?  It turns out that if you compute the second integral in \eqref {misleading}, you will find that it has value $\frac{q}{\epsilon_{0}}$, which accounts for the discrepancy between the left and right hand sides of \eqref {Wrong!}.  But, then, what of our application of Gauss's Theorem here?  It must have been incorrect, as it assigned a zero value to something that was not zero.  This, in fact, was our problem, and this is the reason why care must be taken with Gauss's Theorem.  The reason why we cannot apply Gauss's theorem to \eqref {misleading} lies in the \textit{type} of derivative operator that appears in \eqref {misleading}--that $\nabla_{a}$ is the derivative operator that is compatible with the \textit{bulk} 3-metric, but not necessarily with the \textit{boundary} two metric--in fact, since areas increase as we increase the value of r, the boundary two metric \textit{cannot} be compatible with $\nabla_{a}$--the bulk derivative operator should be able to record information about the increase of areas.  Gauss's Theorem would only be applicable to \eqref {misleading} if the derivative operator appearing there were the two connection compatible with the two metric.  Therefore, care must be taken at all times with these operators, lest unwanted errors creep into calculations.  
$\newline$

\section{Boundary Charges in Maxwell Theory} \label{sec: Maxwell Theory}
Now, let us extend this calculation to the full Maxwell Hamiltonian.  We start with the Maxwell Lagrangian with arbitrary external current $j^{a}$:

\begin{equation}
\mathscr{L} = \frac{1}{4}F^{ab}F_{ab}+j^{a}A_{a}\, ,
\,. \end{equation}

\noindent here, as is usual, $F_{ab} \equiv \partial_{a}A_{b}-\partial_{b}A_{a}$, and indices are raised and lowered using the Minkowski metric $\eta_{ab}$.  Due to the antisymmetry of $F_{ab}$, it should be clear that there are no time derivatives of $A_{t}$ appearing in this action.  Therefore, we should treat $A_{t}$ as a Lagrange multiplier here.  Now, we will 3+1 split the Maxwell Lagrangian along surfaces of constant $t$, defining $A_{a} = (\phi, A_{i})$ and $j^{a}  = (\rho, j^{i})$.  We wish to define a Hamiltonian from the given Lagrangian, using $A_{i}$ and its canonical momentum $\Pi^{i}$ as canonical variables.  First, we solve for $\Pi^{i}$:

\begin{equation}
\Pi^{i} \equiv \frac{\delta \mathscr{L}}{\delta \dot A_{i}} = \frac{1}{2}F^{ab} \frac{\delta}{\delta( \partial_{t} A_{i})}F_{ab}=F^{ti}=\partial^{t}A^{i} - \partial^{i}A^{t}
\,. \end{equation}

So, now, the Maxwell Lagrangian can be rewritten as:

\begin{align}
\mathscr{L} =& \frac{1}{4}\;F^{ab}F_{ab} + j^{a}A_{a} \nonumber\, ,  \\
&=\frac{1}{2}F^{ti}F_{ti} + \frac{1}{4}F^{ij}F_{ij} + j^{t}A_{t}+ j^{i}A_{i} \nonumber \, ,  \\
=&-\frac{1}{2}\Pi^{i}\Pi_{i} +\frac{1}{4}F^{ij}F_{ij} + \rho \phi + j^{i}A_{i}
\,. \end{align}

Now, we perform the Legendre transformation to arrive at the Maxwell Hamiltonian density:

\begin{align}
\mathscr{H} =& \Pi^{i}\dot A_{i} - \mathscr{L} \nonumber \, ,  \\
=& \Pi^{i}\left(-\Pi_{i} + \partial_{i}\phi\right) - \left(-\frac{1}{2}\Pi^{i}\Pi_{i} + \frac{1}{4}F^{ij}F_{ij} + \rho \phi + j^{i}A_{i}\right) \nonumber \, ,  \\
=&-\frac{1}{2}\Pi^{i}\Pi_{i} -\frac{1}{4}F^{ij}F_{ij} + \Pi^{i}\partial_{i}\phi - \rho \phi - j^{i}A_{i} \label{Maxwell Hamiltonian A}
\,. \end{align}

Now, let us take a variation of the phase space action with respect to $\phi$.

\begin{align}
\left.\delta\right|_{\phi} \mathscr{H} =&\int d^{4}x \Pi^{i}\partial_{i}\delta \phi - \rho \delta \phi \nonumber \, ,  \\
=&\int dt \int d^{3}x \partial_{i}\left(\Pi^{i}\phi\right) - \delta \phi \partial_{i}\Pi^{i} - \rho \delta \phi \nonumber \, ,  \\
=& \int dt \left[\oint d^{2}x\bar r_{i}\Pi^{i}\delta \phi - \int d^{3}x \left(\rho + \partial_{i}\Pi^{i}\right)\delta \phi\right]
\,. \end{align}

\noindent the term under the $\int d^{3}x$ is, upon realization that $\Pi^{i} = -E^{i}$, easily recognizable as Gauss's Law.  The boundary term, however, is the integral form of Gauss's Law, only without an appropriate charge term to balance it.  Furthermore, the value of $\phi$ in this formalism is pure gauge--there is no abstract reason it has to take any value at all.  Therefore, the presence of the boundary term gives us no reason at all to believe that the action derived from \eqref{Maxwell Hamiltonian A} is stationary with respect to variations of $\phi$.

Therefore, in this theory, the behaviour of the bulk can create problems with the maintenance of a stationary action, by the production of gauge-dependent boundary terms.  This behaviour, however, can be controlled.  

Consider an ansatz to resolve this inconsistency--we define the Hamiltonian as the 4-dimensional integral of the Hamiltonian density plus a boundary term dependent only on $\phi$ and an as-yet undetermined constant:

\begin{align}
H =& \int d^{3}x{}^{}\left(-\frac{1}{2}\Pi^{i}\Pi_{i} - \frac{1}{4}F^{ij}F_{ij} + \Pi^{i}\partial_{i}\phi - \rho \phi - j^{i}A_{i}\right) \nonumber\, ,  \\
&+ \oint d^{2}x{}^{}Q\,\phi + \oint d^{2}x J^{i}A_{i}
\,. \end{align}

The variation of this Hamiltonian with respect to $A_{i}$ and $\Pi^{i}$ can be shown, relatively easily, to give Faraday's law and the Amp\`ere-Maxwell law\footnote{Since there is a gauge invariance built into $F_{ij}$, a similar boundary term to the one described above is generated by the variation with respect to $A_{i}$.  It is dealt with in the same way as the variation with respect to $\phi$, by the addition of a set of constants $J^{i}$ that are determined in order to cancel the boundary integrals.  Since this treatment isn't difficult, and is near-identical in form to the boundary terms arising from the variation of $\phi$, these terms are not explicitly treated here}.  $\nabla_{i}B^{i} = 0$ is given by the fact that $F_{ij}$ is the exterior derivative of $A_{i}$ and the second exterior derivative of any quantity is zero.  We therefore expect the variation of the Hamiltonian with respect to the only remaining variable, $\phi$, to give Gauss's law:

\begin{align}
\frac{\delta H}{\delta \phi} =& \frac{\delta}{\delta \phi} \left[\int d^{3}x\left(\Pi^{i} \partial_{i} \phi - \rho \phi\right)+Q\oint d^{2}x\left(\phi\right)\right]\nonumber \, ,  \\
=& \frac{\delta}{\delta \phi}\left[\int d^{3}x\left(\partial_{i}\left(\Pi^{i}\phi\right)-\phi\partial_{i}\Pi^{i}- \rho \phi\right) + \oint d^{2}x \left(Q\phi\right)\right] \nonumber \, ,  \\
=& \frac{\delta}{\delta \phi}\left[\int d^{3}x\left(-\phi \partial_{i}\Pi^{i} - \rho \phi\right) + \oint d^{2}x\left(Q\phi + r_{i}\Pi^{i}\phi\right)\right] \nonumber   \\
=& -\int d^{3}x\left(\partial_{i}\Pi^{i} + \rho\right) + \oint d^{2} x\left(Q + r_{i} \Pi^{i}\right) \nonumber \, ,  \\
=&\int d^{3}x\left(\rho - \bar \nabla_{i}E^{i}\right) +\oint d^{2}x\left(Q - r_{i}E^{i}\right)\label{Gauss's New Boundary}
\,. \end{align}

Since $F^{ti} = \Pi^{i} = -E^{i}$, we recognize the integrand in the first term above as simply being the electric field for the given charge configuration.  Since this equation has no time derivatives, we can therefore consider this first term as a constraint upon the equations of motion given by varying the Hamiltonian with respect to $A_{i}$ and $\Pi^{i}$.  Once we have solved the appropriate equations of motion, we can then evaluate the second term in the boundary integral in equation \eqref {Gauss's New Boundary}, and then assign the value to Q that cancels that boundary term.  This process is not unlike the renormalization process used in quantum field theory--where certain terms are computed using prescribed Hamiltonian dynamics, and then appropriate counterterms are assigned the appropriate values at the end of the calculation in order to yield consistent equations of motion.  

This aside being taken, we now discuss the topic of boundary charges in the ADM formulation.  
\subsection{ADM Formulation}

Now, consider the ADM Hamiltonian:
\begin{align}
\mathbf{H} =& \frac{1}{16\pi G}\int d^{3}x_{}\left[-\alpha \sqrt{\gamma} \bar R + \frac{\alpha}{\sqrt{\gamma}}(\Pi^{ab}\Pi_{ab}-\frac{1}{2}\Pi^{2})\right.\nonumber   \\
&\left.-\sqrt{\gamma}(\bar \nabla^{2}\alpha +\alpha \nabla_{a}(n^{a}K))+2\Pi^{ab}\bar\nabla_{a}\beta_{b}\right] +\mathbf{H}_{m} \label{FirstHamiltonianBC}
\,. \end{align}

The $\nabla_{a}(n^{a}K)$ term is acted on with a 4-dimensional divergence operator.  Therefore, it contributed a quantity of action $\int \sqrt{\gamma}d^{3} x K$ over the initial and final time slices, but nowhere else, provided that the spacelike normal to the outer boundary is normal to the timelike normal\footnote{Note that Appendix \ref{sec: variousfoliations} shows that the inner product between the radial vector and the timelike normal to the 3+1 slices is given by a boost parameter at infinity.  We can therefore appropriately choose a slicing that gives a zero boost parameter between the two at infinity.  In the asymptotically flat case, this is equivalent to the condition that $\beta^{a} ~ \frac{1}{r}$}.  We therefore do not concern ourselves with this term.  Now, let us do some massaging to the second to last term, involving the shift vector $\beta_{a}$:

\begin{align}
\frac{1}{8 \pi G}\int d^{3} x_{} \Pi^{ab}\bar \nabla_{a}\beta_{b} =& \frac{1}{8\pi G} \int d^{3} x_{} \frac{\sqrt{\gamma}}{\sqrt{\gamma}}\Pi^{ab}\bar \nabla_{a}\beta_{b} \nonumber\, ,  \\
=& \frac{1}{8\pi G}\int d^{3}x \sqrt{\gamma}\left(\bar \nabla_{a}(\frac{1}{\sqrt{\gamma}}\beta_{b}\Pi^{ab})-\beta_{b}\bar \nabla_{a}(\frac{1}{\sqrt{\gamma}}\Pi^{ab})\right) \nonumber
\,. \end{align}

We group the second term above along with the first and second terms in \eqref {FirstHamiltonianBC} and the matter Hamiltonian into the generic term $\mathbf{H}_{b}$, and refer to them as the bulk Hamiltonian density.  Substituting the above result into \eqref {FirstHamiltonianBC} gives us:

\begin{equation}
\mathbf{H} = \mathbf{H}_{b} +\frac{1}{8 \pi G}\int d^{3}x_{} \sqrt{\gamma} \bar \nabla_{a}(\frac{1}{\sqrt{\gamma}}\beta_{b}\Pi^{ab})-\frac{1}{16\pi G} \int d^{3}x_{} \sqrt{\gamma}\bar \nabla^{2}\alpha
\,. \end{equation}

Now, since we have two integrals of divergences over the bulk manifold, we can convert these integrals into integrals over the boundary of the spacelike slice:

\begin{equation}
\mathbf{H} = \mathbf{H}_{b} + \frac{1}{8\pi G} \int d^{2} x_{} \sqrt{q}r_{a}\beta_{b}(\gamma^{ab}K - K^{ab}) -\frac{1}{16 \pi G} \int d^{2} x_{} \sqrt{q}_{}r_{a}\bar \nabla^{a}\alpha \label{BoundaryHamiltonian}
\,. \end{equation}

Now, we might be tempted to deal with the $\bar \nabla^{a} \alpha$ term by pulling the $\bar \nabla^{a}$ through, integrating by parts, reapplying Gauss's Theorem, and relying on the vanishing of the boundary of the boundary.  Hopefully, the example from Equation \eqref{Gauss's Theorem} provides sufficient intuition to show that, since we have a $\bar \nabla$ and not a $\hat \nabla$ appearing in \eqref{BoundaryHamiltonian}, this path will lead us inexorably to nonsense.  Instead, we must consider the Boundary Hamiltonian to have the form provided above, and to take this seriously as we take our variations, which we will do below.  

\subsection{Variation of the Boundary Hamiltonian} \label{sec: ADM Boundary}

When we take the variation of the Hamiltonian with respect to the lapse function, we obtain:

\begin{align}
\frac{\partial \mathbf{H}}{\partial \alpha} =& 0 = \frac{\partial \mathbf{H}_{b}}{\partial \alpha} +\frac{1}{16\pi G}\int d^{2}x_{} \sqrt{q}r^{a}\frac{\partial \bar \nabla_{a}\alpha}{\partial \alpha} \nonumber \, ,  \\
=& \mathscr{C} -\frac{1}{16\pi G} \int d^{2}x_{} \bar \nabla_{a}(\sqrt{q}r^{a})
\,. \end{align}

Taking $\frac{\partial \mathbf{H}}{\partial \alpha}$  simply gives the Hamiltonian constraint 
$$\int d^{3}x_{} \sqrt{\gamma}\left(\rho -\frac{1}{16 \pi G}(\bar R -K^{ab}K_{ab} + K^{2})\right)$$
\noindent where $\rho \equiv \frac{\partial \mathscr{H}_{matter}}{\partial \alpha}$.  Now, in order to simplify this term, we are going to have to work through some somewhat intricate mathematics.  First, we take the boundary to be the zero of some function $f$, as above.  Then, we define $\beta_{\perp} = \frac{1}{\sqrt{\bar \nabla_{a}f\bar \nabla^{a}f}}$.  Then, $r_{a} = \beta_{\perp} \bar \nabla_{a}f$, is manifestly normal to the boundary and of unit length.  Then, assisted by the identity $\sqrt{\gamma} = \beta_{\perp} \sqrt{q}$, we can work out:

\begin{align}
\frac{\partial \mathbf{H}}{\partial \alpha} - \mathscr{C}=&-\frac{1}{16\pi G} \int d^{2}x_{} \bar \nabla_{a}(\sqrt{q}r^{a})  \\
=& -\frac{1}{16\pi G} \int d^{2}x_{}\bar \nabla_{a} (\sqrt{q} \beta_{\perp}\bar \nabla^{a}f)\, ,  \\
=& -\frac{1}{16\pi G}\int d^{2}x_{} \bar \nabla_{a} (\sqrt{\gamma}\bar \nabla^{a}f)\, ,  \\
=& -\frac{1}{16 \pi G} \int d^{2}x_{} \left( \bar \nabla^{a} f \bar \nabla_{a}\sqrt{\gamma}+ \sqrt{\gamma}\bar \nabla_{a}\bar \nabla^{a} f \right)\, ,  \\
=&-\frac{1}{16\pi G}\int d^{2}x_{}\left[ \frac{1}{\beta_{\perp}}r^{a}\bar \nabla_{a}\sqrt{\gamma}+ \sqrt{\gamma}\gamma^{ab}\left(\partial_{a}\partial_{b}f - \bar \Gamma_{ab}{}^{c}\bar \nabla_{c}f\right)\right]\, ,  \\
=&-\frac{1}{16\pi G}\int d^{x}x \left( \frac{1}{\beta_{\perp}}\sqrt{\gamma}r^{a}\bar \Gamma_{ab}{}^{b}+\beta_{\perp}\sqrt{q} \gamma^{ab} (\partial_{a}\partial_{b} f - \bar \Gamma_{ab}{}^{c} \frac{r_{c}}{\beta_{\perp}}\right)\, ,  \\
=&\frac{1}{16 \pi G}\int d^{2}x{}\sqrt{q}\left(-r^{c}\bar \Gamma_{ca}{}^{a}+\gamma^{ab}r_{c}\bar \Gamma_{ab}{}^{c} - \beta_{\perp}\gamma^{ab}\partial_{a}\partial_{b}f \right) \label{coordinate condition}\, ,  \\
=& \frac{1}{16\pi G} \int d^{2}x{} \sqrt{q} \left(\frac{1}{2} \gamma^{ab}r^{c}\left[\left(\partial_{b}\gamma_{ac}+\partial_{a}\gamma_{cb} - \partial_{c}\gamma_{ab}\right)-\left(\partial_{b}\gamma_{ac}+\partial_{c}\gamma_{ab} \right.\right.\right.\nonumber  \\
&\left.\left.\left.\phantom{abc\frac{1}{16\pi G}d^{2}x\sqrt{q}(} - \partial_{a}\gamma_{bc}\right)\right]\right) \nonumber\, ,  \\
=& \frac{1}{16\pi G} \int d^{2}x\sqrt{q}\left(\gamma^{ab}r^{c}\left[\partial_{a}\gamma_{cb} - \partial_{c}\gamma_{ab}\right]\right)\,. \label{ADM Mass}
\,. \end{align}

We discarded the term $\gamma^{ab}\partial_{a}\partial_{b}f$ in \eqref {coordinate condition} by requiring that we define the surface on the boundary by a function for which this term vanishes there.  This can always be done, since we only require that f vanish on the boundary in order to define $f$.  Therefore, we can always multiply $f$ by another function, and then solve the appropriate Lapace problem in order to find a new $f$ which has a zero on the boundary, but which also satisfies $\gamma^{ab}\partial_{a}\partial_{b}f = 0$.  This then shows us that the variation of the Hamiltonian with respect to $\alpha$ leaves us with only the term \eqref {ADM Mass}, which is known as the ADM Mass.  For asymptotically Cartesian coordinate systems, it is relatively easy to show that this term will very easily yield the mass parameter for simple examples such as the Schwarzschild, Kerr, and Vaidya metrics.  

However, it is very quickly evident that this term is in no way coordinate independent.  In particular, if one were to simply write down the metric for Euclidean space in spherical coordinates, and na\''{i}vely compute the ``mass'', one would find the following, starting by computing the 3-metric:

\begin{equation}
\gamma_{ab} =\left( \begin{tabular}{l c r}
1&0&0 \\
0&$r^{2}$&0  \\
0&0&$r^{2}sin^{2}\theta$  \\ \end{tabular} \right) \nonumber
\,. \end{equation}

We can define $f = r -r_{0}$, which makes our surface one of constant r.  This then yields $\beta_{\perp} =1$, $r^{a} = (1,0,0)$, and $\sqrt{q} = r^{2}sin\theta$.  Putting all of this data into \eqref {ADM Mass}, we get the result:

\begin{equation}
\frac{1}{16\pi G} \int d^{2}x\sqrt{q}\left(\gamma^{ab}r^{c}\left[\partial_{a}\gamma_{cb} - \partial_{c}\gamma_{ab}\right]\right) = \frac{1}{16\pi G} \int d^{2} x_{}\left(r^{2}sin\theta\left[-\frac{4}{r}\right]\right) = -\frac{r_{0}}{G}
\,. \end{equation}

To say the least, this is a silly expression for the amount of mass or energy contained within a sphere of radius $r$ in the Euclidean space!  In particular, as we take the limit $r \rightarrow \infty$, it should be clear that this expression diverges.  Euclidean space, however, is empty, so we would expect that the value we derive would be zero.  Furthermore, if we were to write the Euclidean metric in Cartesian coordinates, it should be manifest that \eqref {ADM Mass} \textbf{is} equal to zero.  We therefore see that this problem is, in fact, a problem with our coordinate system.  Older references such as \cite{MTW} and \cite{Wald}, and  \cite{ADMPaper} will deal with this apparent contradiction by demanding that formula \eqref {ADM Mass} is only applicable for asymptotically flat metrics in which the coordinate system asymptotically approaches the Minkowski metric at an appropriate pace as $r \rightarrow \infty$.  Following work by Hawking \cite{HawkingBoundaryPaper}, newer references (see \cite{aaa}) account for this oddity by including a counterterm in the action equal to $+\frac{r_{0}}{G}$ in the case above, or, more generally, equal to negative one times the ADM Mass of an empty space in the appropriate coordinate system.  This fixes this problem by making the formula for the ADM mass coordinate invariant, but it adds the problem of requiring that a fiducial flat metric be introduced into the formalism in order to define the counterterm.  Below, an alternative technique to this counterterm will be introduced.  It will still retain the problem of requiring a fiducial flat metric, but it will have the advantage of being more compactly defined, and will also have the advantage of being manifestly coordinate invariant.  

So, begin by defining the fiducially flat metric $\tilde \gamma_{ab}$.  It should be defined in such a way that its coordinates are adapted to the spacetime in question--at the minimum, it should have the same coordinate transformation tensor as the true metric does.  For example, if you are dealing with a Kerr solution to Einstein's equation, the fiducially flat metric can be defined by setting the mass parameter of the solution equal to zero.  Then, the idea is to treat the true metric as if it were a tensor field living in the flat metric space.  Then, we can define a torsion-free derivative operator $\mathscr{D}_{a}$ that is compatible with the fiducial flat metric (i.e., so that $\mathscr{D}_{a}\tilde \gamma_{bc} = 0$).  Once this is done, we use the same function $f$ to define the surface at the boundary of our space, only now, we use  $\tilde \gamma_{ab}$ to define a $\tilde \beta_{\perp}$ and  $\tilde r^{a}$.  We then do the integral in \eqref {ADM Mass} over the fiducially flat space, replacing all of the $\partial_{a}$ terms with $\mathscr{D}_{a}$ terms.  The final result is:

\begin{equation}
M_{ADM, new} = \frac{1}{16 \pi G}\int\sqrt{\tilde q}_{} d^{2}x  \tilde \gamma^{ab} \tilde r^{c}\left(\mathscr{D}_{a}\gamma_{bc} - \mathscr{D}_{c}\gamma_{ab}\right) \label{Alternate ADM}
\,. \end{equation}

This expression has several advantages.  In the case of asymptotically flat coordinate systems, all of the $\tilde \Gamma_{ab}{}^{c}$'s are all zero, the metric is equal to $\delta_{ab} + \mathscr{O}(\frac{1}{r})$ at the surface at infinity, and therefore, it should be clear that \eqref {Alternate ADM} gives the same result as \eqref {ADM Mass}.  It should also be clear, moreover, that the integrand of expression \eqref {Alternate ADM} transforms as a scalar under coordinate changes that affect the fiducially flat metric, provided, as stipulated when we defined $\tilde \gamma_{ab}$ above, that both metrics have the same coordinate transformation tensor.  Therefore, it should be clear that \eqref {Alternate ADM} makes sense if the bulk metric is defined in \textit{any} coordinate system, so long as that coordinate system covers an appropriate two-surface at the spacetime's spacelike boundary.  

Furthermore, as a concrete example, consider the Kerr spacetime in the spheroidal coordinates given in Appendix \ref{sec: Kerr}.  Here, the 3-metric is directly obtainable from the 4-metric, and we can find the fiducial metric simply by setting M = 0 \footnote{As always in this work, defining $A = r^{2} + a^{2}$ and $B = r^{2} + a^{2}cos^{2}\theta$, and using the spheroidal coordinates defined by taking $x = \sqrt{A}sin\theta cos \psi$, $y = \sqrt{A}sin\theta sin\psi$, and $z = r_{}{}cos\psi$, followed by a second coordinate transformation $d\phi = d\psi -\frac{a}{A}dr$.  Coordinates are then rendered in the form $(r,\phi,\theta)$.  This form for the metric tremendously simplifies the expression for the metric when compared to the original form, which is a dense 4x4 matrix in both its covariant and contravariant forms. \label{coordinate footnote}}:

\begin{align}
\gamma_{ab} =& \left( \begin{tabular}{l c r} 
$(1 + \frac{2Mr}{B})$&$-(1+\frac{2Mr}{B})a{}^{}sin^{2}\theta$&0  \\
$-(1+\frac{2Mr}{B})a{}^{}sin^{2}\theta$&$\left(\frac{AB+2Mra^{2}sin^{2}\theta}{B}\right)sin^{2}\theta$&0  \\
0&0&$B$ \end{tabular}\right) \nonumber   \\ 
\tilde \gamma_{ab} =& \left( \begin{tabular}{l c r}
1&$-a{}^{}sin^{2}\theta$&0  \\
$-a{}^{}sin^{2}\theta$&$A{}sin^{2}\theta$&0  \\
0&0&$B$ \end{tabular}\right)
\,. \end{align}

Then, it is simple enough to compute the inverses of these metrics, and follow through with the standard prescription given above, and to compute \eqref {Alternate ADM}.\footnote{Of course, we are free to compute \eqref {ADM Mass}, too.  Since, however, this metric reduces to a spherical expression of the Schwarzschild Metric for $a=0$, we would find that this would be beset with the exact same singularity discussed above}  Doing this gives us $\tilde r_{a} = (\sqrt{\frac{B}{A}},0,0)$, and then it only requires the basic definitions of the covariant derivative to compute (dropping the factor of G now, since the metric is written in geometricized units):

\begin{align}
M_{ADM, new}& =\nonumber \, ,  \\
=& \frac{1}{16 \pi} \int d^{2}x \left( \frac{M{}^{}sin\theta\left[ 16r^{4} + 4a^{2}r^{2}(\cos(2\theta)+3)+a^{4}(\cos(4\theta)-1)\right]}{4\left( r^{2}+a^{2}cos^{2}\theta \right)^{2}} \right)
\,. \end{align}

The $\phi$ integration, of course, is easy.  All that it does is cancel a factor of $2\pi$ in the numerator against the factor of $16\pi$ in the denominator.  The $\theta$ integration can be computed with a package such as Mathematica, yielding, amazingly, simply $M$, independently of which radius is chosen for the integration.  This indicates that \eqref {Alternate ADM} has a flavour much more similar to that of Gauss's Law than the expression that you would get from \eqref {ADM Mass}, which would only give you the correct value of $M$ if you were to do the integral on the sphere infinity with the metric expressed in the original Kerr coordinates\footnote{see footnote number \eqref {coordinate footnote}.}, or another asymptotically Cartesian form.  This technique, however, is both manifestly coordinate invariant and works at an arbitrary radius\footnote{while this slicing is invariant under a 3-dimensional coordinate transformation, and always yields the ADM mass at conformal infinity, at other radii, it is not {\it slicing} independent--this alternate ADM mass, when computed in a Boyer-Lindquist slicing, gives simply M at infinity, but will monotonically increase to infinity as one approaches the horizon} without adding explicit counterterms to the action.  Furthermore, it automatically yields a zero result for an empty space, simply by the construction of $\mathscr{D}_{a}$.  The only disadvantage is that it required a massaging of terms after taking a variation, rather than explicitly coming out of a variational principle, in the manner that the terms derived in \cite{ADMPaper} and \cite{aaa} were.  It also retains the problem, inherent to all of these approaches, of defining some sort of flat metric with which to compare the true metric, whether it is done implicitly by choice of coordinates in the style of the original ADM Paper, or more explicitly, as shown here or in the counterterm approach.  
$\newline$
$\newline$
We can apply a similar procedure to the shift vector $\beta_{a}$.  Taking only the pure gravity terms in the ADM Hamiltonian involving $\beta_{a}$, we have:

\begin{equation}
H_{involving \; \beta} = \frac{1}{8\pi G}\left(-\int d^{3}x \sqrt{\gamma} \beta_{b} \bar \nabla_{a} \frac{1}{\sqrt{\gamma}}\Pi^{ab}+ \oint d^{2}x \sqrt{q}\frac{1}{\sqrt{\gamma}}r_{a}\beta_{b}\Pi^{ab}\right)
\,. \end{equation}

Varying this portion of the action with respect to $\beta_{a}$, and remembering that $\Pi^{ab} = \sqrt{\gamma}\left(K^{ab} - \gamma^{ab}K\right)$ gives us:

\begin{equation}
\frac{\delta H}{\delta \beta_{a}} = -\frac{1}{8 \pi G}\left[\int d^{3}x \sqrt{\gamma} \bar \nabla_{a}\left(K^{ab} - \gamma^{ab}K\right) - \oint d^{2}x \sqrt{q} r_{a}\left(K^{ab} - \gamma^{ab}K\right)\right] \label{almost beta}
\,. \end{equation}

The first term above is easily recognized as the 3-dimensional integral of the Momentum constraint, which must therefore vanish.    The second term above, however, does not, in general vanish.  In particular, it can be shown to be equal to $Mv^{i}$ for boosted black holes, and equal to $M\,a\hat \phi^{i}$ for Kerr black holes.  Therefore, this term cannot be expected to vanish for a general solution to Einstein's equations.  Just as in the Maxwell case above, and for the charged case, however, this can be treated simply by adding a boundary counterterm  $\oint d^{2}x \sqrt{q} \left(-P^{a}\beta_{a}\right)$ to the Hamiltonian.  Then, after the bulk equations of motion have been solved, we can then solve for the appropriate value of $P^{a}$ such that it cancels the second term in the variation given in equation \eqref {almost beta}, thereby giving an overall variation of the Hamiltonian that is equal to zero.  Note that each of these counterterms are, in fact, charges corresponding to conserved quantities arising from symmetries in the Hamiltonian/Lagrangian.  The Lapse function is associated with time reparameterization invariance of General Relativity, and therefore, corresponds to an asymptotic energy.  Meanwhile, the shift is associated with the choice of 3-dimensional coordinates on the 3+1 fibers.  Therefore, the shift vector encodes information about asymptotic translations and rotations.  Therefore, the charge, $P^{i}$ associated with it can be associated with linear and angular momentum.

\chapter{Null Geometry} \label{NullVectorSection} \label{sec: NullVectors}
In the following chapters, we are going to be investigating several properties of null submanifolds of general spacetimes.  Null spaces have quite a few special properties that are somewhat nonintuitive if one is used to dealing with nondegenerate vector spaces.  In particular, there is no natural connection between the intrinsic vector space tangent to a null manifold and the equivalent covector space.  Consequently, one must take care in constructing the induced geometry from an enveloping spacetime.  For this reason, explicit examples are worked out in Appendix \ref{sec: Kerr}

First, remember the techniques derived for projecting onto a surface derived from section \ref{sec: Definitions}, summarized here for convenience:
\begin{enumerate}
\item{Find a function $f$ such that $f$ = constant on the surface in question}
\item{choose $f$ as a coordinate in the spacetime in question}
\item{Eliminate the $df$ terms from all relevant forms, and in particular, the lowered version of the metric tensor.  Call this new metric tensor the $\mathbf{induced}$ $\mathbf{metric}$ on the surface}
\item{Take the inverse of the induced metric, and use this as the raising operator on the surface}
\item{If there are vectors that need to be projected, you can use one of the following (equivalent) techniques:}
\begin{enumerate}
\item{If the vector already satisfies $v^{a}df_{a} = 0$, then the appropriate zero component can simply be dropped, and the vector can just be considered to be a vector living in the fiber.  Otherwise:}
\item{lower the vector using the 4-metric, use the above procedure to project, and then raise using the inverse three-metric}
\item{define $n_{a} \equiv \frac{\nabla_{a}f}{\sqrt{|\nabla_{a}f\nabla^{a}f|}}$.  Raise $n_{a}$ using the four-metric.  Then project vector indices using the operator $\gamma_{a}{}^{b} \equiv \delta_{a}{}^{b} - \xi n_{a}n^{b}$, as $v^{a}\gamma_{a}{}^{b}$\footnote{Recall that $\xi$ is the sign of $n_{a}n^{a}$} will, by construction, have zero components along the appropriate direction, which can then simply be dropped.}
\end{enumerate}
\item{If it is necessary to project back into the 4-dimensional space, 3-dimensional vectors can simply be included with the appropriate zeros added}
\item{If you wish to project one-forms back into the 4-dimensional space, however, it is necessary to operate on them using the $\gamma_{a}{}^{b}$ operator}
\end{enumerate}

Now, we are about to need to do the above for the case of a null submanifold of a 4-dimensional space.  It might seem trivial to just follow the above steps that supposedly would allow us to seamlessly go back and forth from the 4-space to the 3-space.  But on closer glance, step (iv) requires that we be able to invert the induced 3-metric.  If the 3-space is null, then the 3-metric necessarily will have an eigenvector with zero eigenvalue, and therefore, will not be invertible.  Also, in the null case, the vector $n_{a}$ defined in (v)(b) will be singular, as $\nabla_{a}f\nabla^{a}f = 0$, which means that the projection/inclusion operator $\gamma_{a}{}^{b}$ cannot be defined.  Therefore, in the null case, some extra care must be taken in the above procedures when working out the appropriate induced geometry. 

So, first, we start by noting that, until the fourth step, there was nothing problematic at all.  In particular, we still consider our surface to be a solution set for some equation.  In particular, for a Kerr spacetime in spheroidal Kerr coordinates, we consider the horizon to  be the solution to the equation $r^{a}+a^{2}-2Mr=0$.   This makes the first few steps easy enough to follow.  In particular, this means that we can map forms in the 4-dimensional spacetime to forms on the null submanifold using the same procedure (i.e., by dropping the $r$ component of the form in the  Kerr spacetime).  Note, however, that this technique will already create some anomalies for null submanifolds.  In particular, it causes the lowered version of the outgoing null normal to the Kerr horizon to be projected to the zero vector on the horizon.   This might lead to some mystery regarding what the three covectors that span the dual tangent space to the Kerr horizon are, but before this mystery is settled, let us soldier along.  

First, let us denote the 3-metric by $q_{ab}$\footnote{There is some trickiness here--it is typical practice to {\it also} define the spacelike metric of the 2-dimensional sections of the horizon defined by constant null parameter as $q_{ab}$.  We follow normal convention here, which is at least aided by the fact that $q_{ab}$ is degenerate with a zero eigenvalue being precisely the difference between the 2-space and the 3-space.  Regardless, the reader is advised to be careful when interpreting something written down as $q_{ab}$.}.  If the form that we omitted in order to define $q_{ab}$ was null according to the 4-metric, then it is the case that $q_{ab}$ can be shown to have zero determinant.  This then means that there is a three vector that is mapped to the zero form by $q_{ab}$.  To ascertain the nature of this vector, note that, by the fact that $df_{a}$ is null, that $df_{a}g^{ab}$ will have no component along the $\partial_{f}$ direction.  We therefore can infer that the vector that is mapped to the zero covector by $q_{ab}$ is, in fact, $g^{ab}df_{b}$, times a function, which we will henceforward refer to as the outgoing null normal $\ell^{a}$.  

Now, let us return to looking at the basis of the covector space.  If we use $q_{ab}$ to evaluate the norms of the vectors in the tangent plane to the null surface, we can see, relatively quickly, that the tangent plane to the null surface is spanned by $\ell^{a}$ and two other spacelike vectors.  Therefore, we would expect to have the covector space spanned by three vectors that, when acting upon unit vectors in the tangent space, will give you Kroneker delta functions as output.  Since the raised versions are already spacelike under the action of $q_{ab}$, for the spacelike vectors, we can simply raise and lower indices using the metric of the horizon just like any other vector.  $q_{ab}\ell^{b}=0$, however, so we cannot do this.  And we already established that the 4-dimensional covector $\ell_{a}$ has a vanishing pullback onto the null surface.  However, if we specify two spacelike vectors and $\ell_{a}$, then there is a unique null covector $k_{a}$ such that it is orthogonal to the two spacelike vectors and $g^{ab}\ell_{a}k_{b} = -1$.  The pullback of this vector onto the horizon (denoted $\underset{\leftarrow}{k_{a}}$) will be the third one-form spanning the covector space.  Then, we will require that $q^{ab} \underset{\leftarrow}{k_{a}}=0$, which guarantees that $q^{ab}$ is degenerate, as well as uniquely specifying $q^{ab}$.  Finally, we conclude this section by pointing out that we now can define the projection operator of vectors onto the null surface as: $P_{a}{}^{b} \equiv -k_{a}\ell^{b} + q^{ac}q_{cb}$, which manifestly has no effect upon the two spacelike vectors normal to $k_{a}$ and $\ell_{a}$, and has no effect on $\ell^{a}$, but maps $k^{a}$ to zero.  

Now, with both $q^{ab}$ and $q_{ab}$ uniquely specified, we can define the intrinsic connection $\tilde \nabla_{a}$ to the null surface according to (for $v^{b}$ tangent to the null surface) $\tilde \nabla_{a}v^{b} \equiv \partial_{a} v^{b} + \tilde \Gamma_{ab}{}^{c}$, where $\tilde \Gamma_{ab}{}^{c} \equiv \frac{1}{2}q^{cm}(q_{am,b}+q_{bm,a}-q_{ab,m})$.  This concludes our section on the embeddings of null submanifolds.  We now progress to analyze Ashtekar's Isolated and Dynamical Horizon formalism.  For the sake of clarity, in the appendix, we work out this decomposition for two null surfaces: a null cone in Minkowski spacetime and the event horizon of a Kerr spacetime.

\section{Null Expansions}
\label{sec: ExpansionAreaProof}
Now that we have defined our two null normals and come up with an unambiguous way to define induced geometries on null surfaces, let us consider a few properties of these null surfaces.  
One quantity that will repeatedly be of critical importance in the following proofs will be the notion of the expansion of a null normal.  As was shown above, the covector space to the horizon will contain one null one-form, labeled $k_{a}$, while the vector space will contain a vector $\ell^{a}$ that is not metrically related to $k_{a}$ in the enveloping spacetime.  We will denote $k_{a}$ as the ingoing null normal, and $\ell_{a}$ as the outgoing null normal.  Note that $\ell_{a}$ is the null generator of the horizon.  
Now, consider the null expansions of these two vectors, defined by $\theta = q^{ab} \nabla_{a} \ell_{b}$ and $\theta_{(k)} = q^{ab} \nabla_{a}k_{b}$.  These will have a few interpretations in this work, but for now, we will interpret them as the time rate of change of two surfaces along flowlines of the relevant vector.  
If one doubts this interpretation, it is easy to show that this is in fact valid:
\begin{align}
£_{\ell}A=&£_{\ell}\int \sqrt{q} = \int \frac{1}{2\sqrt{q}}£_{\ell}q\nonumber \, ,  \\ 
=& \frac{1}{2}\int \sqrt{q} q^{ab}£_{\ell}q_{ab}=\frac{1}{2}\int\sqrt{q}q^{ab}\left(\ell^{c}\partial_{c}q_{ab} + 2 q_{ac}\partial_{b}\ell^{c}\right)\nonumber \, ,  \\
=&\frac{1}{2}\int\sqrt{q}\left(q^{ab}\ell^{c}\partial_{c}q_{ab}+2q^{ab}q_{ac}\left(\nabla_{b}\ell^{c}-\Gamma_{bd}{}^{c}\ell^{d}\right)\right)\nonumber \, ,  \\
=&\frac{1}{2}\int\sqrt{q}\left(2\theta_{(\ell)}+q^{ab}\ell^{c}\partial_{c}q_{ab}-q^{ab}q_{ac}\ell^{d}g^{cm}\left(g_{bm,d}+g_{dm,b}-g_{bd,m}\right)\right)\nonumber \, ,  \\
=&\frac{1}{2}\int\sqrt{q}\left(2\theta_{(\ell)}+q^{ab}\ell^{c}\partial_{c}q_{ab}-q^{mb}\ell^{d}\left(g_{bm,d}+g_{dm,b}-g_{bd,m}\right)\right)\nonumber \, ,  \\
=&\frac{1}{2}\int\sqrt{q}\left(2\theta_{(\ell)}+q^{ab}\ell^{c}\partial_{c}q_{ab}-q^{mb}\ell^{d}\partial_{d}\left(q_{bm}-\ell_{b}k_{m}-\ell_{m}k_{b}\right)\right)\nonumber \, ,  \\
=&\int\sqrt{q}\left(\theta_{(\ell)}\right)
\,. \end{align}
 Therefore, if $\theta_{(\ell)}$ is constant on the two-surface, which will typically be the case in this work, then the rate of change of that surface's area as one Lie drags along $\ell^{a}$ is simply $\theta_{(\ell)}A$, and if $\theta_{(\ell)}=0$, then the surfaces have a constant area.  In the dynamical case, this argument will break down, but the reason for that is that $\ell^{a}$ will cease to be a tangent vector of the dynamical horizon, which will be spacelike, rather than null.
\section{Raychaudhuri's equation}
In this section, we will derive Raychaudhuri's equation for the case of a null vector.  This result will be used frequently throughout this text, so it is useful to show its origin here.  

Consider a non-affinely parameterized null vector $\ell^{a}$.  Define its affine parameter by $\ell^{a} \nabla_{a}\ell^{b} = \kappa \ell^{b}$.  Furthermore, define a second associated null vector $k^{a}$ such that $g_{ab}k^{a}\ell^{b}=-1$, and raise and lower indices on $\ell^{a}$ and $k^{a}$ freely using the metric tensor and its inverse.  Then, the operator $q_{ab} \equiv g_{ab} + \ell_{a}k_{b} + k_{a} \ell_{b}$ annhilates both $\ell^{a}$ and $k^{a}$, as can be verified by inspection.  We wish to consider $q_{ab}$ to, in a sense, be the metric of a two surface, but note that it is degenerate\footnote{Having, after all, two linearly independent null vectors!}, and therefore, has no natural inverse.  We therefore must raise and lower indices on $q_{ab}$ using the full metric tensor, as is customarily the case when dealing with objects in a null geometry.  

Now, consider the tangent gradients of $\ell_{a}$ given by $q_{a}{}^{c}q_{b}{}^{d}\nabla_{c}\ell_{d}$.  Obviously, this is a tensor with two indices, and therefore, can be decomposed into an  antisymmetric part, a traceless symmetric part, and a trace, as shown below:

\begin{equation}
q_{a}{}^{c}q_{b}{}^{d}\nabla_{c}\ell_{d}\equiv \frac{1}{2}q_{ab} \theta + \sigma_{ab} + \omega_{ab}
\,. \end{equation}

\noindent where the factor of $\frac{1}{2}$ is included so that $q^{ab}\nabla_{a} \ell_{b} = \theta$.  It will be of great interest to evaluate the evolution of quantities along flowlines of $\ell^{a}$, which will usually be taken to be the null tangent vector of the horizon.  Furthermore, the quantity $\theta$ will prove to be the critical central element in the definition of Isolated and Dynamical horizons.  Therefore, it is natural to want to calculate the quantity $£_{\ell}\theta$.  So, let us now do so\footnote{Note that $g^{ab}R_{acbd}\ell^{c}\ell^{d} = q^{ab}R_{acbd}\ell^{c}\ell^{d} = R_{ab}\ell^{a}\ell^{b}$, since all of the terms by which $g^{ab}$ and $q^{ab}$ differ include factors of $\ell^{a}$, and the antisymmetry of the Riemann tensor will guarantee that these terms vanish.}.  Note that similar derivations can be found in several sources, including \cite{Wald} \cite{aaa}

\begin{align}
£_{\ell}\theta =& £_{\ell}\left(q^{ab}\nabla_{a} \ell_{b}\right) = q^{ab}£_{\ell}\nabla_{a}\ell_{b} + \left(\nabla_{a}\ell_{b}\right)£_{\ell}q^{ab}\nonumber \, ,  \\
=&q^{ab} \ell^{c}\nabla_{c}\nabla_{a} \ell_{b} + q^{ab} \left(\nabla_{c}\ell_{b}\right)\nabla_{a}\ell^{c} + q^{ab}\left(\nabla_{a}\ell_{c}\right)\nabla_{b}\ell^{c}\nonumber   \\
&+\left(\nabla_{a}\ell_{b}\right)\left(\ell^{c}\nabla_{c}q^{ab}-q^{cb}\nabla_{c}\ell^{a} - q^{ac}\nabla_{c}\ell^{b}\right)\nonumber \, ,  \\
=&q^{ab} \ell^{c}R_{cab}{}^{d}\ell_{d} + q^{ab} \ell^{c} \nabla_{a}\nabla_{c}\ell_{b}+q^{ab} \left(\nabla_{c}\ell_{b}\right) \nabla_{a}\ell^{c}+q^{ab}\left(\nabla_{a}\ell_{c}\right)\nabla_{b}\ell^{c} \nonumber   \\
&+\left(\nabla_{a}\ell_{b}\right)\left(\ell^{c}\nabla_{c}\left(\ell^{a}k^{b}+\ell^{b}k^{a}\right)-q^{cb}\nabla_{c}\ell^{a} - q^{ac}\nabla_{c}\ell^{b}\right)\nonumber \, ,  \\
=&-R_{ab}\ell^{a}\ell^{b} + q^{ab} \nabla_{a}\left(\ell^{c}\nabla_{c}\ell_{b}\right)-q^{ab} \left(\nabla_{a}\ell^{c}\right)\nabla_{c}\ell_{b}+q^{ab}\left(\nabla_{a}\ell^{c}\right)\nabla_{c}\ell_{b} \nonumber   \\
&+ q^{ab}\left(\nabla_{b}\ell^{c}\right)\nabla_{a}\ell_{c}+\left(\nabla_{a}\ell_{b}\right)\ell^{a}\ell^{c}\nabla_{c}k^{b}+\left(\nabla_{a}\ell_{b}\right)k^{b}\ell^{c}\nabla_{c}\ell^{a}\nonumber   \\
&+\left(\nabla_{a}\ell_{b}\right)\ell^{c}k^{a}\nabla_{c}\ell^{b}-q^{cb}\left(\nabla_{a}\ell_{b}\right)\nabla_{c}\ell^{a}-q^{ac}\left(\nabla_{a}\ell_{b}\right)\nabla_{c}\ell^{b} \nonumber \, ,  \\
=&-R_{ab}\ell^{a}\ell^{b}+q^{ab} \nabla_{a}\left(\kappa \ell_{b}\right)+q^{ab}\left(\nabla_{b}\ell^{c}\right)\nabla_{a}\ell_{c}+\kappa \ell_{b}\ell^{c}\nabla_{c}k^{b}\nonumber   \\
&+\kappa \ell^{a}k^{b}\nabla_{a}\ell_{b}+\kappa \ell^{b}k^{a}\nabla_{a}\ell_{b} - q^{ab}\left(\nabla_{c}\ell_{b}\right)\nabla_{a}\ell^{c}-q^{ab}\left(\nabla_{a}\ell_{c}\right)\nabla_{b}\ell^{c}\nonumber \, ,  \\
=&-R_{ab}\ell^{a}\ell^{b}+ \kappa \theta -\kappa k^{b}\ell^{c}\nabla_{c}\ell_{b} + \kappa \ell^{a}k^{b}\nabla_{a}\ell_{b} -q^{ab}\left(\nabla_{c}\ell_{b}\right)\nabla_{a}\ell^{c} \nonumber \, ,  \\
=&-R_{ab}\ell^{a}\ell^{b} + \kappa \theta - \left(\frac{1}{2}\theta q^{ba} + \sigma^{ba} + \omega^{ba}\right)\left(\frac{1}{2}\theta q_{ab} +\sigma_{ab}+\omega_{ab}\right) \nonumber   \\
=&-8\pi T_{ab}\ell^{a}\ell^{b} + \kappa \theta -\frac{1}{2}\theta^{2} - \sigma^{ab}\sigma_{ab} +\omega^{ab}\omega_{ab} \label{Null Raychaudhuri Equation}
\,. \end{align}

\noindent where the Einstein equation was used in the last line in order to convert the Ricci tensor into a stress energy tensor.  

Equation \eqref {Null Raychaudhuri Equation} is the null Raychaudhuri equation.  It is of critical importance in several applications involving various horizons in relativity.  In particular, its use was critical in the singularity proofs of Hawking and Penrose that showed that the existence of a closed trapped surface\footnote{i.e., one for which, given a closed 2-surface and its two associated null vectors, $\ell^{a}$ and $k^{a}$, you have  $\theta_{(\ell)} \leq 0$  and $\theta_{(k)} < 0$.} along with a reasonable restriction on the matter content of spacetime, you will necessarily have a spacetime singularity.

\chapter{Isolated Horizons}
\begin{figure}
\begin{tikzpicture} 
\draw[dashed](0,0)--(2.83,-2.83)--(5.67,0); 
\draw(-1.5,-2.5)--(7.2,-2.5); 
\draw(-1.7,-1.7)--(7.4,-1.7); 
\draw[x=3.141ex,y=1ex] (0,0) sin (1,1) cos (2,0) sin (3,-1) cos (4,0) sin (5,1) cos (6,0) sin (7,-1) cos (8,0) sin (9,1) cos (10,0);
\draw(5.67,0)--(7.67,-2)--(2.8,-6.8)--(-2,-2)--(0,0); 
\draw(1.4,-1.4)--(2.83,-2.2)--(4.3,-1.4); 
\draw [white](0,0)--node[above=2pt,black,fill=white] {$T_{1}$} (0.1,-5.8);
\draw [white](0,0)--node[above=2pt,black,fill=white] {$T_{2}$} (0.1,-4.1);
\draw [white](0,0)--node[above=2pt,black] {$i_{0}$} (15.7,-4.7);
\draw [white](0,0)--node[above=2pt,black] {$i_{0}$} (-4.3,-4.8);
\draw [white](0,0)--node[above=2pt,black] {$i_{-}$} (6,-15);
\draw [white](0,0)--node[above=2pt,black] {$i_{+}$} (-0.5,-0.2);
\draw [white](0,0)--node[above=2pt,black] {$i_{+}$} (12.1,-0.5);
\draw [white](0,0)--node[above=2pt,black,fill=white] {$\mathscr{J}^{+}$} (13.2,-2.3);
\draw [white](0,0)--node[above=2pt,black,fill=white] {$\mathscr{J}^{+}$} (-1.8,-2.5);
\draw [white](0,0)--node[above=2pt,black,fill=white] {$\mathscr{J}^{-}$} (10.7,-9.5);
\draw [white](0,0)--node[above=2pt,black,fill=white] {$\mathscr{J}^{-}$} (0.7,-9.2);
\end{tikzpicture}
\vspace{-230pt}
\caption{A Penrose-Carter Diagram of a Vaidya spacetime with a mass function that is linear for a finite advanced time.  The dashed line represents the event horizon, and the solid line in the interior of the diagram represents the dynamical horizon.  $T_{1}$ represents a spacelike slice which intersects the event horizon, but has zero extrinsic curvature and a Euclidean metric.  $T_{2}$ represents a spacelike slice that intersects the dynamical horizon.  All slices after the dynamical horizon merges with the event horizon are identical to slices of Schwarzschild.}
\label{fig:VaiydaDiagram}
\end{figure}
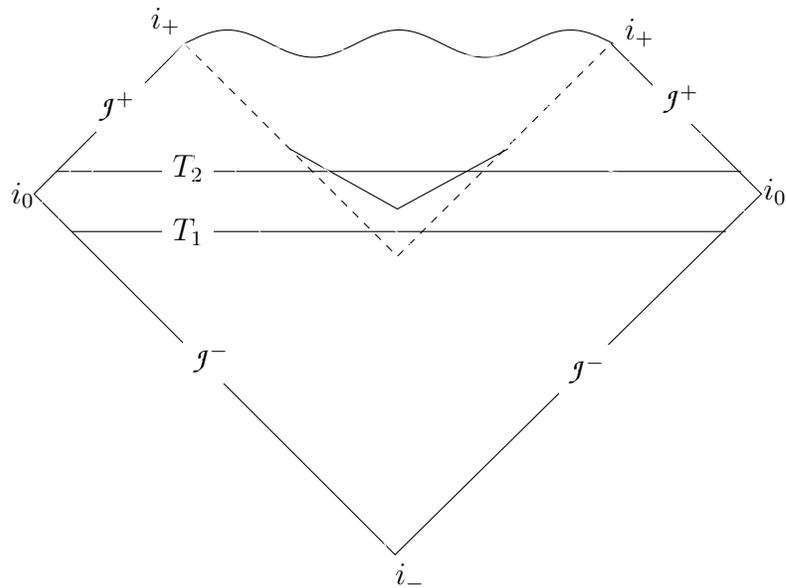

Now, let us consider the Isolated and Dynamical Horizon\footnote{Henceforward, we will use the abbreviation IH to denote Isolated Horizon, and DH to denote Dynamical Horizon.} framework as developed by Ashtekar and others.  Previous attempts to define black holes have been fraught with difficulty.  The most typical definition of the boundary of the black hole was the event horizon, which is defined as the past development of timelike future infinity.  While this definition has a convenient feature of capturing the nature of `the point of no return' that is typically associated with black holes, it also faces the significant problem that it is defined only globally in a spacetime.   Therefore, locating a spacetime's event horizon requires knowledge of the entire future development of that spacetime.  In particular, there are simple analytical examples (i.e., the Vaidya metric where M(v) is nonconstant for a compact subdomain of v\footnote{As a concrete example, take $M(v) = 0$ for $v < 0$, $M(v) = M_{0}v$ for $0<v<1$ and $M(v) =M_{0}$ for $v>1$, as shown in figure \ref{fig:VaiydaDiagram}.  There will be regions in the past domain of dependence of the late-time singularity (More succinctly, points inside of the event horizon) that will have completely flat intrinsic and extrinsic geometries.  Observers at these times will have no way of knowing that they are inside of an event horizon.}) where event horizons exist in flat regions of spacetime.  Therefore, local observers will have no means by which to identify whether or not an event horizon is present.  Furthermore, beyond this conceptual `epistemological' objection, there is the much more practical problem that the global definition of the event horizon makes it difficult to locate its intersection with a particular Cauchy slice of a spacetime, and therefore, very difficult to locate numerically  (and in fact, for the Vaidya example discussed above, the event horizon \textit{cannot} be located using only quantities defined locally on that particular Cauchy slice, as the points for which $v<0$ were exactly isomorphic to finitely sized regions of Minkowski spacetime.).  However, having at our disposal all of the machinery of isometric embeddings and manifold theory in the DH case, while the behaviour of the horizon can still be, in the abstract, wild and superluminal, we now know something of how to track the motion of the DH across the Cauchy slice.

Therefore, the focus began to be shifted toward finding a definition of the `point of no return' that could be defined locally, or at least quasi-locally.  The first attempt to do so was done by Hawking, with his notion of Apparent Horizon, which he defined as the boundary of a `trapped region'.  This was an improvement over the event horizon as a notion, as it is definable at a moment of time.  It still proved problematic, however, as it does require some use of the non-local notion of a `trapped region'.  In particular, this problem would lead to several cases where the behaviour of the apparent horizon proved to be `wild'--apparent horizons would move and evolve discontinuously in certain spacetimes for certain slicings.  Still, they have proven useful, and have been implemented numerically.

All of these problems, however, have been removed by the closely related notions of Hayward's trapping horizon, and Ashtekar et al.'s notion of Isolated and Dynamical Horizons.  These notions require merely that the notion of non-expansion be confined to a single null or spacelike submanifold of the full spacetime.  In particular, this guarantees that the wild behaviour discussed above can be avoided--since the horizon is defined to be a smooth manifold, the behaviour of its slicings is guaranteed to also be smooth.  It is therefore considered to be quite compatible with a great number of approaches in Relativity, including numerical ones.  Using the machinery developed in the previous chapter, we now work to develop this framework in detail.  We start with the much simpler and restrictive notion of Isolated Horizon, and move on to the Dynamical horizon.  Finally, this chapter will conclude with a proof of the Laws of black hole dynamics using this formalism, and then with a concrete application to the Vaiyda spacetime.  

Now, consider a 3-manifold that is a subspace of a Lorentzian 4-manifold.  Following Ashtekar and Krishnan, this three manifold will be considered to be a non-expanding horizon if the following three conditions are met:
\begin{enumerate}
\item{The 3-manifold is null, topologically $\mathbb{R} \times \mathbb{S}^{2}$ or $\mathbb{R} \times \mathbb{T}^{2}$ and has the null vector $\ell^{a}$ as a tangent vector, and associated (degenerate) metric $q_{ab}$}
\item{On the three-mainifold, the expansion $\theta_{(\ell)} = q^{ab}\nabla_{a}\ell_{b}$ vanishes}
\item{The 4-metric $g_{ab}$ satisfies Einstein's equation, and all relevant matter equations of motion are satisfied, and any matter involved obeys the Dominant Energy Condition.}
\end{enumerate}

The third condition just specifies that we are in fact talking about General Relativity, and the Dominant Energy Condition guarantees the vanishing of certain terms in the Raychaudhuri equation, as will be seen below.  Meanwhile, the first condition guarantees that you have a static, non-evolving horizon, as translation along the null vector $\ell^{a}$ leaves one on the three-manifold.  The essential condition, therefore, is the second one, which captures the essence of the Horizon being the outermost `point of no return':  The null geodesics that have tangent vector $\ell^{a}$ simply transport null observers along a horizon of constant area\footnote{One might wonder about this statement regarding area.  See \eqref {sec: ExpansionAreaProof}}, rather than allow them to move spatially, as would typically be the case for a `typical' null hypersurface, like a null-cone in Minkowski spacetime.  In particular, this guarantees that, for some function $\kappa$, defined on the horizon, $\ell^{a}\nabla_{a}\ell^{b} = \kappa \ell^{b}$.   There will have much more to say about $\kappa$ further along.

Now, consider the Null Raychaudhuri equation:

\begin{equation}
\ell^{a}\tilde \nabla_{a} \theta_{(\ell)} = -\frac{1}{2}\theta_{(\ell)}^{2}-\sigma_{ab}\sigma^{ab} + \omega_{ab}\omega^{ab} - 8\pi T_{ab}\ell^{a}\ell^{b}\
\,. \end{equation}

Where $\sigma_{ab} \equiv P_{a}{}^{m}P_{b}{}^{n}\nabla_{(m}\ell_{n)}- \frac{1}{2}q_{ab}\theta_{(\ell)}$, and $\omega_{ab} \equiv  P_{a}{}^{m}P_{b}{}^{n}\nabla_{[m}\ell_{n]}$.  Since we know that the horizon is defined by the fact that $\theta_{(\ell )} = 0$, and that it has $\ell^{a}$ as a null tangent vector, we know that the left hand side of this equation vanishes on the horizon.  Furthermore, we know that $\omega_{ab}$ vanishes, since the pullback of $\ell_{a}$ onto the horizon also vanishes.  Thus, the above equation reduces to:

\begin{equation}
0 = \sigma_{ab}\sigma^{ab} + 8 \pi T_{ab}\ell^{a}\ell^{b}
\,. \end{equation}

Since both of these terms are manifestly positive (the horizon metric is non-negative-definite, and the null energy condition is satisfied), this means that they both must independently vanish.  Now, since the decomposition of $P_{a}{}^{m}P_{b}{}^{n} \nabla_{a}\ell_{b} \equiv \tilde \nabla_{a}\ell_{b}$ is:

\begin{align}
\tilde \nabla_{a}\ell_{b} =& \frac{1}{2}q_{ab} \theta_{(\ell)} + \sigma_{ab} + \omega_{ab}\, ,  \\
=&\sigma_{ab}\,,
\,. \end{align}

\noindent and since $\ell_{b}\ \nabla_{a}\ell^{b}=0$ along with $\omega_{ab} = \sigma_{ab} = \theta_{(\ell)} = 0$, it must be the case that all contractions of $q^{ma}q_{nb}\nabla_{a}\ell^{b}=\tilde \nabla_{m}\ell^{n}=0$. Therefore, we must have, on the horizon only, $$\nabla_{a}\ell^{b} = \omega_{a}\ell^{b} + \tilde \omega_{a}k^{b}\,.$$  Since $\ell^{a}$ has a fixed norm, however, we know that $\ell_{a}\nabla_{b}\ell^{a} = 0$, since it is equal to minus itself by the product rule.  Therefore, contracting the above expression for $\nabla_{a}\ell^{b}$ onto $\ell_{b}$ gives us the result that $\tilde \omega_{a} = 0$.  Therefore, on the horizon, we have the condition:

\begin{equation}
\nabla_{a}\ell^{b} = \omega_{a}\ell^{b}\,.
\,. \end{equation}

Where $\omega_{a}$ is called the H\'{a}ji\v{c}ek one-form, and should be taken as a three component object intrinsic to the horizon.  Also, it should be clear that $\kappa = \ell^{a}\omega_{a}$.  In computations, it is often simpler to take advantage of the fact that $\ell^{a}k_{a} = -1$, and work with $\omega_{a} = - \displaystyle \underset{\leftarrow}{k_{b}\nabla_{a}\ell^{b}}$ since, among other things, the meaning of the pullback operator is clear, as one can compute the action of the derivative operator, perform the index contraction and $\mathbf{then}$ perform the pullback operation.  

Now, note that all of the above is completely invariant if we replace $(\ell_{a},k_{a})$ with $(L_{a},K_{a})$ using the rescaling transformation:
\begin{equation}
L_{a} = f \ell_{a} \;\;\;\;\;\;\;\;\;\; K_{a} = \frac{1}{f}k_{a} \label{rescale}
\,. \end{equation}

Since all results above were merely dependent on the inner product $\ell_{a}k^{a}$ being equal to minus one, and each of them being a null vector, and both of these are true of you rescale $\ell_{a}$ by an arbitrary function f.  Note, however, that when you apply this rescaling transformation, you induce a change in $\omega_{a}$:

\begin{align}
K_{b}\nabla_{a}L^{b} =& \frac{1}{f} k_{b}\nabla_{a}(f \ell^{b})\, ,  \\
=&k_{b}\nabla_{a} \ell^{b} -\frac{1}{f}\nabla_{a}f
\,. \end{align}

Which, after pulling back onto the horizon, gives us the result 

\begin{equation}
\omega_{a} \rightarrow \omega_{a} + \underset{\leftarrow}{\nabla}{}_{a} ln(f)
\,. \end{equation}

And, since $\kappa = \ell^{a}\omega_{a}$, we also get $\kappa \rightarrow f\kappa + £_{\ell}f$.  This rescaling freedom gives us the ability to, without loss of generality, impose more stringent conditions upon our horizon.  In particular, if we choose an appropriate function, we can impose the condition that $\omega_{a}$ not evolve along flowlines of $\ell^{a}$.  More formally, if we make this choice, we can declare that the horizon be a ``Weakly Isolated Horizon'' (WIH) by saying that it satisfies:

\begin{equation}
0=[£_{\ell}, \underset{\leftarrow}{\nabla}{}_{a}]\ell^{b} =£_{\ell} (\omega_{a}\ell^{b}) = \ell^{b}£_{\ell}\omega_{a} \label{WIH}
\,. \end{equation}

Where we twice used the condition that $£_{\ell}\ell^{a} = 0$.  Now, the role of the rescaling freedom becomes clear.  If $\omega_{a}$ does not satisfy \eqref {WIH}, then we simply choose to rescale $\ell_{a}$ and $k_{a}$ by some $f$ according to \eqref {rescale}, and solve the differential equation for $f$ in such a way that \eqref {WIH} is satisfied.  This will then fix the function $f$ up to a positive constant (since $\ell^{a}$ needs to remain future-pointing and nonzero).  Once this choice is made, it will impose the condition that $\kappa$ be constant on the Horizon (though it does not fix the nonzero value of $\kappa$).  This result is called the Zeroth Law of Black Hole Dynamics.  

\section{The First Law of Black Hole Dyanmics}
\subsection{A Simple Proof of the First Law}
We can use the isolated and dynamical horizon rules to derive the first law of black hole dynamics.  The proof will be somewhat laborious, however, and rather than directly going into it, we will instead derive the first law quickly here, and then generate the full proof.  The below proof will have the advantage of being direct and easy.  It will have the disadvantage of being very dependent on the details of the Kerr solution.  This simpler proof is outlined in \cite{aaa}

Now, consider the Kerr spacetime, with metric as given in Appendix \ref {sec: Kerr}.  We define the horizon to be the surface at which $A-2\,M\,r =r^{2}+a^{2}-2\,M\,r =0$.  Each closed 2-dimensional section will also have a constant $t$ coordinate.  Using the decomposition formalism built up  in Chapter \ref{sec: NullVectors}, it is apparent that the 2-dimensional metric of this surface is given by:

\begin{equation}
q_{AB} = \left( \begin{tabular}{l r}
$\frac{A_{+}^{2}sin^{2}\left(\theta\right)}{B_{+}}$&0  \\
0&$B_{+}$\end{tabular}\right) \label{TwoDimensionalHorizon}
\,. \end{equation}

Where the coordinates are $(\phi, \theta)$ and $A_{+}$ and $B_{+}$ denote that the functions $A$ and $B$ as defined in Appendix \ref{sec: Kerr} take the values of $r$ required by the equation $r^2-2\,M\,r-a^{2} =0$.  It is now easy to compute the area of the 2-dimensional surface:

\begin{align}
A =& \int d^{2}x\,\sqrt{q} = \int_{0}^{\pi} d\theta\int_{0}^{2\pi}d\phi A_{+}\sin\left(\theta\right)=4\pi\,A_{+}\nonumber \, ,  \\
=& 4\pi\left(r_{+}^{2}+a^{2}\right)=4\pi\left(M^{2}+2\,M\sqrt{M^{2}-a^{2}}+M^{2}-a^{2}+a^{2}\right)\nonumber \, ,  \\
=&4\pi \left(2\,M^{2} + 2\,M\sqrt{M^{2}-a^{2}}\right) \label{Kerr Black Hole Area}
\,. \end{align}

Meanwhile, it is easy to calculate the angular momentum in the Kerr spacetime.  Remembering the boundary formulation given in \eqref {sec: ADM Boundary}, we see that it is necessary to add a term $\oint d^{2}x \, P^{i}\beta_{i}$ to the ADM Hamiltonian.  Where $P^{a}$ is given by the solution valued integral:

\begin{equation}
P^{i}=\oint d^{2} x\, \bar r_{j}\left(K^{ij}-\gamma^{ij}K\right)
\,. \end{equation}

In order to find the angular momentum, we look for the $\phi$ component of this vector.  Due to the vagaries of vector valued integrals, 
we must multiply by a unit vector in the $\phi$ direction under the integral.  After multiplying by this vector, and then taking the limit to infinity, we get:


\begin{equation}
J = P^{\phi} = M\,a \rightarrow a = \frac{J}{M}
\,. \end{equation}

Now, we insert this relationship into the expression \eqref {Kerr Black Hole Area} to eliminate the dependence on $a$ and replace it with dependence on $J$.  The result is:

\begin{equation}
\frac{1}{8\pi}A = M^{2} + \sqrt{M^{4}-J^{2}}
\,. \end{equation}

Now, we take the variation of this equation, treating $M, J,$ as independent variables with independent variations:

\begin{align}
\frac{1}{8\pi} \delta A &= 2 M \delta M + \frac{1}{2}\left(\frac{4 M^{3} \delta M - 2 J \delta J}{\sqrt{M^{4} - J^{2}}}\right) \nonumber\, ,  \\
\frac{1}{8\pi} \delta A + \frac{J \delta J}{\sqrt{M^{4}-J^{2}}} &=\delta M\left( \frac{2M\sqrt{M^{4}-J^{2}} + 2M^{3}}{\sqrt{M^{4}-J^{2}}}\right) \nonumber \, ,  \\
\delta M =& \frac{1}{8\pi}\delta A\left(\frac{\sqrt{M^{4} - J^{2}}}{2M\sqrt{M^{4}-J^{2}}+ 2M^{3}}\right) \nonumber \, ,  \\
&+ \delta J \left(\frac{J}{2M\sqrt{M^{4}-J^{2}} + 2M^{3}}\right) \nonumber   \\
=&\frac{1}{8\pi}\delta A \left(\frac{\sqrt{M^{2} - a^{2}}}{2M^{2} + 2M\sqrt{M^{2}-a^{2}}}\right)\nonumber   \\
&+ \delta J \left(\frac{a}{2M^{2}+2M\sqrt{M^{2}-a^{2}}} \right) \nonumber \, ,  \\
=& \frac{1}{8 \pi} \kappa \delta A + \Omega \delta J \label{Kerr Proof of First Law}
\,. \end{align}

Where the multiplier $\kappa$ of $\frac{1}{8\pi} \delta A$ is the surface gravity of the horizon, defined from the two null normals $\ell^{a}$ and $k^{a}$ to the horizon by $\kappa = - \ell^{a}k^{b} \nabla_{a}\ell_{b}$, while $\Omega$ is the rotation parameter to the horizon, most directly defined as the value that $\frac{g^{r \phi}}{g^{\phi \phi}}$ takes on the horizon\footnote{And, thus, in a na\''ive sense, the ``rate of rotation'' of the space around the horizon.}.  The generalization to the Kerr-Reissner-Nordstrom case is straightforward, and in the end will only involve the addition of a $\Phi \delta Q$ term, where $Q$ is the charge parameter of the black hole, and $\Phi$ is the electrostatic potential at the horizon.  

Several things should be noted about this derivation.  First, it was heavily dependent on the Kerr solution.  We had to explicitly calculate the area of the Kerr horizon, and then take its variation.  Therefore, this can really only be interpreted as a statement about the phase space of Kerr (Reissner-Nordstrom) spacetimes, and perturbations thereof, and not a general statement about black holes.  Furthermore, it required some odd mixtures of terms--the mass, charge and angular momentum, in standard formulations, are properly defined by integrals at infinity.  We performed one of these above in order to obtain the expression $J=a\,M$.  On the other hand, $\Omega$, $\kappa$, $\Phi$, and A are properties inherently intrinsic to the horizon itself.  They would, in principle, be computable even in spacetimes without boundary so long as those spacetimes contained closed trapped surfaces.  It seems weird to have a single equation that multiplies these objects together haphazardly without much regard to which surface they belong.  Older, more general formulations of the first law also have found themselves beset by this problem.  And it is the task of the Isolated and Dynamical Horizon formulation to remove these problems by generating a formulation of the First Law that is both not dependent on the details of a particular solution and also in terms of quantities wholly defined on the horizon.  In order to do this, however, we must continue further into the Isolated/Dynamical Horizon rabbit hole in order to build up some tools.

\chapter{Decomposition of quantities on the horizon into ADM quantities}
Another key use of the IH/DH formalism is the ability that it gives you to decompose 3+1 quantities into quantities well defined on the horizon.  Now, consider the case of either an isolated or dynamical horizon.  In general, a spacelike section of a 3+1 split will intersect an isolated or dynamical horizon in 2-dimensional spacelike slices.  Each of these slices will therefore have a spacelike normal $\bar s^{a}$ lying in the 3+1 slice, and will also be perpendicular to the timelike normal $n^{a}$ used to perform the 3+1 slicing.  Therefore, we can make the usual choice $\ell^{a} =\alpha\left(n^{a}+ \bar s^{a}\right)$ to decompose the outgoing null vector to the horizon, where the proportionality to $\alpha$ is chosen so that $\ell^{a}$ mirrors the time vector $t^{a}$ as closely as possible.  We will first develop some generic formalism that works for all zero expansion surfaces.  Then, we will work out some quantities specialized to the Isolated and Dynamical cases.
\section{Doubly decomposing ADM quantities on the horizon}

It should be clear that that the project described above, where one defines an ADM splitting, and then, within the context of that splitting, divides the 3-space into the tangents and normals to an intersection of the horizon with the particular 3+1 slice, is exactly the sort of the double foliation discussed in Appendix \ref{sec: variousfoliations}.  The formalism developed above can now be used to derive powerful relationships between the intrinsic horizon quantities and the intrinsic 3+1 quantities.  In order to accomplish this, we are going to need to work out a few identities.  First, recall from equation \eqref {4GradientN} that 
$$\nabla_{a}n_{b} = -n_{a} \bar \nabla_{b} ln\left(\alpha\right) -K_{ab}  $$
It is our goal to decompose the quantity $\nabla_{a}\bar s_{b}$ in the same way.  In order to do this, we write down the quantity, and then expand the 4-metric in terms of projection operators and normal operators:
\begin{align}
\nabla_{a} \bar s_{b} =& g_{a}{}^{c}g_{b}{}^{d}\nabla_{c}\bar s_{d} \nonumber \, ,  \\
=& \left(\gamma_{a}{}^{c}-n_{a}n^{c}\right)\left(\gamma_{b}{}^{d}-n_{b}n^{d}\right)\nabla_{c}\bar s_{d} \nonumber \, ,  \\
=& \gamma_{a}{}^{c}\gamma_{b}{}^{d}\nabla_{c}\bar s_{d} -\gamma_{a}{}^{c}n_{b}n^{d}\nabla_{c}\bar s_{d} - n_{a}n^{c}\gamma_{b}{}^{d}\nabla_{c}\bar s_{d} +n_{a}n_{b}n^{c}n^{d}\nabla_{c}\bar s_{d} \nonumber \, ,  \\
=&\bar \nabla_{a}\bar s_{b} + \gamma_{a}{}^{c}n_{b}\bar s_{d}\nabla_{c}n^{d}-n_{a}n_{b}n^{c}\bar s_{d}\nabla_{c}n^{d} - n_{a}n^{c}\gamma_{b}{}^{d}\nabla_{c}\bar s_{d}\nonumber \, ,  \\
=&\bar \nabla_{a}\bar s_{b} -n_{b}\bar s_{d}K_{a}{}^{d}-n_{a}n_{b}\bar s_{d} \bar \nabla^{d}ln\left(\alpha\right)-n_{a}\gamma_{b}{}^{d}n^{c}\nabla_{c}\bar s_{d}
\,. \end{align}

So, we have decomposed the first three terms relatively easily in terms of 3+1 quantities.  Furthermore, since the argument used in deriving equation \eqref {4GradientN} made no assumptions about the dimension of the space involved nor the signature of the normal vector, we can further decompose the first term into $-H_{ab} - \bar s_{a} \hat \nabla_{b} ln\left(\bar B\right)$, where $H_{ab}$ is the extrinsic curvature of the horizon section in the 3+1 slice, and $\bar B$ is the generalized Lapse function of the 2+1 foliation in the neighborhood of the horizon\footnote{Less formally, $\bar B = \frac{1}{\sqrt{\gamma^{RR}}}$, where $R=$constant determines the location of the horizon, and is chosen as a coordinate on the 3+1 slice}.  The last term above, however, requires a more intricate analysis.  First, we need to calculate the inner product of the unbarred radial vector with the normal vector\footnote{See the first section of this chapter for more clarity here.  In essence, the unbarred radial vector is $s_{a} =B \nabla_{a}R$, where $B$ is chosen in such a way as to make $s_{a}$ a unit normal relative to $g^{ab}$ while $\bar s_{a} = \bar B \bar \nabla_{a}R$ is a unit normal relative to $\gamma^{ab}$.} in terms of 3+1 variables:

\begin{align}
n^{a}s_{a} =& g^{ab}n_{a}s_{b} = \alpha \,B g^{ab}\left(\nabla_{a}\tau\right)\left(\nabla_{b}R\right)\nonumber \, ,  \\
=&\alpha\,B\left(g^{ab}\nabla_{a}\tau\right)\nabla_{b}R \nonumber \, ,  \\
=&\alpha\,B\left(-\frac{1}{\alpha^{2}}n^{b} + \frac{1}{\alpha^{2}}\beta^{b}\right)\left(\frac{1}{\bar B}\bar s_{b}\right)\nonumber\, ,  \\
=&\left(\frac{B}{\alpha\,\bar B}\right)\beta^{a}\bar s_{a}
\,. \end{align}

Now, we compute the value of $-\gamma_{b}{}^{d}n^{c}\nabla_{c} \bar s_{d}$.  Before moving on, note that the fixed norm of $\bar s^{a}$ means that the contraction of this term onto $\bar s^{b}$ vanishes.  Therefore, we can replace the 3-metric appearing in this term with the 2-metric.  Now, we compute this term, taking advantage of the lack of torsion to interchange the order of derivatives of a scalar:
\begin{align}
-\gamma_{b}{}^{d}\,n^{c}\nabla_{c} \bar s_{d} =& -q_{bd}\,n^{c}\nabla_{c}\left(\bar B \gamma^{de}\nabla_{e}R\right)\nonumber \, ,  \\
=&-q_{bd}\,n^{c}\bar B \gamma^{de}\nabla_{c}\nabla_{e}R -q_{bd}\,n^{c} \bar B\left(\nabla_{e}R\right)\nabla_{c}\gamma^{de}\nonumber  \\
& -q_{bd}\,n^{c}\gamma^{de}\left(\nabla_{e}R\right)\nabla_{c}\bar B\nonumber \, ,  \\
=&-q_{b}{}^{e}\,n^{c}\bar B\nabla_{c}\nabla_{e}R -q_{bd}\,n^{c}\bar B\left(\nabla_{e}R\right)\left(n^{d}\nabla_{c}n^{e}+n^{e}\nabla_{c}n^{d}\right)\nonumber   \\
&-q_{bd}\,n^{c}\bar s^{d}\nabla_{c}ln\left(\bar B\right)\nonumber \, ,  \\
=&-q_{b}{}^{e}n^{c}\bar B \nabla_{e}\nabla_{c}R -\bar B\,q_{bd}\,n^{c}n^{e}\left(\nabla_{e}R\right)\nabla_{c}n^{d} \nonumber \, ,  \\
=&-\bar B\,q_{b}{}^{e}\nabla_{e}\left(n^{c}\nabla_{c}R\right)+\bar B \,q_{b}{}^{e}\left(\nabla_{e}n^{c}\right)\nabla_{c}R - \bar B\,q_{bd}\,\left(n^{e}\nabla_{e}R\right)n^{c}\nabla_{c}n^{d} \nonumber \, ,  \\
=&-\bar B \,q_{b}{}^{e}\nabla_{e}\left(\frac{1}{B}n^{c}s_{c}\right)-q_{b}{}^{c}\bar s^{d}K_{cd}-\frac{\bar B}{B}\,q_{bd}\,n^{c}s_{c}\bar \nabla^{d}ln\left(\alpha\right)\nonumber \, ,  \\
=&-\bar B q_{b}{}^{c}\nabla_{c}\left(\frac{1}{\alpha \bar B}\beta^{a}\bar s_{a}\right)-q_{b}{}^{c}\bar s^{d}K_{cd} -\frac{1}{\alpha}\beta^{a}\bar s_{a}\hat \nabla_{b}ln\left(\alpha\right) \nonumber \, ,  \\
=&-\bar B q_{b}{}^{c}\left[-\frac{\nabla_{c}\alpha}{\alpha^{2}\bar B}\beta^{a}\bar s_{a} -\frac{\nabla_{c}\bar B}{\alpha \bar B^{2}}\beta^{a}\bar s_{a} +\frac{1}{\alpha \bar B}\left(\beta^{a}\bar s_{a}\right)\right]\nonumber  \\
&-q_{b}{}^{c}\bar s^{d}K_{cd}-\frac{1}{\alpha^{2}}\beta^{a}\bar s_{a}\hat \nabla_{b}\alpha \nonumber \, ,  \\
=&\frac{1}{\alpha}\beta^{a}\bar s_{a}\hat \nabla_{b}ln\left(\bar B\right)-\frac{1}{\alpha}\hat \nabla_{b}\left(\beta^{a}\bar s_{a}\right)-q_{b}{}^{c}\bar s^{d}K_{cd}\nonumber \, ,  \\
=&\frac{\beta_{\perp}}{\alpha}\hat \nabla_{b}ln\left(\bar B\right)-\frac{1}{\alpha}\hat \nabla_{b} \beta_{\perp} - q_{b}{}^{c}\bar s^{d}K_{cd}
\,. \end{align}

\noindent where we made the definition $\beta_{\perp} = \beta^{a} \bar s_{a}$.  
Putting all of the above expressions together, we find that

\begin{align}
\nabla_{a}\bar s_{b} =& -H_{ab} -\bar s_{a} \hat \nabla_{b}ln\left(\bar B\right)-n_{b}\bar s^{c}K_{ac}-n_{a}n_{b}\bar s^{c}\bar \nabla_{c}ln\left(\alpha\right)\nonumber   \\*
&+\frac{\beta_{\perp}}{\alpha}n_{a}\hat \nabla_{b}ln\left(\bar B\right)-\frac{1}{\alpha}n_{a}\hat \nabla_{b}\beta_{\perp}-n_{a}q_{b}{}^{c}\bar s^{d}K_{cd}
\,. \end{align}

So, this allows us to completely decompose $\nabla_{a} \bar s_{b}$ into ADM variables, while equation \eqref{4GradientN} enables us to decompose $\nabla_{a} n_{b}$ into ADM variables.  We can now use this formalism to express the intrinsic horizon quantities in terms of ADM variables.  Before fully doing this, we follow the notation of \cite{CookPaper}\footnote{Though it should be noted that Cook does not define $K_{\perp}$ and instead just writes it as $K-J$.  Also, note that Cook works under a gauge where $\bar B = \beta_{\perp}=\bar s_{a}\beta^{a}$.  This is not consistent with a na\''ive decomposition of the induced geometry on the horizon from the 4-geometry, since if $t$ is the time coordinate and $r$ is the radial coordinate of the horizon, then $\bar B = \frac{1}{\sqrt{\gamma^{rr}}}$ and $\beta_{\perp} = \bar B \alpha^2 g^{tr}$, which are clearly two inequivalent expressions in the general case.} and decompose the extrinsic curvature into components normal and tangential to the horizon:

\begin{align}
J_{ab} \equiv& q_{a}{}^{c}q_{b}{}^{d}K_{cd} \quad \quad\quad J_{a} \equiv q_{a}{}^{b}\bar s^{c}K_{bc}\quad\quad\quad K_{\perp}\equiv \bar s^{a}\bar s^{b}K_{ab}\nonumber \, ,  \\*
K_{ab} =& \gamma_{a}{}^{c}\gamma_{b}{}^{d}K_{cd}=\left(q_{a}{}^{c}+\bar s_{a}\bar s^{c}\right)\left(q_{b}{}^{d}+\bar s_{b}\bar s^{d}\right)K_{cd}\nonumber \, ,  \\*
=&J_{ab} + J_{a}\bar s_{b} + \bar s_{a}J_{b} + \bar s_{a}\bar s_{b}K_{\perp} \label{KDecomposed}
\,. \end{align}

Now, we are ready to work through the relevant intrinsic and extrinsic quantities.  We start with the expansions, as they are the easiest.  

\begin{align}
\theta =& q^{ab}\nabla_{a}\ell_{b} = q^{ab}\nabla_{a}\left[\alpha\left(n_{a}+ \bar s_{a}\right) \right] \nonumber\, ,  \\
=&\alpha\left(q^{ab}\nabla_{a}n_{b} + q^{ab}\nabla_{a}\bar s_{b}\right)\nonumber \, ,  \\
=&\alpha \left(-J - H\right)
\,. \end{align}

A nearly identical derivation will show that $\theta_{(k)} = \frac{1}{2\,\alpha}\left(H-J\right)$.  Immediately, we can therefore see that, in 3+1 language, the Isolated/Dynamical horizon condition is a requirement that $H=-J$ and $J >0$ on the horizon.  We can perform a similar decomposition to the shear:

\begin{align}
\sigma_{ab} =& q_{a}{}^{c}q_{b}{}^{d}\nabla_{c}\ell_{d} - \frac{1}{2}q_{ab} \theta \nonumber \, ,  \\
=& q_{a}{}^{c}q_{b}{}^{d}\nabla_{c}\left[\alpha\left(n_{d}+\bar s_{d}\right)\right]+\frac{1}{2}\alpha\,q_{ab}\left(J+H\right)\nonumber \, ,  \\
=&\alpha\left(-J_{ab} - H_{ab}\right)+\frac{1}{2}\alpha\,q_{ab}\left(J+H\right)
\,. \end{align}

This then shows us that the isolated horizon condition that $\sigma_{ab}$ vanish is equivalent to the condition that $H_{ab} = -J_{ab}$.  Tracing this equation will then automatically give you the $H=-J$ requirement for $\theta =0$.  Note that the twist $\omega_{ab}\equiv q_{a}{}^{c}q_{b}{}^{d}\nabla_{[c}\ell_{d]}$ is automatically zero.  This should not be surprising since the 2-dimensional horizon is smoothly embedded in both the 3-surface and the 4-surface, and therefore must have a vanishing twist by Frobenius's theorem\footnote{See Appendix B.3 in \cite{Wald}, amongst other sources}.  Alternately, it is easy to substitute the definitions of $n_{a}$ and $\bar s_{a}$ in terms of gradients of functions and then take both of their twists, and then show that the twists are proportional to $\nabla_{[a}\nabla_{b]}f$ for some function in both cases, and are therefore zero.   The Raychaudhuri equation 
\eqref{Null Raychaudhuri Equation} is then satisfied if, in addition to $J_{ab} = -H_{ab}$, we also have $T_{ab}\ell^{a}\ell^{b} = \alpha^{2}\left(T_{ab}n^{a}n^{b} + 2T_{ab}n^{a}\bar s^{b} +T_{ab}\bar s^{a}\bar s^{b}\right) =\alpha^{2}\left(\rho - 2 j^{a}\bar s_{a} + S_{ab}\bar s^{a} \bar s^{b}\right)=0$.  These conditions are then sufficient to ensure that the horizon is a WIH\cite{AshtekarKrishnan} as defined in equation \eqref{WIH}.  
\subsection{Dynamical Horizon case} \label{sec: DH stuff}
Now, let us consider the case of a dynamical horizon (DH).  These horizons are still defined as a topologically compact level surface of $\theta =0$.  If the horizon is expanding, however, it is necessarily the case that outgoing null vectors can only momentarily have zero expansion relative to infinity, because the expansion of the horizon will make the old horizon radius lie in the interior of the horizon at later times.  In other words, the light ray hovering stationary on the horizon at time $t$ will be falling inward toward $r=0$ at time $t+\delta t$.  Therefore, the outgoing null vector is no longer a tangent to the Dynamical Horizon, which goes from being a null surface to a spacelike surface\footnote{Contracting horizons will be timelike surfaces.  They have several properties, however, that are not horizon-like.  In particular, timelike $\theta =0$ surfaces will not be trapping surfaces--some null and timelike objects from inside these sorts of horizons will be able to escape to infinity.  
Furthermore, their existence necessarily requires the presence of matter violating the dominant energy condition.  This may in fact be the case for black holes emitting Hawking radiation, but is otherwise not considered physically realistic.  Therefore, we will not consider these surfaces here.}.  By this argument, then, the formerly null generators of the dynamical horizon must now be spacelike vectors.  It is logical to adjust these generators in such a way as to make them as close to the $\ell^{a}$ as possible.  We don't expect there to be dependence on the angular variables on the horizon, so the most logical choice for the generating vector $v^{a}$ is given by the ansatz:
\begin{equation}
v^{a}=\ell^{a} + \psi \bar s^{a}
\,. \end{equation}
It is easy to show that $v_{a}v^{a} = 2\,\alpha \psi + \psi^{2}$, which is positive so long as $-2\,\alpha < \psi$.  So, therefore, for positive $\psi$, $v^{a}$ is a spacelike vector that corresponds to $\ell^{a}$ for $\psi=0$.  Now, we enforce the condition that $v^{a}$ generates a surface of $\theta = 0$:

\begin{align}
£_{v}\theta =& 0 =v^{c}\nabla_{c}\theta =\left(\ell^{c}+ \psi \bar s^{c}\right)\nabla_{c}\theta \nonumber \, ,  \\
=&£_{\ell}\theta + \psi \bar s^{c}\nabla_{c}\left(q^{ab}\nabla_{a}\ell_{b}\right)\nonumber\, ,  \\
\frac{1}{\psi}\left(£_{v}\theta - £_{\ell}\theta\right) =&\bar s^{c}\nabla_{c}\left[-\alpha \left(J+H\right)\right] \nonumber \, ,  \\
=& -\bar s^{c}\left(J+H\right)\bar \nabla_{c}\alpha - \alpha \bar s^{c}\bar \nabla_{c}J - \alpha \bar s^{c} \bar \nabla_{c}H \nonumber \, ,  \\
=& \bar s^{c}\theta \bar \nabla_{c}ln\left(\alpha\right) - \alpha \bar s^{c}\bar \nabla_{c}J - \alpha \bar s^{c} \bar \nabla_{c}H \label{DynamicVectorFirstStep}
\,. \end{align}

Further simplification of the second and third terms on the right hand side above is involved enough that each term will be dealt with individually.  Before we begin, recall the Gauss-Codazzi condition from equation \eqref{Hamiltonianequationtwo} tells us, in the case of a 2-surface embedded in a 3-space, we have, after using the Hamiltonian constraint \eqref{HamiltonianConstraint}:

\begin{align}
\hat R =& \bar R -2\bar s^{a}\bar s^{b}\bar R_{ab} - H^{ab}H_{ab} + H^{2} \nonumber \, ,  \\
2\bar s^{a}\bar s^{b}\bar R_{ab} =&\bar R -\hat R -H^{ab}H_{ab}+H^{2}  \nonumber \, ,  \\
=&\left(16\pi\rho +K^{ab}K_{ab} - K^{2}\right) - \hat R -H^{ab}H_{ab} + H^{2} \nonumber \, ,  \\
=&16\pi \rho + J^{ab}J_{ab} + 2 J^{a}J_{a} + K_{\perp}^{2} - \left(K_{\perp}+J\right)^{2} - \hat R -H^{ab}H_{ab} + H^{2}\nonumber \, ,  \\
\bar s^{a}\bar s^{b}\bar R_{ab}=&\frac{1}{2}\left(16\pi \rho + J^{ab}J_{ab} - H^{ab}H_{ab} + 2 J^{a}J_{a} -J^{2}+H^{2}-2K_{\perp}J-\hat R\right) \label{2DRdecomp}
\,. \end{align}

Now, knowing this identity, we work toward decomposing $\bar s^{a}\bar \nabla_{a}H$:

\begin{align}
\bar s^{a}\bar \nabla_{a} H =& -\bar s^{c}\bar \nabla_{c}\left(q^{ab} \bar \nabla_{a} \bar s_{b}\right)\nonumber \, ,  \\
=&-\bar s^{c}q^{ab}\bar \nabla_{c}\bar \nabla_{a}\bar s_{b} - \bar s^{c}\left(\bar \nabla_{c}q^{ab}\right)\bar \nabla_{a}\bar s_{b} \nonumber \, ,  \\
=&-\bar s^{c}q^{ab}\bar R_{cab}{}^{d}\bar s_{d} - \bar s^{c}q^{ab}\bar \nabla_{a}\bar \nabla_{c}\bar s_{b} + \bar s^{a}\bar s^{c}\left(\bar \nabla_{a}\bar s^{b}\right)\bar \nabla_{a} \bar s_{b} \nonumber   \\
&+ \bar s^{c}\bar s^{b}\left(\bar \nabla_{c}\bar s^{a}\right)\bar \nabla_{a}\bar s_{b} \nonumber \, ,  \\
=&\bar s^{c}\bar s^{d} \bar R_{cd}-q^{ab} \bar \nabla_{a}\left(\bar s^{c}\bar \nabla_{c}\bar s_{b}\right) + q^{ab}\left(\bar \nabla_{a}\bar s^{c}\right)\bar \nabla_{c}\bar s_{b}\nonumber   \\
&+\left[\hat \nabla^{b}ln\left(\bar B\right)\right]\hat \nabla_{b} ln \left(\bar B\right) \nonumber \, ,  \\
=&\frac{1}{2}\left(16\pi \rho + J^{ab}J_{ab} - H^{ab}H_{ab} + 2 J^{a}J_{a} -J^{2}+H^{2}-2K_{\perp}J-\hat R\right) \nonumber   \\
&+q^{ab}\bar \nabla_{a}\left[\hat \nabla_{b} ln\left(\bar B\right)\right]+H^{ab}H_{ab} + \frac{1}{\bar B^{2}}\left(\hat \nabla^{b}\bar B\right) \hat \nabla_{b}\bar B \nonumber \, ,  \\
=&8\pi \rho + \frac{1}{2}\left(J^{ab}J_{ab} + H^{ab}H_{ab} +H^{2}-J^{2}-\hat R\right)+J^{a}J_{a} -K_{\perp}J \nonumber   \\
&+\frac{1}{\bar B} \hat \nabla^{2}\bar B \label{RadialHdecomp}
\,. \end{align}

This takes care of the third term in equation \eqref{DynamicVectorFirstStep}.  In order to take care of the second term, we first start with the momentum constraint, equation \eqref{MomentumConstraint}, and contract it onto $\bar s_{a}$\footnote{Here, we make use of the decomposition in equation \eqref{KDecomposed}, and drop any terms that vanish either by the fact that two tensors are normal to each other or that vanish by virtue of $\bar s_{a}$ having a fixed norm.}:

\begin{align}
8\pi j^{a}\bar s_{a} =& \bar s_{a}\bar \nabla_{b}\left(K^{ab} - \gamma^{ab}K\right)\nonumber \, ,  \\
=&\bar s_{a}\bar \nabla_{b}\left(J^{ab} + \bar s^{a} J^{a} + \bar s^{b}J^{a} + \bar s^{a}\bar s^{b} K_{\perp}\right)-\bar s^{a}\bar \nabla_{a}\left(q^{bc}K_{bc} + \bar s^{b}\bar s^{c}K_{bc}\right) \nonumber \, ,  \\
=&-J^{ab}\bar \nabla_{b} \bar s_{a}  + \bar \nabla_{b}J^{b} + \bar s_{a}\bar s^{b} \bar \nabla_{b} J^{a}+ K_{\perp}\bar \nabla_{b}\bar s^{b} + \bar s^{b} \bar \nabla_{b} K_{\perp} \nonumber   \\
&- \bar s^{a} \bar \nabla_{a}J - \bar s^{a} \bar \nabla_{a}K_{\perp} \nonumber \, ,  \\
=& J^{ab}H_{ab} + \hat \nabla_{a}J^{a} + 2 \bar s^{a}\bar s^{b} \bar \nabla_{a}J_{b} -K_{\perp}H -\bar s^{a} \bar \nabla_{a}J \nonumber \, ,  \\
\bar s^{a} \bar \nabla_{a}J =&J^{ab}H_{ab} + \hat \nabla_{a}J^{a}-2 \bar s^{a}J^{b} \bar \nabla_{a}\bar s_{b}-K_{\perp}H - 8\pi j^{a}\bar s_{a}\nonumber \, ,  \\
=& J^{ab}H_{ab} + \hat \nabla_{a} J^{a} +2J^{b}\hat \nabla_{b} ln\left(\bar B\right)-K_{\perp}H - 8 \pi j^{a}\bar s_{a}\label{RadialJdecomp}
\,. \end{align}

Therefore, putting equations \eqref{RadialJdecomp} and \eqref{RadialHdecomp} into equation \eqref{DynamicVectorFirstStep}, we get

\begin{align}
\frac{1}{\psi}\left(£_{v}\theta - £_{\ell}\theta\right) =& \bar s^{c}\theta \bar \nabla_{c}ln\left(\alpha\right)-\alpha\left[\left(J^{ab}H_{ab} + \hat \nabla_{a} J^{a} +2J^{b}\hat \nabla_{b} ln\left(\bar B\right)-K_{\perp}H \right.\right.\nonumber   \\
&\left.\left.- 8 \pi j^{a}\bar s_{a}\right)+\left(8\pi \rho + \frac{1}{2}\left(J^{ab}J_{ab} + H^{ab}H_{ab} +H^{2}-J^{2}-\hat R\right)\right.\right.\nonumber   \\
&\left.\left.+J^{a}J_{a} -K_{\perp}J + \frac{1}{\bar B}\hat \nabla^{2}\bar B \right)\right]\nonumber \, ,  \\
=& \bar s^{c}\theta \bar \nabla_{c}ln \left(\alpha\right)-\alpha \left[\frac{1}{2}\left(J^{ab}+H^{ab}\right)\left(J_{ab}+H_{ab}\right)-K_{\perp}\left(J+H\right)\right.\nonumber   \\
&\left.+\frac{1}{2}\left(H+J\right)\left(H-J\right)+\hat \nabla_{a}J^{a}+ J^{a}J_{a}-\frac{1}{2}\hat R +8\pi \left(\rho-j^{a}\bar s_{a}\right) \right. \nonumber   \\
&\left.+ 2 J^{b}\hat \nabla_{b}ln\left(\bar B\right) +  \frac{1}{\bar B}\hat \nabla^{2}\bar B\right] \label{3+1VectorDiff}   \nonumber \, ,  \\
=&\bar s^{c}\theta \bar \nabla_{c} ln\left(\alpha\right) - \alpha \left[\frac{1}{2\alpha^{2}}\left(\sigma^{ab}\sigma_{ab} + \frac{1}{2}\theta^{2}\right)+\frac{1}{\alpha}K_{\perp}\theta-\theta\theta_{(k)} \right. \nonumber   \\
&\left.-\frac{1}{2}\hat R+ \hat \nabla_{a}J^{a}+J^{a}J_{a}+8\pi \left(\rho-\bar s^{a}j_{a}\right) + 2J^{a}\hat \nabla_{a} ln\left(\bar B \right)\right. \nonumber   \\ 
&\left.+ \frac{1}{\bar B}\hat \nabla^{2}\bar B\right]
\,. \end{align}

And finally, requiring that $£_{v}$ vanish, and that the surface is a dynamical horizon where the outgoing expansion $\theta$ vanishes, we get

\small
\begin{align}
\psi =& \frac{£_{\ell}\theta}{\alpha\left[\frac{1}{2\alpha^{2}}\left(\sigma^{ab}\sigma_{ab}\right)-\frac{1}{2}\hat R + \hat \nabla_{a}J^{a} + J^{a}J_{a} + 8\pi\left(\rho-j_{a}\bar s^{a}\right) + 2 J^{a} \hat \nabla_{a} ln\left(\bar B\right) + \frac{1}{\bar B}\hat \nabla^{2}\bar B\right]}\nonumber \, ,  \\
=& \frac{-8\pi T_{ab}\ell^{a}\ell^{b}-\sigma_{ab}\sigma^{ab}}{\alpha\left[\frac{1}{2\alpha^{2}}\left(\sigma^{ab}\sigma_{ab}\right)-\frac{1}{2}\hat R + \hat \nabla_{a}J^{a} + J^{a}J_{a} + 8\pi\left(\rho-j_{a}\bar s^{a}\right) + 2 J^{a} \hat \nabla_{a} ln\left(\bar B\right) + \frac{1}{\bar B}\hat \nabla^{2}\bar B\right]}\label{PsiEquation}
\,. \end{align}
\normalsize

Several things should be noted about the above equation.  The first thing is that $\psi$ has a strong dependence on angular variations on the horizon.  The traditional definition of the horizon angular momentum\cite{ADMPaper}\cite{RTboundary}\cite{HaywardAM} is given by $L = \frac{1}{8\pi}\oint d^{2}x\, \sqrt{q} K_{ab} \bar r^{a} \hat \phi^{a} = \frac{1}{8\pi}\oint d^{2}x\,\sqrt{q}J_{a}\hat \phi^{a}$.  Secondly, note that the only derivatives of the 2+1 `lapse' function $\bar B$ appearing in \eqref{PsiEquation} are 2-dimensional derivatives--apparently $\psi$ is not sensitive to radial dependencies in $\bar B$\footnote{This is something that we might expect, considering that $\bar B$ is defined as $\frac{1}{\sqrt{\bar \nabla_{i} R\bar \nabla^{i}R}}$.  Any radial variation in $\bar B$ is gauge, as it can be adjusted merely by a r paramaterization of the function $R$.}.  

Finally, one might be concerned by the presence of $\hat R$ in the denominator of this equation, since its value is heavily restricted by the topology of the horizon by the Gauss-Bonet theorem.  But, in fact, the presence of $\hat R$ is actually quite fortunate, as in the case of the Schwarzschild metric, it is the only term in equation \eqref{PsiEquation} that is not explicitly zero, and therefore, without the presence of $\hat R$, we would have no control over $\psi$ approaching zero as our spacetime approached the spherically symmetric, non-dynamical case.\footnote{For example, in the analysis of the Vaidya metric below, note that the $\dot M \rightarrow 0$ limit would be very poorly behaved without the 2-curvature term.}   

\subsection{Area Balance Law}
We now show how this formalism can generate an area balance law, and in turn, how this area balance law dictates conditions upon the lapse function and shift vector, as originally proved by Ashtekar and Krishnan \cite{AshtekarKrishnan}.  The key insight in this derivation is to note that the Dynamical Horizon is a spacelike 3-surface locally defined by the level surface $\theta = 0$ of the locally defined function $\theta$.  Therefore, it is a well-defined Cauchy surface that can be chosen for the 3+1 splitting of spacetime.  Now, having made this definition, the next step is to calculate the mass flow across the horizon, and then to make what have now become commonplace 2+1 decompositions of the relevant geometrical quantities.  As we go through the following derivation, we will freely use the dynamical horizon condition $\theta = 0$ to eliminate any terms of the form $J+H$\footnote{Note that the derivation to follow is done {\it intrinsically to the Dynamical Horizon}, using the constraint equations {\bf on} the horizon.  Therefore, it is best to think of the $\bar s^{a}$ that appears below to be the unit normal parallel to the $v^{a}$ discussed in this section, and for the unit timelike normal to be the timelike normal perpendicular to this--namely, the unit normal of a timelike observer falling into the black hole with no velocity transverse to the black hole surface.}:  

\begin{align}
\int\sqrt{\gamma}\, d^{3}x 8\pi T_{ab}\ell^{a}n^{b}=& \int\sqrt{\gamma}\, d^{3}\,x \,\alpha\left(8\pi T_{ab} n^{a}n^{b}+ 8\pi T_{ab} n^{a}\bar s^{b}\right)\nonumber \, ,  \\
=&\int\sqrt{\gamma}\, d^{3}x\,\alpha \left[\frac{1}{2}\left(\bar R -K^{ab}K_{ab} + K^{2}\right)\right.\nonumber\nopagebreak   \\
&\left.-\bar s_{a}\bar \nabla_{b}\left(K^{ab}-\gamma^{ab}K\right)\right]\nonumber
\end{align}
\begin{align}
=& \int\sqrt{\gamma}\, d^{3}x \,\alpha\left[\frac{1}{2}\left(\hat R + 2 \bar s^{a}\bar s^{b} \bar R_{ab} + H^{ab}H_{ab}-H^{2}-J^{ab}J_{ab} - 2 J^{a}J_{a} \right.\right.\nonumber   \\
&\left.\left.-K_{\perp}^{2} + K_{\perp}^{2}+J^{2}-2K_{\perp}J\right)-\bar s_{a}\bar \nabla_{b}\left(J^{ab} + J^{a}\bar s^{b} + J^{b} \bar s^{a} + \bar s^{a} \bar s^{b} K_{\perp}\right)\right.\nonumber  \\
&\left.+ \bar s^{a} \bar \nabla_{a}\left(J+K_{\perp}\right)\right]\nonumber \, ,  \\
=&\int\sqrt{\gamma}\, d^{3}x \,\alpha\left[\bar s^{a}\bar s^{b} \gamma^{cd}\bar R_{cadb} -K_{\perp}J-J_{a}J^{a} + \frac{1}{2}\left(\hat R + H^{ab}H_{ab} -J^{ab}J_{ab}\right)\right.\nonumber  \\
&\left.-J^{ab}H_{ab}-\bar s_{a}\bar s^{b} \bar \nabla_{b}J^{a} -\bar \nabla_{b} J^{b} - \bar s^{b}\bar \nabla_{b} K_{\perp} -K_{\perp}H + \bar s^{a} \bar \nabla_{a}J + \bar s^{a}\bar \nabla_{a}K_{\perp}\right] \nonumber \, ,  \\
=&\int\sqrt{\gamma}\, d^{3}x \,\alpha\left[\bar s^{a} \bar \nabla_{c}\bar \nabla_{a} \bar s^{c} - \bar s^{a} \bar \nabla_{a} \bar \nabla_{c} \bar s^{c}-J^{a}J_{a} -\hat \nabla_{a}J^{a} \right.\nonumber  \\
&\left.+ \frac{1}{2}\left(\hat R + H_{ab}H^{ab} - J^{ab}J_{ab} -2 J^{ab}H_{ab}\right) -2\bar s_{a}\bar s^{b} \bar \nabla_{b}J^{a}+ \bar s^{a}\bar \nabla_{a}J \right]\nonumber \, ,  \\
=&\int\sqrt{\gamma}\, d^{3}x\,\alpha\left[ \bar \nabla_{c}\left(\bar s^{a}\bar \nabla_{a}\bar s^{c}\right) + \bar s^{c}\bar \nabla_{a}H+  \left(\bar \nabla_{a}\bar s^{a}\right)^{2} - \left(\bar \nabla_{a}\bar s^{c}\right)\bar \nabla_{c}\bar s^{a}-J^{a}J_{a} \right.\nonumber   \\
&\left.- \hat \nabla_{a}J^{a} + \frac{1}{2}\left( \hat R + H_{ab}H^{ab} - J_{ab}J^{ab} - 2J^{ab}H_{ab}\right)+ 2 \bar s^{b}J^{a}\bar \nabla_{b} \bar s_{a} \right.\nonumber \\
&\left.+  \bar s^{a}\nabla_{a}J\right]\nonumber\, ,  \\
=& \int \sqrt{\gamma}\,d^{3}x\,\alpha\left[\bar \nabla_{a}\left(\bar s^{b}\bar \nabla_{b}\bar s^{a}\right) - J^{a}J_{a} - \hat \nabla_{a}J^{a}-2J^{a}\hat \nabla_{a}ln\left(\bar B\right)\right.\nonumber  \\
&\left.+ \frac{1}{2}\left(\hat R -H^{ab}H_{ab}-J^{ab}J_{ab}-2J^{ab}H_{ab}\right)+ \bar s^{a}\bar \nabla_{a}\left(J+H\right)\right] \label{AreaBalanceMidTerm}
\,. \end{align}

\noindent And now, we are nearly done.  To further simplify this expression, we follow Ashtekar et al. \cite{AshtekarKrishnan}, and make a few simplifying assumptions.  First, we remember that $\bar s^{a}$ is tangent to the DH.  Therefore, $\bar s^{a}\nabla_{a}\left(J+H\right) =0$, and we can ignore the last term in \eqref{AreaBalanceMidTerm}.  Second, we make a coordinate choice on the DH that makes $\bar B$ depend only upon the radial coordinate.  This eliminates the total divergence term as well as $\hat \nabla_{a} \bar B$.  Finally, we make a choice of the lapse function so that $\alpha = \frac{1}{\bar B}$, which makes $\alpha \sqrt{\gamma} = \sqrt{q}$.  This enables us to rewrite the 3-integral $\int d^{3}x$ as $\int dr \,\oint \sqrt{q}d^{2}x$.  The integral of $\hat \nabla_{a}J^{a}$ becomes an integral of a total divergence over a boundary, while the integral of $\hat R$ becomes the curvature invariant $\chi$ of the 2-dimensional section of the DH.  Consequently, we can now rewrite \eqref{AreaBalanceMidTerm} as

\begin{align}
\int d^{3}x\,8\pi\,T_{ab}\ell^{a}n^{b} =& \int dr\,\oint \sqrt{q}\,d^{2}x\left[-J_{a}J^{a} + \frac{1}{2}\hat R - \frac{1}{2}\sigma^{ab}\sigma_{ab}\right]\, \nonumber \, ,  \\
\chi \int dr =& \int d^{3}x\left( 16 \pi T_{ab}\ell^{a}n^{b} + 2J^{a}J_{a} + \sigma_{ab}\sigma^{ab}\right)\,. \label{Area Balance Law}
\,. \end{align}

\noindent The right hand side of \ref{Area Balance Law} is known as the Hawking energy, which has several interesting properties:  

First, if $T_{ab}$ satisfies an energy condition such that $T_{ab}\ell^{a}n^{b}\geq 0$, as one would expect for a fluid with a positive density greater in magnitude than its pressure, then the right hand side of the equation is explicitly positive-definite, which then immediately tells us that r must be increasing and that $\chi >0$, meaning that the apparent horizons that make up the DH must individually have spherical topologies, and that the DH satisfies the law of nondecreasing area as one would expect from the second law of black hole dynamics.  

Second, it is already apparent that \eqref{Area Balance Law} also contains the content of the first law of black hole dynamics.  The left hand side term describes an increase in areal radius, which can be taken to be an infinitesimal area increase that can be identified with $\delta A$.  The stress energy tensor term exactly describes the amount of mass entering the black hole, and thus can certainly be identified with $\delta M$.  Meanwhile, the last two terms involve nondiagonal terms of $K_{ab}$, which are known to be associated with black hole angular momentum.  Therefore, it shouldn't be very surprising that with some minor tweaks, \eqref{Area Balance Law} can be modified to produce a proof of the First Law of Black Hole dynamics, but with the advantage of being defined explicitly on the horizon, and that it is proved without the dependence on the details of the Kerr solution used in the proof given in \eqref{Kerr Proof of First Law}.

\section{Dynamics of the Vaiyda solution}
Now, let us apply this formalism to one of the simplest dynamical examples:  the Vaidya solution \cite{Vaidya}.  This solution describes a spherically symmetric null dust in the background of a Schwarzschild black hole.  In order to make the relationship between the 3+1 split and these boundary conditions more explicit, we are going to explicitly provide a 3+1 split, and then compute the necessary value for $\psi$ such that \eqref{DynamicVectorFirstStep} is satisfied.  

Start with the Vaidya metric as it is commonly given, where $M(v)$ is an arbitrary function of $v$:

\begin{equation}
ds^{2} = -\left(1-\frac{2\,M(v)}{r}\right)dv^{2} + 2\,dv\,dr + r^{2} d\theta^{2} + r^{2}sin^{2}\left(\theta\right)d\phi^{2}
\,. \end{equation}

While this form gives us the simplest and most direct description of the metric, it is not ideally suited for a 3+1 split, as a computation of the inverse metric tensor will show us that $dv_{a} dv_{b} g^{ab}=0$.  Therefore, we transform this to Kerr-like coordinates by making the coordinate transformation $v=t+r$.  Upon completing this transformation, the metric tensor takes the form (in $(t,r,\theta, \phi)$ coordinates):

\begin{equation}
g_{ab} =\left(\begin{tabular}{l c c r}
$-\left(1-\frac{2\,M(t+r)}{r}\right)$&$\frac{2\,M(t+r)}{r}$&0&0  \\
$\frac{2\,M(t+r)}{r}$&$\left(1+\frac{2\,M(t+r)}{r}\right)$&0&0  \\
0&0&$r^{2}$&0  \\
0&0&0&$r^{2}sin^{2}\left(\theta\right)$
\end{tabular}\right)
\,. \end{equation}

Where $M$ is now an arbitrary function of $t+r$, rather than $v$.  This metric has the exact same form as the Schwarzschild metric does in Kerr coordinates.  A labourious but straightforward computation gives that the Ricci scalar of this metric is equal to zero, and that:

\begin{equation}
G_{ab} = R_{ab} = \frac{2\,\dot M}{r^{2}}\left( \begin{tabular}{l c c r}
$1$&$1$&0&0  \\
$1$&$1$&0&0  \\
0&0&0&0  \\
0&0&0&0  \end{tabular}\right)
\,. \end{equation}

Furthermore, we can compute two null vectors that have only $t$ and $r$ components, and after fixing their inner product with each other to be equal to $-1$, we find that they are:

\begin{align}
k_{a} dx^{a} =& \frac{1}{2\,c}\left(-dt - dr\right)\, , and \nonumber \\\
ell_{a}dx^{a} =& c\left[-dt\left(1-\frac{2\,M}{r}\right)+dr\left(1+\frac{2\,M}{r}\right)\right]
\,, \end{align}

\noindent where $c$ is an as of now arbitrary constant.  It is then easy to show that $q_{ab} =g_{ab} + \ell_{a}k_{b} + k_{a}\ell_{b} = r^{2}d\Omega^{2}$, which then gives:

\begin{align}
q_{ab} = &\left(\begin{tabular}{l c c r}
$-\left(1-\frac{2\,M(t+r)}{r}\right)$&$\frac{2\,M(t+r)}{r}$&0&0  \\
$\frac{2\,M(t+r)}{r}$&$\left(1+\frac{2\,M(t+r)}{r}\right)$&0&0  \\
0&0&$r^{2}$&0  \\
0&0&0&$r^{2}sin^{2}\left(\theta\right)$
\end{tabular}\right)\nonumber   \\
&+\left(\begin{tabular}{l c c r}
$\left(1-\frac{2\,M(t+r)}{r}\right)$&$-\frac{2\,M(t+r)}{r}$&0&0  \\
$-\frac{2\,M(t+r)}{r}$&$-\left(1+\frac{2\,M(t+r)}{r}\right)$&0&0  \\
0&0&0&0  \\
0&0&0&0
\end{tabular}\right)\nonumber\, ,  \\
=&r^{2}\,d\theta^{2} + r^{2}\sin^{2}\left(\theta\right)d\phi^{2}
\,. \end{align}

After doing this, we then can easily find the trace against $q^{ab}$ the gradient of any one-form that has only temporal and radial dependence, since the contraction on $q^{ab}$ will eliminate anything but angular terms, and the angular terms depend only on $r$:

\begin{align}
q^{ab}\nabla_{a}v_{b} =& q^{ab} \partial_{a} v_{b} -q^{ab}\Gamma_{ab}{}^{c}v_{c} \nonumber \, ,  \\
=&0-\frac{1}{2}q^{ab}v^{c}\left(g_{ac,b}+g_{bc,a}-g_{ab,c}\right) \nonumber \, ,  \\
=&\frac{1}{2}q^{ab}v^{c}g_{ab,c} \nonumber \, ,  \\
=&\frac{2}{r}v^{r} \label{VaiydaExpansions}
\,. \end{align}

\noindent Now, we can use \eqref{VaiydaExpansions} to essentially read off the expansions of the two null vecors.  The result is that $\theta_{(k)} = -\frac{1}{r\,c}$ and $\theta =\frac{2\,c}{r}\left(1-\frac{2\,M}{r}\right)$.  Since, at the point where $r=2M(t,r)$ we have $\theta_{(k)}$ negative and $\theta$ zero, it is clear that this surface represents a trapped surface.  Note that it is also clearly dynamical, due to the explicit inclusion of the time coordinate above.  

Now, we wish to 3+1 split this metric tensor.  We will make the obvious choice of $t=$constant slices and proceed from there.  With this choice, we clearly have $\alpha = \frac{1}{\sqrt{1+\frac{2\,M}{r}}}$, $\beta_{a}dx^{a} = \frac{2\,M}{r}dr$, and $n_{a} = (-\alpha,0,0,0)$.  The 3-metric tensor is given by:

\begin{equation}
\gamma_{ab} = \left(\begin{tabular}{l c r}
$1+\frac{2\,M}{r}$&0&0  \\
0&$r^{2}$&0  \\
0&0&$r^{2}sin^{2}\left(\theta\right)$
\end{tabular}\right)
\,. \end{equation}

\noindent while the extrinsic curvature can be calculated according to the rule 
\begin{align}
K_{ab} = &- \gamma_{a}{}^{c}\nabla_{c} n_{b} \nonumber\, ,  \\
=& -\left(\gamma_{a}{}^{c}\gamma_{b}{}^{d}\partial_{c}n_{d} - \gamma_{a}^{d}\gamma_{b}{}^{e}\Gamma_{de}{}^{c}n_{c}\right) \nonumber \, ,  \\
K_{ij}=&0- \Gamma_{ij}{}^{t}\alpha
\,. \end{align}

\noindent which gives 

\begin{align}
K_{rr} =& \sqrt{1+\frac{2\,M}{r}}\left(\frac{\dot M}{r} - \frac{4\,M\,\dot M}{r^{2}} - \frac{2\,M}{r^{2}}+\frac{6\,M^{2}}{r^{3}}\right)\nonumber \, ,  \\
K_{\theta \theta}=&\frac{K_{\phi \phi}}{sin^{2}\left(\theta\right)} = \frac{2\,M}{\sqrt{1+\frac{2\,M}{r}}}
\,. \end{align}

With all other extrinsic curvature components equal to zero.  We now 2+1 decompose onto the horizon.  Our radial vector is given by

\noindent $\bar r_{a} = \sqrt{1+\frac{2\,M}{r}}(1,0,0)$, therefore giving us a spacelike ``lapse'' of $\sqrt{1+ \frac{2\,M}{r}}$.  The 2-metric is simply the ordinary metric of a 2-sphere.  Since $K_{ab}$ only has diagonal terms, we get $J_{A} = 0$, $K_{\perp}= \frac{1}{1+\frac{2\,M}{r}}K_{rr}$ and $J_{AB} =  \frac{2\,M}{\sqrt{1+\frac{2\,M}{r}}}d\Omega^{2}$.  Finally, we have the extrinsic curvature of the horizon in the 3-surface:
\begin{align}
H_{ab} =& - q_{a}{}^{c}q_{b}{}^{d}\bar \nabla_{c} \bar r_{d}\nonumber \, ,  \\
H_{AB}=&\bar \Gamma_{AB}{}^{c}\bar r_{c} \nonumber \, ,  \\
=&\frac{1}{2}\bar r^{c} \left(\gamma_{Ac,B}+\gamma_{Bc,A}-\gamma_{AB,c}\right)\nonumber \, ,  \\
=&\frac{1}{2}\frac{1}{\sqrt{1+\frac{2\,M}{r}}}\left(-\frac{\partial}{\partial r}\gamma_{AB}\right) \nonumber \, ,  \\
=&-\frac{r}{\sqrt{1+\frac{2\,M}{r}}}d\Omega^{2}
\,. \end{align}

Therefore, on the horizon, we have $H_{AB} = - J_{AB}$, which tells us, automatically, that $\sigma_{ab} = 0$ and $\theta = 0$, the latter fact we, of course, already derived in the full 4-space.  The vanishing of the expansion and of the shear tremendously simplifies \eqref{3+1VectorDiff}, which, on the horizon, now becomes:

\begin{equation}
\frac{1}{\psi}\left(£_{v}\theta-£_{\ell}\theta\right)=-\alpha\left[-\frac{1}{r^{2}} + 8\pi \left(\rho-j^{a}\bar r_{a}\right)\right]
\,. \end{equation}

A glimpse at the Raychaudhuri equation \eqref{Null Raychaudhuri Equation}, coupled with all of these computations, will show one that the only term remaining in $£_{\ell}\theta$ is $-8\pi T_{ab}\ell^{a}\ell^{b}$.  Setting $£_{v}\theta = 0$, and solving for $\psi$,

\begin{align}
\psi=& \frac{8\pi T_{ab}\ell^{a}\ell^{b}}{\alpha\left(\frac{1}{r^{2}}-8\pi T_{ab}n^{a}n^{b} - 8\pi T_{ab}n^{a} \bar r^{b}\right)} \label{AlmostDone}
\,. \end{align}
\noindent We now remember that $G_{ab} =  \frac{2\, \dot M}{r^{2}}\left(dt + dr\right)^2$, which gives us $G_{ab} n^{a}n^{b} =\frac{ \frac{2\,\dot M}{r^{2}}}{1+\frac{2\, M}{r}}$, $G_{ab}n^{a} \bar r^{b} = - G_{ab} n^{a}n^{b}$ and $G_{ab}\ell^{a}\ell^{b} = \frac{8\,c\,\dot M}{r^{2}}$.  Putting all of this together into \eqref{AlmostDone}, we get the result 

\begin{equation} 
\psi =\frac{8\,c\,\dot M}{\alpha}
\,. \end{equation}

Now, all that is left is to set the value of $c$ by requiring that $\ell_{a} = \alpha \left(n_{a} + \bar r_{a}\right)$, so that $\ell^{a}$ becomes the best approximation possible of the time evolution vector $t^{a} = \alpha n^{a} + \beta^{a}$ on the horizon.  To do this, we simply calculate $\ell_{a}-\alpha n_{a}$, and then require that the answer be orthogonal to $n^{a}$

\begin{align}
n_{a}\left(\ell^{a} - \alpha n^{a}\right) =& \ell_{a}n_{b}g^{ab} + \alpha \nonumber \, ,  \\
0=&-\frac{c}{\sqrt{1+\frac{2M}{r}}}\left[\left(1+\frac{2M}{r}\right)\left(1-\frac{2M}{r}\right) +\frac{2M}{r}\left(1+\frac{2M}{r}\right)\right]\nonumber   \\
&+\frac{1}{\sqrt{1+\frac{2M}{r}}}\nonumber\, ,  \\
=&-2\,c+1\longrightarrow c=\frac{1}{2}
\,. \end{align}

So, after all of this, we find that the value $\psi$ takes on the horizon is ${4\dot M}{\sqrt{2}}$.  We can now use this quantity to calculate the time rate of change of the horizon's area:

First, we calculate the tangent vector to the dynamical horizon:

\begin{align}
v^{a} =&\ell^{a}+\psi \bar r^{a}= c\left[\left(1+\frac{2\,M}{r}\right)\partial_{t} + \left(1-\frac{2\,M}{r}\right)\partial_{r}\right]+ \psi\left(\frac{1}{\sqrt{1+ \frac{2\,M}{r}}}\right)\partial_{r}\nonumber \, ,  \\
\hat =& \partial_{t} + 4 \dot M \partial_{r}
\,. \end{align}

\noindent where the hat indicates that the equality is only valid on the horizon.

Now, it is easy enough to directly calculate the rate of area increase of the black hole:

\begin{align}
\delta A =& £_v \oint d^{2}x\, \sqrt{q} \nonumber \, ,  \\
=& \oint d^{2}x\,\frac{1}{2} \sqrt{q}q^{ab}v^{a}\partial_{a} q_{ab} \nonumber \, ,  \\
=& \oint d^{2}x\, 4 \dot M \frac{2}{r}\sqrt{q}\nonumber \, ,  \\
=& \oint d^{2}x\, 4 \dot M \frac{2}{r}r^{2}\sin \theta \nonumber \, ,  \\
=& 16\pi \dot M \left(2\,M\right) \nonumber \, ,  \\
=& 32 \pi \dot M \, M
\,. \end{align}

Which is a logical conclusion, considering that the straightforward, direct computation of the black hole area gives us the value $A=4\,\pi r^2 = 16\,\pi M^2$, whose first derivative is clearly $32\,\pi M \dot M$.  Clearly, this method was massive overkill for a situations such as the Vaiyda metric with a known analytical solution.  However, in a dynamical spacetime, where the location of the horizon is not known, this technique gives an exact way to trace not only the horizon, but also the normals to the horizon as it expands.\footnote{Note that the area of the dynamical horizon is a slicing dependent quantity.  See \cite{VaidyaSlicing} for a direct analysis of the slicing dependence of the area of the Vaiyda DH.}

\appendix
\chapter{Null decomposition of the Minkowski and Kerr spacetimes}
\label{sec: Kerr}
In order to further clarify the above procedure, here, we will work out a couple of examples of null decompositions.  First, let us examine the Minkowski spacetime, given in spherical coordinates $(t,r,\theta$,$\phi$):

\begin{equation}
g_{ab}dx^{a}dx^{b} = -dt^{2} + dr^{2} +r^{2}d\theta^{2} +r^{2}sin^{2}(\theta)d\phi^{2}
\,. \end{equation}

We wish to examine the null geometry of the surface whose vector space is tangent to the outgoing vector $\ell^{a} = (1,1,0,0)$.  To do this, we define $v = -t + r$, which gives $g^{ab}dv_{a} = \ell^{b}$.  Then, we do a coordinate transformation, replacing t with $v$, which gives the metric:

\begin{equation}
\displaystyle g_{ab} =  \left( \begin{array}{cccc}
-1 & 1 & 0& 0  \\
1 & 0 & 0&0   \\
0 & 0 & r^{2}& 0   \\
0&0&0&r^{2}sin^{2}(\theta) \end{array} \right)  g^{ab} =  \left( \begin{array}{cccc}
0 & 1 & 0& 0  \\
1 & 1 & 0&0   \\
0 & 0 & \frac{1}{r^{2}}& 0   \\
0&0&0&\frac{1}{r^{2}sin^{2}(\theta)} \end{array} \right)  
\,. \end{equation}

Where the first row/column represents $\alpha$, and the next three coordinates are $(r,\theta, \phi$) as they normally would be.  Obviously, $d\alpha_{a} = (1,0,0,0)$, and action upon this form with $g^{ab}$ gives $\ell^{a} = (0,1,0,0)$.  We can see that, at any point in the spacetime, the tangent space is spanned by $\ell^{a}$, $(0,0,\displaystyle \frac{1}{r},0)$ and $(0,0,0,\displaystyle \frac{1}{r \sin(\theta)})$, along with one additional vector, normal to the second two vectors.  We shall call this vector $k^{a}$, and a simple computation shows that there is only one choice of $k^{a}$ that is null, normal to the two angular directions, and satisfies $k^{a}d\alpha_{a} = -1$.  This choice is given by $k^{a} = (-1,-\frac{1}{2},0,0)$, which, upon lowering with $g_{ab}$, is equivalent to $k_{a} = (\frac{1}{2},-1,0,0)$

Now, we define the outgoing light cone at some time by setting $v = constant$, which means that $dv = 0$.  Eliminating the appropriate columns from the above equation then gives the induced metric:

\begin{equation}
q_{ab} =  \left( \begin{array}{ccc}
0 & 0 & 0 \\
0  & r^{2}&0   \\
0&0&r^{2}sin^{2}(\theta) \end{array} \right)
\,. \end{equation}

Which is manifestly degenerate.  The general matrix $q^{ab}$ satisfying $q_{am}q^{mn}q_{nb} = q_{ab}$ is given by:

\begin{equation}
q^{ab} =  \left( \begin{array}{ccc}
a & b & c  \\
b  & \frac{1}{r^{2}}&0   \\
c&0&\frac{1}{r^{2}sin^{2}(\theta)} \end{array} \right)
\,. \end{equation}

But, we now require that $q^{ab}$ also satisfies $q^{ab}\underset{\leftarrow}{k_{a}}=0$.  By inspection, we can see that $\underset{\leftarrow}{k_{a}} = (-\frac{1}{2},0,0)$.  Therefore, we can see that $q^{ab}\underset{\leftarrow}{k_{a}} = -\frac{1}{2}(a,b,c)$, which then gives us:

\begin{equation}
q^{ab} =  \left( \begin{array}{ccc}
0 & 0 & 0  \\
0  & \frac{1}{r^{2}}&0   \\
0&0&\frac{1}{r^{2}sin^{2}(\theta)} \end{array} \right)
\,. \end{equation}

Which is, of course, the answer that one would have na\''{i}vely guessed to be the correct one.  As before, one should carefully note that the the first index of the lowered induced metric acts upon what was the r index in the 4-dimensional spacetime.  As shown above, this means that the corresponding vector is $\ell^{a}$.  Meanwhile, the raised indices also have an index that is labeled as r.  This index, however, acts upon $\underset{\leftarrow}{k_{a}}$, and $\mathbf{not}$ upon $\ell_{a}$, which has zero pullback onto the null surface.  Note (as stated above) that this is, in fact necessary, since the covector space to a vector space is defined precisely by the condition that the covector basis $e_{a}$, when acting on the vector basis $e^{b}$ must give the result $\delta_{a}{}^{b}$, which is impossible if $\ell^{a}$ is a vector and if $\ell_{a}$ is a covector.  The simplicity of the above Minkowski example was chosen to show the inevitability of this fact.

Now, let us turn our attention the Kerr solution to the Einstein Equation.  This solution is an algebraically special solution, and therefore, can be given in the form of a fiducial Minkowski metric plus one of the principal null vectors times itself.  In spheroidal coordinates $(t,r,\phi,\theta)$, we have:

\begin{equation}
g_{ab} = \eta_{ab} + C k_{a}k_{b} \;\;\;\;\;\;\;\;\;\;\;\;\;\; g^{ab} = \eta^{ab} - C k^{a}k^{b}
\,. \end{equation}

\begin{equation}
\eta_{ab} = diag(-1, \frac{B}{A}, Asin^{2}\theta, B) \;\;\;\;\;\;\;\;\;\;\; k_{a} = (-\frac{A}{B}, -1 , \frac{A a sin^{2}\theta}{B},0)
\,. \end{equation}

\begin{equation}
C = \frac{2MrB}{A^{2}} \;\;\;\;\;\;\;\;\;\;\;\;\;\;\;\;\; A = r^{2} + a^{2} \;\;\;\;\;\;\;\;\;\;\;\;\;\;\;\;\;\; B = r^{2}+a^{2}cos^{2}\theta \label{KerrSolutionDefinition}
\,. \end{equation}

Where a and M are taken to be two parameters, with $|a| \leq M$.  It is easy to check that $k_{a}$ is null according to the Minkowski background metric, and therefore, it is also easy to verify that $k_{a}$ is null relative to the full metric as well.  Therefore, either $\eta_{ab}$ or $g_{ab}$ can be used to raise or lower indices on $k_{a}$, and we take $\eta^{ab}$ to be the metric inverse of $\eta_{ab}$.  We are now going to investigate the properties of this spacetime's horizon, and the properties of its embedding into the full 4-dimensional spacetime.  Before we do so, it is useful to introduce the other principal null vector of this spacetime:

\begin{equation}
\ell_{a} =  (-( \frac{A-2Mr}{2A}),\frac{B(A+2Mr)}{2A^{2}},\frac{(A-2Mr)a sin^{2}\theta}{2A},0)
\,. \end{equation}

Which, when raised using $g^{ab}$ gives:

\begin{equation}
\ell^{a} = (\frac{A+2Mr}{2A},\frac{A-2Mr}{2A},\frac{(A+2Mr)a}{2A^{2}},0)
\,. \end{equation}

It can be verified that this vector has zero norm, and that $\ell_{a}k^{a} = \ell^{a}k_{a}=-1$.  Now, note that this vector seems to have some quite strange behaviour when $A-2Mr = 0$.  Noting from the definition above that $A = r^{2} + a^{2}$, it is easy enough to prove that $r^{2}-2\,M\,r+a^{2}=0$ is satisfied when:

\begin{equation}
r= r_{±} \equiv M ± \sqrt{M^{2}-a^{2}}
\,. \end{equation}

Note that, for these values of r, $\ell_{a} \propto dr$, while $\ell^{a}$ has a vanishing coordinate in the $\partial_{r}$ direction.  We therefore expect that for these values of r, this is, in fact, a null submanifold with associated null vector $\ell^{a}$.  Using the above condition, and taking $r = r_{+}$ as the associated function (and thereby setting dr = 0), we take the 4-metric, do some algebra, and find the following induced metric:

\begin{equation}
q_{ab} =  \left( \begin{array}{ccc}
\frac{a^{2}sin^{2}\theta}{B_{+}} & \frac{-A_{+}asin^{2}\theta}{B_{+}} & 0  \\
\frac{-A_{+}asin^{2}\theta}{B_{+}}  & \frac{A_{+}^{2}sin^{2}\theta}{B_{+}}&0   \\
0&0&B_{+} \end{array} \right)
\,. \end{equation}

Where a subscript of + indicates that a function of r is taking on the value $r=r_{+}$ in its r argument.  This is a more complicated looking metric than the one that we derived for the Minkowski spacetime, but, once again, it is easy to verify that it has vanishing determinant, is nonnegative definite, and that its zero eigenvector is $(1,\frac{a}{A_{+}},0)= (\ell^{t}{}_{+},\ell^{\phi}{}_{+},\ell^{\theta}{}_{+})$, and, since $\ell^{r}{}_{+}=0$, we can therefore just project this vector onto our null space and call it $\ell^{a}$.  We therefore have a null tangent space to our horizon spanned by $\ell^{a}$ and two spacelike angular directions.  Finally, we pull back $k_{a}$ on the horizon with result $\underset{\leftarrow}{k_{a}}= (-\frac{A_{+}}{B_{+}},\frac{A_{+}sin^{2}\theta}{B_{+}},0)$.  

Solving the equation $q_{am}q^{mn}q_{nb} = q_{ab}$ gives the following matrix, for undetermined functions X, Y, Z:

\begin{equation}
q^{ab} =  \left( \begin{array}{ccc}
\frac{B_{+}+2A_{+}a(sin^{2}\theta)X -A^{2} ( sin^{2}\theta)Y}{a^{2}sin^{2}\theta} & X & Z  \\
X&Y&\left( \frac{a}{A} \right) Z   \\
Z&\left( \frac{a}{A} \right) Z&\frac{1}{B_{+}} \end{array} \right)
\,. \end{equation}

Contracting this onto $\underset{\leftarrow}{k_{a}}$ and requiring that the answer be the zero vector then gives us the solution:

\begin{equation}
X = \frac{a}{B_{+}} \;\;\;\;\;\;\;\;\;\;\; Y = \frac{1}{B_{+}sin^{2}\theta} \;\;\;\;\;\;\;\;\;\;\;\;\; Z=0
\,. \end{equation}

Substituting this answer into the above expression for $q^{ab}$ gives us:

\begin{equation}
q^{ab} =  \left( \begin{array}{ccc}
\frac{a^{2}sin^{2}\theta}{B_{+}} & \frac{a}{B_{+}}& 0  \\
\frac{a}{B_{+}}&\frac{1}{B_{+}sin^{2}\theta}&0   \\
0&0&\frac{1}{B_{+}} \end{array} \right)
\,. \end{equation}

Which we can then take to be the inverse 3-metric on the Kerr Horizon.  As a final note, it is often easier to work with the so-called ``untwisted'' coordinates on the horizon.  These coordinates are defined by the coordinate transformation:

\begin{equation}
\phi = \varphi +\frac{at}{A_{+}}
\,. \end{equation}

Under this transformation, we can follow the above procedure, and obtain the following 3-metric (in $(t,\varphi, \theta)$ coordinates):

\begin{equation}
q_{ab} =  \left( \begin{array}{ccc}
0 & 0 & 0  \\
0  & \frac{A_{+}^{2}sin^{2}\theta}{B_{+}}&0   \\
0&0&B_{+} \end{array} \right)\;\;\;\;\;\;\;\;\;\;\;\; q^{ab} =  \left( \begin{array}{ccc}
\frac{a^{2}sin^{2}\theta}{B_{+}} & \frac{a}{A_{+}} & 0  \\
\frac{a}{A_{+}}  & \frac{B_{+}}{A_{+}^{2}sin^{2}\theta}&0   \\
0&0&\frac{1}{B_{+}} \end{array} \right) 
\,. \end{equation}

As one might guess, this transformation makes $\ell^{a} = (1,0,0)$ while leaving $\underset{\leftarrow}{k_{a}}$ equal to the more complicated expression $(-1, \frac{A_{+}asin^{2}\theta}{B_{+}},0)$.  Having done all of this work, it is now simple enough to proceed using quantities defined intrinsically on the Kerr Horizon.  Note that the above procedures are nontrivial in the Kerr case--the inverse of the $(\theta,\phi)$ part of $q_{ab}$ is {\bf not} equivalent to $q^{ab}$.  

Now, having completed this Appendix, it should be clear how to go about inducing enveloping geometries onto null subspaces.  In the end, the process is the same as for non-null subspaces, only with an extra step, because the tangent space and the cotangent space have to be treated separately.

\chapter{Hamiltonian of Klein-Gordon Field in an external gravitational field}\label{sec: Klein-Gordon 3+1}
Here, I will decompose the action for a Klein-Gordon field coupled to an external gravitational field using a 3+1 formalism, and then derive the Hamiltonian and the associated equations of motion.  In particular, we will show that while you can get boundary terms in the Hamiltonian, you get none of these boundary terms in the equations of motion.  While it is not explicitly included here, the entire argument below is essentially unaltered if one were to add a self-interaction term $V(\phi)$ to the action, so long as it dependent only on the value of the field $\phi$ and none of its derivatives.  

First, consider the Klein-Gordon action:

\begin{equation}
S = \int d^{4}x\sqrt{|g|}\mathscr{L}=\int d^{4}x\sqrt{|g|}\left(-\frac{1}{2}\nabla_{a}\phi \nabla^{a}\phi -\frac{1}{2} m^{2}\phi^{2}\right)
\,. \end{equation}

Since we are concerning ourselves with only the dynamics of the scalar field in the background metric, and therefore holding the external metric fixed, and since we furthermore are not computing the stress-energy tensor, we will treat the square root of the determinant as part of the volume element, and treat the Lagrangian density as a true geometric density.  This will simplify the calculation of boundary terms and does not change the content of any results.  Therefore, from here on, the term involving the square root of the metric will be factored into the $d^{4}x$ term, which will be understood to have the appropriate weight to be a volume element.

Knowing the Lagrangian density, it is easy to compute the momentum conjugate to the field:

\begin{equation}
\Pi \equiv \frac{\delta \mathscr{L}}{\delta \dot \phi} = \frac{\delta}{\delta \dot \phi}\left(-\frac{1}{2}\nabla_{a}\phi \nabla^{a}\phi -\frac{1}{2} m^{2}\phi^{2}\right) = -\nabla^{t}\phi \label{Momentum}
\,. \end{equation}

If this calculation were being done in Minkowski spacetime, then it would be direct to equate $\nabla^{t}\phi$ to $-\dot \phi$.  Since we are instead in a 3+1 split, it is necessary to insert a factor of the inverse metric and decompose, yielding:

\begin{align}
-\nabla^{t}\phi =& -g^{ta}\nabla_{a}\phi = -\left(-\frac{1}{\alpha^{2}}\delta_{t}{}^{a} + \frac{1}{\alpha^{2}}\beta^{i}\right)\nabla_{a}\phi \nonumber\, ,  \\
=&\frac{1}{\alpha^{2}}\dot \phi -\frac{1}{\alpha^{2}}\beta^{i}\bar \nabla_{i}\phi
\,. \end{align}

Where $\alpha$ is the lapse function and $\beta^{i}$ is the shift vector for the particular slicing in question, $i, j, k...$ indicate 3-dimensional indices, and $\bar \nabla_{i}$ denotes the connection compatible with the 3-dimensional metric $\gamma_{ij}$.  Putting this all together, we get:

\begin{equation}
\Pi = \frac{1}{\alpha^{2}}\left(\dot \phi - \beta^{i}\bar \nabla_{i}\phi\right)
\,. \end{equation}

So, having done this, now we can see that:

\begin{align}
\mathscr{L} =& \frac{1}{2} \Pi \dot \phi -\frac{1}{2} \nabla^{i}\phi \nabla_{i} \phi - \frac{1}{2}m^{2}\phi^{2}\nonumber \, ,  \\
=& \frac{1}{2} \Pi \dot \phi -\frac{1}{2} g^{ai}\nabla_{a}\phi \nabla_{i}\phi -\frac{1}{2}m^{2}\phi^{2}\nonumber \, ,  \\
=& \frac{1}{2} \Pi \dot \phi -\frac{1}{2} g^{ti}\dot \phi\nabla_{i}\phi -\frac{1}{2}g^{ij}\nabla_{i}\phi \nabla_{j}\phi -\frac{1}{2}m^{2}\phi^{2} \nonumber\, ,  \\
=& \frac{1}{2}\Pi \dot \phi -\frac{1}{2 \alpha^{2}}\beta^{i} \dot \phi \bar \nabla_{i} \phi - \frac{1}{2}\left(\gamma^{ij} - \frac{1}{\alpha^{2}}\beta^{i}\beta^{j}\right)\bar \nabla_{i}\phi \bar \nabla_{j}\phi - \frac{1}{2} m^{2}\phi^{2} \nonumber\, ,  \\
=& \frac{1}{2}\Pi \dot \phi - \frac{1}{2 \alpha^{2}} \left(\beta^{i}\bar \nabla_{i}\phi\right)\dot \phi+ \frac{1}{2\alpha^{2}}\left(\beta^{i}\bar \nabla_{i} \phi \right)^{2}-\frac{1}{2} \bar \nabla^{i} \phi \bar \nabla_{i}\phi - \frac{1}{2}m^{2}\phi^{2} \label{Rewritten}
\,. \end{align}

Now, we can combine \eqref {Rewritten} with the definition of the Hamiltonian density $\mathscr{H} \equiv \Pi \dot \phi - \mathscr{L}$ to obtain the Hamiltonian density, and then we can use \eqref {Momentum} to eliminate all dependence on $\dot \phi$:

\begin{align}
\mathscr{H} =& \Pi \dot \phi - \mathscr{L} \nonumber \, ,  \\
=& \frac{1}{2} \Pi \dot \phi + \frac{1}{2 \alpha^{2}} \left(\beta^{i}\bar \nabla_{i}\phi\right)\dot \phi- \frac{1}{2\alpha^{2}}\left(\beta^{i}\bar \nabla_{i} \phi \right)^{2}+\frac{1}{2} \bar \nabla^{i} \phi \bar \nabla_{i}\phi + \frac{1}{2}m^{2}\phi^{2} \nonumber\, ,  \\
=& \frac{1}{2} \alpha^{2} \Pi^{2} + \frac{1}{2} \Pi \beta^{i}\bar \nabla_{i}\phi + \frac{1}{2} \Pi \beta^{i} \bar \nabla_{i}\phi + \frac{1}{2\alpha^{2}}\left(\beta^{i}\bar \nabla_{i} \phi\right)^{2}-\frac{1}{2\alpha^{2}}\left(\beta^{i}\bar \nabla_{i}\phi\right)^{2}  \nonumber  \\
&+ \frac{1}{2} \bar \nabla^{i}\phi \bar \nabla_{i}\phi+ \frac{1}{2}m^{2}\phi^{2}\nonumber \, ,  \\
=& \frac{\alpha^{2}}{2}\Pi^{2} + \Pi \left(\beta^{i}\bar \nabla_{i}\phi\right) + \frac{1}{2} \bar \nabla^{i}\phi \bar \nabla_{i}\phi + \frac{1}{2} m^{2} \phi^{2}
\,. \end{align}

This then immediately lets us integrate the Hamiltonian density to get the Hamiltonian (with the factor of $\sqrt{\gamma}$ once again an implied subcomponent of the $d^{3}x$:

\begin{equation}
H \equiv \int \mathscr{H} d^{3}x = \int d^{3}x \left(  \frac{\alpha^{2}}{2}\Pi^{2} + \Pi \left(\beta^{i}\bar \nabla_{i}\phi\right) + \frac{1}{2} \bar \nabla^{i}\phi \bar \nabla_{i}\phi + \frac{1}{2} m^{2} \phi^{2}\right) \label{Hamiltonian}
\,. \end{equation}

Which gives the equivalent phase space action

\begin{equation}
S = \int d^{4}x \left[\Pi\dot \phi-\left(  \frac{\alpha^{2}}{2}\Pi^{2} + \Pi \left(\beta^{i}\bar \nabla_{i}\phi\right) + \frac{1}{2} \bar \nabla^{i}\phi \bar \nabla_{i}\phi + \frac{1}{2} m^{2} \phi^{2}\right)\right]
\,. \end{equation}

And we therefore take \eqref {Hamiltonian} to define our Hamiltonian for this Klein-Gordon system.  It is now trivial to get the equation of motion for $\phi$:

\begin{align}
\frac{\delta}{\delta \Pi} S =& \frac{\delta}{\delta \Pi}\int d^{4}x \left[\Pi\dot \phi-\left(  \frac{\alpha^{2}}{2} \Pi^{2} + \Pi \left(\beta^{i}\bar \nabla_{i}\phi\right) + \frac{1}{2} \bar \nabla^{i}\phi \bar \nabla_{i}\phi + \frac{1}{2} m^{2} \phi^{2}\right)\right]\nonumber \, ,  \\
=& \int d^{4}x\left[-\dot \Pi + \alpha^{2}\Pi + \beta^{i}\bar \nabla_{i} \phi \right]
\,. \end{align}

Which, since the variation must be equal to zero under an arbitrary variation, gives us the equation of motion $\dot \phi = \alpha^{2} \Pi + \beta^{i} \bar \nabla_{i} \phi$.  The variation with respect to $\phi$ required to obtain the equation of motion for $\dot \Pi$ is somewhat more intricate however:

\begin{align}
\frac{\delta}{\delta \phi} S=&  \int d^{4}x -\dot \Pi +\left(m^{2} \phi - \bar \nabla^{i} \bar \nabla_{i} \phi\right) \nonumber  \\
&+ \frac{\delta}{\delta \phi} \int d^{4}x \left( \bar \nabla_{i}\left(\Pi \beta^{i}\phi\right) - \phi \Pi \bar \nabla_{i}\beta^{i} - \phi \beta^{i}\bar \nabla_{i} \Pi\right) \nonumber\, ,  \\
0+\int d^{4}x\dot \Pi=& \int d^{4}x\left(m^{2} \phi - \bar \nabla^{i}\bar \nabla_{i} \phi - \Pi \bar \nabla_{i}\beta^{i} - \beta^{i} \bar \nabla_{i} \Pi \right) + \oint d^{3} x\left( r_{i} \Pi \beta^{i}\right) \nonumber \, ,  \\
=& \int d^{4} x \left[ m^{2} \phi - \bar \nabla^{i}\nabla_{i} \phi - \bar \nabla_{i}\left(\Pi \beta^{i}\right)\right] + \oint d^{3}x \left(r_{i} \Pi \beta^{i}\right) \label{second}\, ,  \\
=& \int d^{4}x \left(m^{2} \phi - \bar \nabla^{i}\bar \nabla_{i} \phi \right) \label{Pi equation}
\,. \end{align}

Where, in going from \eqref {second} to \eqref {Pi equation} we noted that the divergence in the bulk integral precisely canceled the boundary integral via a simple application of Gauss's theorem.  Therefore, we obtain the equation of motion for $\Pi$, which is simply $\dot \Pi = \bar \nabla^{i}\bar \nabla_{i} - m^{2} \phi$.  We have now found the complete equations of motion for this Hamiltonian system after taking a variation of our dynamical variables, and importantly, we have found that there are no dangling boundary terms remaining after our variation has been taken, meaning that it is not necessary to add any counterterms to our original action--the variation of the phase space action will be zero for any $\left(\Pi, \phi\right)$ that satisfy the equations of motion for the Klein-Gordon field.  

As a final step, it would be remiss to not check our result against the traditional empty-space Klein-Gordon equations of motion in Cartesian coordinates.  This is simply equivalent to setting $\gamma^{ij} = \delta^{ij}$, and making the choices $\beta_{i} = 0$ and $\alpha =1$.  If we do this we have:

\begin{equation}
\dot \phi = \Pi \;\;\;\;\;\;\;\;\;\;\;\;\;\;\;\;\;\;\;\; \dot \Pi = \partial^{i}\partial_{i} \phi - m^{2} \phi
\,. \end{equation}

If we take a time derivative of the equation for $\dot \phi$, and then substitute this answer into the equation for $\dot \Pi$, we get

\begin{align}
\ddot \phi =& \partial^{i}\partial_{i}\phi -m^{2}\phi \nonumber \, ,  \\ 
0=& - \ddot \phi + \partial^{i}\partial_{i} \phi - m^{2} \phi \nonumber \, ,  \\ 
0=& \eta^{ab}\partial_{a}\partial_{b}\phi - m^{2} \phi \label{KGEOM}
\,. \end{align}

And \eqref {KGEOM}, indeed, is the Klein-Gordon equation of motion.

\chapter{Boundary terms in the Schwarzschild-Roberson-Walker spacetime}
As another example, let us consider the case of the Schwarzschild-Robertson-Walker spacetime.  This is furnished by a simple extension of both the Schwarzschild and Robertson-Walker spacetimes, and is given by the line element\footnote{Note: the function $a(t)$ is meant to be evocative of the Robertson-Walker function $a(t)$.  It is {\bf not} meant to have anything to do with the Kerr parameter $a$.}

\begin{align}
g_{ab}dx^{a}{}dx^{b}=&-\left(1 -\frac{2M}{r}\right) dt^{2}  +\left(a(t)\right)^{2}\left[\left(\frac{1}{1-\frac{2M}{r}}\right)dr^{2} \nonumber\right.  \\
&\left.+ r^{2}d\theta ^{2} +r^{2}\sin\left(\theta\right)d\phi^{2}\right] \label{RWmetric}
\,. \end{align}

If we take the limit $a(t) = 1,\dot a(t) \rightarrow \ddot a(t) \rightarrow 0$ we clearly recover the Schwarzschild spacetime, while if we take the limit $M\rightarrow0$ we recover the flat Robertson-Walker model.  Furthermore, direct computation of the curvature generated by this line element shows that its Einstein tensor is:

\begin{align}
G_{ab}dx^{a}{}dx^{b} =& 3\left(\frac{\dot a}{a}\right)^{2}dt^{2}+2dt{}dr\left(\frac{2M}{r^{2}}\right)\left(\frac{\dot a}{a(1-\frac{2M}{r})}\right)-dr^{2}\left(\frac{\dot a^{2}+2a\ddot a }{(1-\frac{2M}{r})^{2}} \right)\nonumber  \\
& -d\theta^{2}r^{2}\left(\frac{\dot a^{2}+2a\ddot a }{(1-\frac{2M}{r})} \right)-d\phi^{2}r^{2}sin^{2}\left(\theta\right)\left(\frac{\dot a^{2}+2a\ddot a }{(1-\frac{2M}{r})^{2}} \right)
\,. \end{align}

Which manifestly satisfies $G_{r}{}^{r}=G_{\theta}{}^{\theta}=G_{\phi}{}^{\phi} =- \left(\frac{\dot a^{2}+2a\ddot a }{(1-\frac{2M}{r})} \right)$.  We can therefore interpret the line element given in equation \eqref {RWmetric} as the gravitational field of a mass M surrounded by a strain-free dynamical fluid with a radial current\footnote{by changing the time coordinate to $\tau = 3 ln (a) + ln\left(1-\frac{2M}{r}\right)$, we can diagonize the Einstein tensor.  This will make the fluid pressure in the r direction different from the fluid pressure in the angular directions as well as making the metric tensor nondiagonal.  Therefore, the fluid is not isotropic in its comoving frame}.  Note that this solution, however, is likely unphysical, as computing the trace of the Einstein tensor above will show that both the Ricci scalar and $R_{ab}R^{ab}$ contain a singularity at what would otherwise be the black hole horizon at $r=2M$.  Assuming a polytropic equation of state for the fluid can produce a concrete form for the function $a(t)$, and given this form, numerical computation of the geodesic equations generated from equation \eqref {RWmetric} will show that this coordinate is accessible in a finite amount of proper time, so the black hole horizon in the McVittie solution is a naked singularity.  Furthermore, computation of the scalar $C_{abcd}C^{abcd}$ out of the metric \eqref{RWmetric} produces a result that is also singular, making it unlikely that this singularity arises solely due to a caustic in the matter\footnote{There is one exception to this: if $a(t)$ satisfies $\frac{\dot a(t)}{a(t)} = - \frac{\dot a(t)}{\dot a(t)}$, which then makes $a(t)$ an exponential function.  In this case, $C_{abcd}C^{abcd} $ and $R$ become finite, although $R_{ab}R^{ab}$ retains its singularity.  So, while the singularity remains, its strength is lessened and its character changed. }.  

Furthermore, varying the line element with respect to t will generate the equation:

\begin{align}
\frac{d E_{Schwarzchild}}{dt}& = -a \dot a\left(\dot r^{2}\frac{1}{1-\frac{2M}{r}} + \dot \theta^{2} r^{2} + \dot \phi^{2} r^{2}sin^{2}\left(\theta\right)\right)\nonumber \, ,  \\
&= -\frac{\dot a}{a}E_{Robertson-Walker}
\,. \end{align}

Here $E_{Schwarzschild}$ is the formerly conserved quantity in the pure Schwarzschild spacetime arising from the former killing vector $\partial_{t}$, and $E_{Robertson-Walker}$ is the conserved energy in the pure Robertson-Walker spacetime that is derived by combining the geodesic equation for the time coordinate with the unit timelike condition.

In particular, this means that for an expanding (contracting) universe, geodesics will tend to have their energy redshifted (blueshifted) as time progresses.  This dynamical effect will make stable circular orbits impossible, and on a timescale governed by the Hubble time and by the mass of the central object, initially circular orbits will tend to fall into the singularity at $r=2M$ (or spiral outward toward infinity and become unbound).  Numerical solutions of the geodesic equations verify this property of geodesics and generalize it to elliptical orbits as well.  A few plots of solutions with polytropic equations of state are plotted below.  The geodesic equation was solved numerically using the Mathematica software package, which also produced the plots.  

\begin{figure}[htb!]
\includegraphics[scale=1.3]{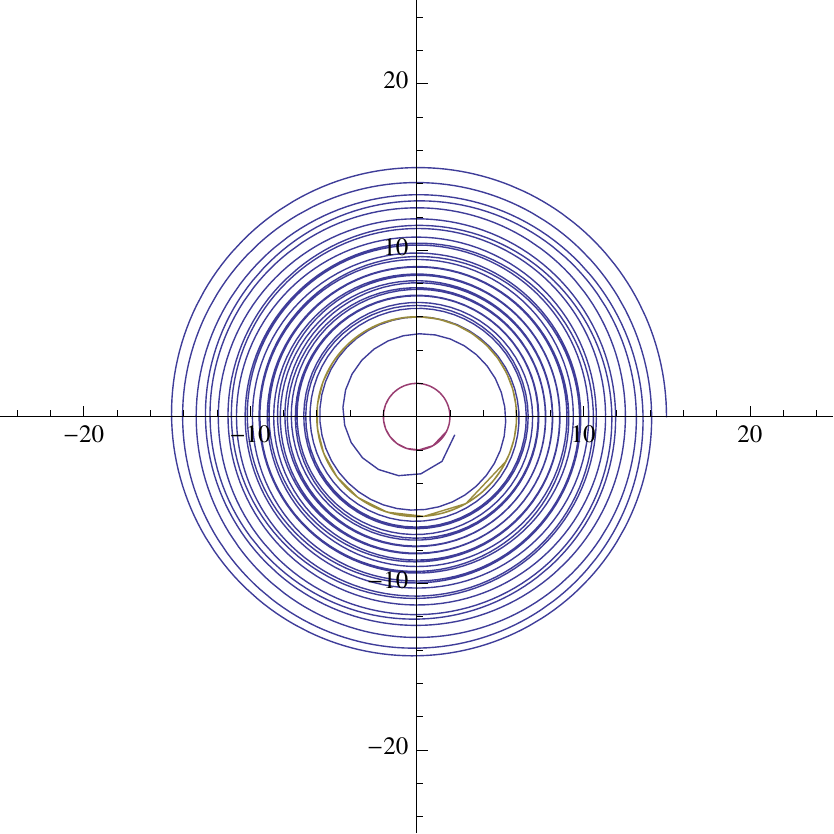}
\caption{A graph of the solution for an equatorial Geodesic with $r_{0}=15M$, $H_{0} = \frac{5.41 \times 10^{-5}}{M}$, and $L=L_{circular}$.  The red circle represents $r=2M$, while the gold circle represents the innermost stable circular orbit at $r=6M$.  The test particle completes 21 full orbits before falling into the black hole.}
\label{fig:SRWSmallH}
\end{figure}

\begin{figure}[htb!]
\includegraphics[scale=1.3]{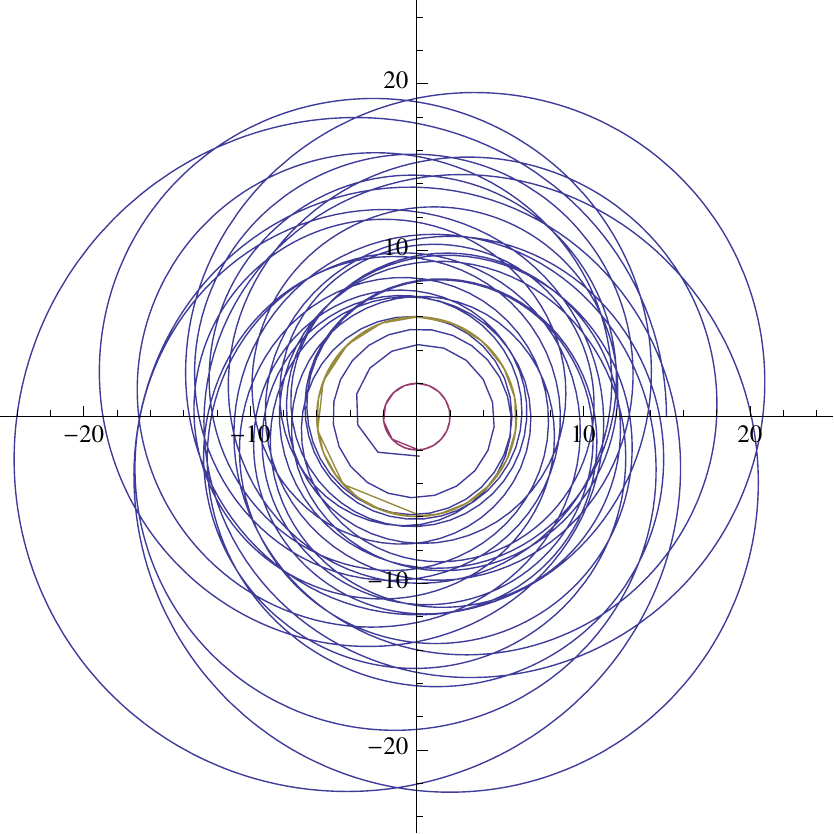}
\caption{A graph of the solution for an equatorial Geodesic with $r_{0}=15M$, $H_{0} = \frac{5.41 \times 10^{-5}}{M}$, and $L=1.1L_{circular}$.  An addition of a relatively small amount of additional angular momentum than this produces a simple escape orbit, approximately at $L=1.35\,M$.  The red circle represents $r=2M$, while the gold circle represents the innermost stable circular orbit at $r=6M$.  The test particle completes 21 full orbits before falling into the black hole.  The test particle completes 28.9 full orbits before plunge happens}

\label{fig:SRWEllipse}
\end{figure}
\begin{figure}[htb!]
\includegraphics[scale=1.3]{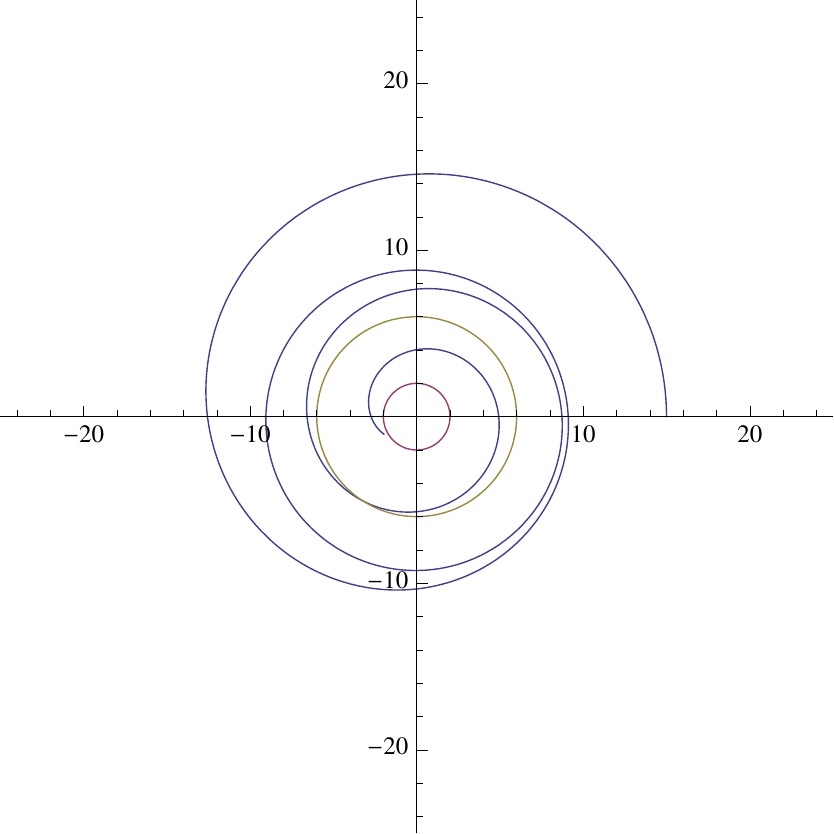}
\caption{A graph of the solution for an equatorial Geodesic with $r_{0}=15M$, $H_{0} = \frac{5.41 \times 10^{-4}}{M}$, and $L=L_{circular}$.  First, note that the Hubble constant for this plot is ten times the Hubble constant for the other two plots.  The red circle represents $r=2M$, while the gold circle represents the innermost stable circular orbit at $r=6M$.  The test particle completes 3 full orbits before falling into the black hole.  Note the degree to which there is already a poor division between the inspiral phase and the plunge phase, although the geodesic does manage to complete three full orbits.  Orbits with values of $H_{0}$ much larger than this tend to not even complete whole orbits before plunge, even when given large amounts of angular momentum.}
\label{fig:SRWLargeH}
\end{figure}

Since a particular orbit is only bound for a finite amount of time in this spacetime, we therefore expect to find some interesting time-dependence in the energetics of this spacetime.  In particular, this example was chosen as a way to show that it is, in fact possible to generate reasonable-seeming spacetimes that have time dependent ADM masses, as shall be shown below.

So now, let us compute this spacetime's ADM mass.  For the slicing coordinate, we choose the coordinate t from the line element above.  Meanwhile, for the fiducially flat comparison background metric $\tilde \gamma_{ij}$, we make the obvious choice of the spherical Euclidean metric given by choosing $a(t) = 1, M = 0$\footnote{One might object to this choice, arguing that the spacetime is not, in fact, asymptotically flat.  However, each timeslice under this choice IS asymptotically flat, and with a rescaling of the time coordinate, this spacetime is conformal to Schwarzschild spacetime, which is asymptotically flat.  Also, since none of the derivatives in the ADM formula is a time derivative, and there are two factors of the lowered metric (inside the parentheses, and in the two-metric determinant), and one factor of the raised metric, this choice yields the same formula as you would obtain by choosing a fiducial metric by defining M = 0}.  Then, denoting covariant differentiation relative to $\tilde \gamma_{ij}$ as $\tilde \nabla_{i}$, we compute the ADM mass:

\begin{align}
16\pi M_{ADM} =& \oint \sqrt{\tilde q} \tilde \gamma^{ij}\tilde r^{k}\left(\tilde \nabla_{i}\gamma_{jk} - \tilde \nabla_{k}\gamma_{ij}\right)\nonumber \, ,  \\
=& \oint \sqrt{\tilde q} \tilde \gamma^{ij}\tilde r^{k}\left(\partial_{i}\gamma_{jk}-\partial_{k}\gamma_{ij} -\tilde \Gamma_{ij}{}^{\ell}\gamma_{\ell k} + \tilde \Gamma_{kj}{}^{\ell}\gamma_{\ell i} \right)\nonumber\, ,  \\
=&\oint \sqrt{\tilde q} \tilde \gamma^{ij}r^{k}\left(\partial_{i}\gamma_{jk}-\partial_{k}\gamma_{ij}\right)+\oint \sqrt{\tilde q} \tilde \gamma^{ij}r^{k}\left(\tilde \Gamma_{jk}{}^{\ell}\gamma_{\ell i} - \tilde \Gamma_{ij}{}^{\ell}\gamma_{\ell k}\right) \nonumber\, ,  \\
=&\oint \sqrt{\tilde q}\left(-\tilde \gamma^{\theta \theta}\partial_{r}\gamma_{\theta \theta} - \tilde \gamma^{\phi \phi}\partial_{r}\gamma_{\phi \phi}\right)+\oint \sqrt{\tilde q}\tilde \gamma^{ij}\left(\tilde \Gamma_{j r}{}^{\ell}\gamma_{\ell i} - \tilde \Gamma_{ij}{}^{\ell}\gamma_{\ell r}\right)\nonumber\, ,  \\
=&\oint d\phi{} d\theta{} r^{2}\sin\left(\theta\right)\left(-\frac{4a^{2}}{r}\right)\nonumber  \\
&+\oint\sqrt{\tilde q}\tilde \gamma^{ij}\left(\left(\frac{1}{r}\right)\left(\delta_{j}{}^{\theta}\gamma_{i \theta} + \delta_{j}{}^{\phi}\gamma_{i \phi}\right) - \frac{a^{2}}{1-\frac{2M}{r}}\left(\frac{1}{r}\right)\left(-\tilde q_{ij}\right)\right)\nonumber \, ,  \\
=&\oint d\phi{}d\theta{}r^{2}\sin\left(\theta\right)\left(-\frac{2a^{2}}{r} + \left(\frac{2a^{2}}{r}\right)\frac{1}{1-\frac{2M}{r}}\right) \nonumber\, ,  \\
=& \oint d\phi{}d\theta{}r^{2}\sin\left(\theta\right)\left(\frac{2a^{2}}{r}\right)\left(\frac{1}{1-\frac{2M}{r}}\right)\left(\frac{2M}{r}-1+1\right)\nonumber \, ,  \\
=&\frac{4Ma^{2}}{1-\frac{2M}{r}}\oint d\phi{}d\theta{}\sin\left(\theta\right)\nonumber \, ,  \\
=&\frac{16\pi M a^{2}}{1-\frac{2M}{r}}
\,. \end{align}

And then, when we take the limit $r \rightarrow \infty$, we obtain the result $M_{ADM}=Ma^{2}$, indicating that the ADM mass of this spacetime is time-dependent.  Furthermore, this behaviour explains the odd behaviour of the geodesics of this spacetime--rather than interpreting the energy of an orbiting particle as being redshifted by the cosmological expansion, we can also interpret the mass concentrated in the central object increasing by an amount proportional to the scale factor over time.  Therefore, initially circular orbital paths eventually find themselves attracted by a mass larger than the one they were attracted by initially, and see the radius of their orbit shrink.  Eventually, they find themselves beyond the stable limit of their orbit, and have no choice but to plunge into the surface of the singularity.  

The conclusion to reach is that while, for asymptotically flat spacetimes, the ADM mass is conserved, there do exist examples where one can find a time-dependence in the ADM quantities.  Relativity's insistence on local mass-energy conservation does not necessarily imply global mass-energy conservation.  This result arises from the fact that this spacetime does not admit a global timelike Killing vector on the ``sphere at infinity''\footnote{In classical mechanics, it is not wholly abnormal for boundary conditions to change the overall value of the Hamiltonian, even if the bulk states stay the same.  Consider a system of a string vibrating in its fundamental mode on a string of length $L$, mass per length $\mu$ and amplitude $A$.  Now, a simple evaluation of the Hamiltonian of this system\cite{LandauLifshitzClassical} will show you that the total energy in the string is given by $\frac{\pi^{2}\mu A^{2}}{4L}$.  Therefore, if one were to lengthen the string without changing its tension, and if $\dot L$ were much smaller than the characteristic wave velocity in the string, one would expect that the string would stay in its new fundamental mode, and have energy $\frac{\pi^{2}\mu A^{2}}{4L(t)}$.  Thus the slow expansion of the string would allow the string to do work, causing it to lose energy.  The proposal in this appendix is that we can similarly interpret time-dependent boundary terms in relativity--they are work terms indicating net \textit{global} energy generated by the gravitational field, just as the factor expanding the classical string creates/removes net energy from that system.}. 

\chapter{On various foliations of a spacelike surface in a 3+1 split} \label{sec: variousfoliations}

Note that, when dealing with boundary conditions in 3+1 foliations, we often have to deal with TWO foliations of boundary terms--one involving the choice a constant $\tau$ slice, giving a spacelike surface and one involving the choice of a constant $R$ slice, which gives the boundary of that spacelike surface.  In this appendix we will work out some consistency problems involving these two choices.  The general scope of this argument is following that given in \cite{ YorkBC}, but the notation chosen in this derivation is chosen so as to match the notation in the rest of this work, as certain results derived in this appendix will be used heavily elsewhere in this work, particularly in Chapters \ref{sec: Brute Force}, \ref{sec: Gauss's Theorem}, and \ref{sec: NullVectors}.  Also, certain details are worked out in a slightly different way.  For the below derivation, unless otherwise noted, we will be using the inclusion operator to map intrinsic 2- and 3-geometries into the enveloping 4-geometry, since we will be comparing geometric quantities to each other.  Also, we will freely raise and lower vectors using the full 4-metric $g_{ab}$ and its inverse.  

Now, consider a 4-manifold with boundary $\mathbb{M}$ with topology $\mathbb{R}\times \mathbf{m}$ with boundary $\mathbf{m}\big{|}_{0} \cup \mathbf{m}\big{|}_{f} \cup \left(\mathbb{R}\times \partial \mathbf{m}\right)$.  We will label the $\mathbb{R}$ portion of the spacetime with the function $\tau$.  Similarly, there is some function $R$ such that $R$ is constant on $\partial \mathbf{m}$ and such that $R$ is not constant on some neighborhood of $\partial \mathbf{m}$.  For now, we will assume that $\nabla_{a}R$ is a spacelike object, and examine the null case at the end of this appendix.  Now, as was done in \eqref {sec: normals}, we can easily define the two normals to this surface as they live in the enveloping 4-space.  And from these normals, we can construct projection operators onto their respective normal spaces:

\begin{align}
n_{a} &= \alpha \nabla_{a} \tau & & s_{a} = B \nabla_{a} R \nonumber \, ,  \\
\phantom{\bigg{|}}\alpha& =\frac{1}{\sqrt{|g^{ab}\left(\nabla_{a}\tau\right)\nabla_{b}\tau|}}&&B = \frac{1}{\sqrt{g^{ab}\left(\nabla_{a}R\right)\nabla_{b}R}}  \, ,  \\
\gamma^{ab} &=g^{ab} + n^{a}n^{b} && {\bar {\bar \gamma}}^{ab}=g^{ab}-s^{a}s^{b}\nonumber
\,. \end{align}

Note, however, that it is {\bf not} the case that $n^{a}s_{a} =0$, since we have done nothing to guarantee that these things are normal to each other.  In particular, if we choose $\tau$ and $R$ as coordinates, so that our coordinate system is $(\tau,R,x^{3},x^{4})$, it should be clear that it should not generally be the case that $g^{\tau R} = 0$ on any particular surface.  Therefore, $n^{a}$ and $s^{a}$ are not a particularly good choice of basis for the normal vector space to $\partial \mathbf{m}$.

So, what to do?  Clearly, it will be necessary to use the projection operators in a Gramm-Schmidt procedure.  Since we are starting with the same 4-manifold and we are projecting down to the same final 2-manifold, it shouldn't matter which vector we start with and which vector we project.  As we will see below, however, there is not complete equivalence\footnote{For an extreme example for why this should be the case, consider the Schwarzschild spacetime in Kerr-Schild coordinates at $r=2M$.  If we slice by Kerr Schild time first, then we get a spacelike 3-manifold that stretches all of the way to the singularity at $r=0$.  The $r=2M$ surface appears like a perfectly well-behaved spacelike sphere.  Null geometry would never have to be used or invoked.  

Now, however, if we were to choose to slice the spacelike slice FIRST, we would get a null 3-surface, and only recover the spacelike 2-geometry after setting the null parameter of the horizon equal to a constant.  Care must be taken in order to ensure consistency between these two approaches}.  So, we can define two different projections of the vectors, first onto the barred space defined by $\gamma^{ab}$ and second, onto the double barred space defined by ${\bar {\bar \gamma}}^{ab}$.  Finally, we define the projection operators defined by these two different methods, and note that they must be the same, as the target 2-space is simply $\partial \mathbf{m}$

\begin{align}
\bar s_{a} &= \bar B \bar \nabla_{a}R=\bar B \gamma_{a}{}^{b}\nabla_{b}R&&{\bar {\bar n}}_{a} = {\bar {\bar \alpha}} {\bar {\bar \nabla}}_{a}\tau ={\bar{\bar\alpha}}\,{\bar{\bar \gamma}}_{a}{}^{b}\nabla_{b}\tau\nonumber\, ,  \\
\bar B &= \frac{1}{\sqrt{\gamma^{ab}\bar s_{a}\bar s_{b}}}&& {\bar {\bar \alpha}} =\frac{1}{\sqrt{| {\bar {\bar \gamma}}^{ab}{\bar {\bar n}}_{a} {\bar {\bar n}}_{b}  {|}}}\label{Decomposition consistency equations}
\,. \end{align}
$$q^{ab} = g^{ab} + n^{a}n^{b} - \bar s^{a} \bar s^{b} = g^{ab} -s^{a}s^{b} + {\bar{\bar n}}^{a}{\bar{\bar n}}^{b}$$

Now, the obvious conclusion that we can get from equations \eqref {Decomposition consistency equations} is that $n^{a}n^{b} - \bar s^{a} \bar s^{b} = -s^{a}s^{b} + {\bar{\bar n}}^{a}{\bar{\bar n}}^{b}$.  We are going to use this fact in order to derive an interesting conclusion regarding the two sets of vectors.  In what follows, note that the constructions of all of our projection operators are made in such a way that they annihilate relevant vectors.  Therefore, $n^{a}\bar s_{a} = {\bar {\bar n}}^{a}s_{a} =0$.  it is {\bf not} the case, however, that the other contractions of these normals are going to  vanish.  Therefore, we will make the definition $n^{a}s_{a} = \psi$.   

Expanding the definition of $\bar s_{a}$:

\begin{align}
\bar s_{a} &= \bar B \gamma_{a}{}^{b}\nabla_{b} R = \bar B\left(\delta_{a}{}^{b} + n_{a}n^{b}\right)\nabla_{b}R \nonumber \, ,  \\
&=\bar B \nabla_{a}R + \bar B n_{a}n^{b}\nabla_{b}R \nonumber \, ,  \\
&=\frac{\bar B}{B}\left(s_{a} + n_{a}\psi\right) \nonumber \, ,  \\
s_{a} &= \frac{B}{\bar B}\bar s_{a} - \psi n_{a} \label{LorentzPreForm1}
\,. \end{align}

Similarly,

\begin{align}
{\bar{\bar n}}_{a} &= {\bar{\bar \alpha}} {\bar{\bar \gamma}}_{a}{}^{b}\nabla_{b} \tau={\bar{\bar \alpha}}\left(\delta_{a}{}^{b} - s_{a}s^{b}\right)\nabla_{b} \tau \nonumber \, ,  \\
&= {\bar{\bar \alpha}}\left(\nabla_{a}\tau - s_{a} s^{b} \nabla_{b}\tau\right)\nonumber \, ,  \\
&= \frac{{\bar{\bar \alpha}}}{\alpha}\left(n_{a} - s_{a} \psi\right) \nonumber\, ,  \\
&= \frac{{\bar{\bar \alpha}}}{\alpha}n_{a} -\psi\,\frac{{\bar{\bar \alpha}}}{\alpha}\left( \frac{B}{\bar B}\bar s_{a} - \psi n_{a} \right)\nonumber \, ,  \\
&= \frac{{\bar{\bar \alpha}}}{\alpha}\left(1+\psi^{2}\right)n_{a}-\left(\frac{{\bar{\bar \alpha}}B}{\alpha \bar B}\right)\psi \bar s_{a}\label{LorentzPreForm2}
\,. \end{align}

So, we have now solved for the reverse order ADM vectors $\bar s_{a}$ and $n_{a}$ in terms of their standard order counterparts, the ``lapse'' functions and the parameter $\psi$.   Next, we enforce the fact that $s_{a}$ and ${\bar {\bar n}}^{a}$ is an orthonormal basis.  Then, we will check to make sure that the consistency relationship derived from equating the $q^{ab}$ is valid.  Starting with equations \eqref {LorentzPreForm2} and \eqref {LorentzPreForm1}, and assuming that all of the 'lapse functions' are positive \footnote{And if they are not, minus signs can be absorbed into the $R$ and $\tau$ in order to enforce this positivity}:

\begin{align}
s_{a}s^{a} =& 1 =\left(\frac{B}{\bar B}\bar s_{a}-\psi n_{a}\right)\left(\frac{B}{\bar B}\bar s^{a}-\psi n^{a}\right) \nonumber \, ,  \\
1=&\left(\frac{B}{\bar B}\right)^2 - \psi^{2} \nonumber \, ,  \\
\frac{\bar B}{B} =& \frac{1}{\sqrt{1+\psi^{2}}}
\,. \end{align}

Continuing this process,

\begin{align}
{\bar {\bar n}}_{a}{\bar {\bar n}}^{a}  =&\left( \frac{{\bar{\bar \alpha}}}{\alpha}\left(1+\psi^{2}\right)n_{a}-\left(\frac{{\bar{\bar \alpha}}B}{\alpha \bar B}\right)\psi \bar s_{a}\right)\left( \frac{{\bar{\bar \alpha}}}{\alpha}\left(1+\psi^{2}\right)n^{a}-\left(\frac{{\bar{\bar \alpha}}B}{\alpha \bar B}\right)\psi \bar s^{a}\right) \nonumber \, ,  \\
-1 =&-\left(\frac{{\bar{\bar\alpha}}}{\alpha}\right)^{2}\left(1+\psi^{2}\right)^{2}+ \left(\frac{{\bar {\bar \alpha}}B}{\alpha \bar B}\right)^{2}\psi^{2} \nonumber \, ,  \\
=& -\left(\frac{{\bar{\bar \alpha}}}{\alpha}\right)^{2}\left[1+2\psi^{2} + \psi^{4}-\left(1+\psi^{2}\right)\psi^{2}\right]\nonumber \, ,  \\
\frac{{\bar {\bar \alpha}}}{\alpha} =& \frac{1}{\sqrt{1+\psi^{2}}} = \frac{\bar B}{B} \label{CrazyYorkShiftResult}
\,. \end{align}

While the condition ${\bar {\bar n}}_{a}s^{a} = 0$ gives no new information.  

Now, substituting equation \eqref {CrazyYorkShiftResult} back into equations \eqref {LorentzPreForm1} and \eqref {LorentzPreForm2} yields 

\begin{align}
{\bar {\bar n}}_{a} = \sqrt{1+\psi^{2}}n_{a} - \psi \bar s_{a} \nonumber \, ,  \\
s_{a} = \sqrt{1+ \psi^{2}}\bar s_{a} - \psi n_{a}
\,. \end{align}

Which is clearly a Lorentz boost with boost parameter $\phi$ given by $\psi = sinh(\phi)$.  Therefore, inverting the order in which one splits in a 1+1+2 split formalism will be equivalent to a boost when done the other way.  

And, as a consistency check, we compute:

\begin{align}
{\bar {\bar n}}^{a}{\bar{\bar n}}^{b} - s^{a}s^{b} =& \left(\sqrt{1+\psi^{2}}n^{a} - \psi \bar s^{a} \right)\left(\sqrt{1+\psi^{2}}n^{b}- \psi \bar s^{b} \right) \nonumber \, ,  \\
&- \left(\sqrt{1+ \psi^{2}}\bar s^{a} - \psi n^{a}\right)\left(\sqrt{1+ \psi^{2}}\bar s^{b} - \psi n^{b}\right) \nonumber \, ,  \\
=&\left(1+\psi^{2}\right)n^{a}n^{b} -\psi\sqrt{1+\psi^{2}}\left(n^{a}\bar s^{b}+n^{b} \bar s^{a}\right) + \psi^{2}s^{a}s^{b} \nonumber   \\
& -\left(1+\psi^{2}\right)\bar s^{a} \bar s^{b} + \psi \sqrt{1+\psi^{2}}\left(\bar s^{a} n^{b} + \bar s^{b} n^{a} \right) - \psi^{2} n^{a} n^{b} \nonumber \, ,  \\
=& n^{a}n^{b} -\bar s^{a} \bar s^{b}
\,. \end{align}

So, as promised, our two projection operators are identical.  

\section{Example: calculating the velocity parameter of Kerr Spacetime}

Now, for the sake of completeness, let us calculate the value of the boost parameter $\psi$ for the Kerr spacetime.  Take the metric to be given by:

\begin{equation}
g_{ab}= \left(\begin{tabular}{l c c r}
$-\left(1-\frac{2\,M\,r}{B}\right)$&$\frac{2\,M\,r}{B}$&$\frac{2\,M\,r}{B}a\,sin^{2}(\theta)$&0  \\
$\frac{2\,M\,r}{B}$&$1+\frac{2\,M\,r}{B}$&$-\left(1+\frac{2\,M\,r}{B}\right)a\,sin^{2}(\theta)$&0  \\
$\frac{2\,M\,r}{B}a\,sin^{2}(\theta)$&$-\left(1+\frac{2\,M\,r}{B}\right)a\,sin^{2}(\theta)$&$\left(A+\frac{2\,M\,r\,a^{2}\,sin^{2}(\theta)}{B}\right)sin^{2}(\theta)$&0  \\
0&0&0&$B$  \\
\end{tabular}\right)
\,. \end{equation}

\begin{equation}
g^{ab} = \left(\begin{tabular}{l c c r}
$-\left(1+\frac{2\,M\,r}{B}\right)$&$\frac{2\,M\,r}{B}$&0&0 \\
$\frac{2\,M\,r}{B}$&$\frac{A-2\,M\,r}{B}$&$\frac{a}{B}$&0  \\
0&$\frac{a}{B}$&$\frac{1}{B\,sin^{2}(\theta)}$&0  \\
0&0&0&B  \\
\end{tabular}\right)
\,. \end{equation}

In coordinates labeled by $(t,r,\phi,\theta)$, and all other parameters labeled as in Appendix \ref {sec: Kerr}.  Take the function $\tau = -t$, and the function $R=r$.  Then, we obtain via a simple reading from the above expression of the inverse to the metric tensor:

\begin{align}
n_{a} =& \left(-\sqrt{\frac{B}{B+2\,M\,r}},0,0,0\right)\qquad s_{a} =\left(0,\sqrt{\frac{B}{A-2\,M\,r}},0,0\right)\nonumber \, ,  \\
\psi =& n^{a}s_{a} = g^{ab}n_{a}s_{b} =-\frac{2\,M\,r}{B}\left(\sqrt{\frac{B}{B+2\,M\,r}}\right)\sqrt{\frac{B}{A-2\,M\,r}}\nonumber \, ,  \\
=&-\frac{2\,M\,r}{\sqrt{\left(B+2\,M\,r\right)\left(A-2\,M\,r\right) }}=sinh(\phi)
\,. \end{align}

Then, solving $1=cosh^{2}(\phi) - sinh^{2}(\phi)$ gives:

\begin{equation}
cosh(\phi) = \sqrt{\frac{A\,B+2\,M\,r\,a^{2}sin^{2}(\theta)}{\left(A-2\,M\,r\right)\left(B+2\,M\,r\right)}}
\,. \end{equation}

Which immediately allows us, in analogy with Special Relativity, to find the velocity parameter $v$, 

\begin{equation}
v=tanh(\phi)=-\frac{sinh(\phi)}{cosh(\phi)}=\frac{2\,M\,r}{\sqrt{A\,B+2\,M\,r\,a^{2}sin^{2}(\theta)}}
\,. \end{equation}

\begin{figure}[htb!]
\centering%
\includegraphics[scale=.45]{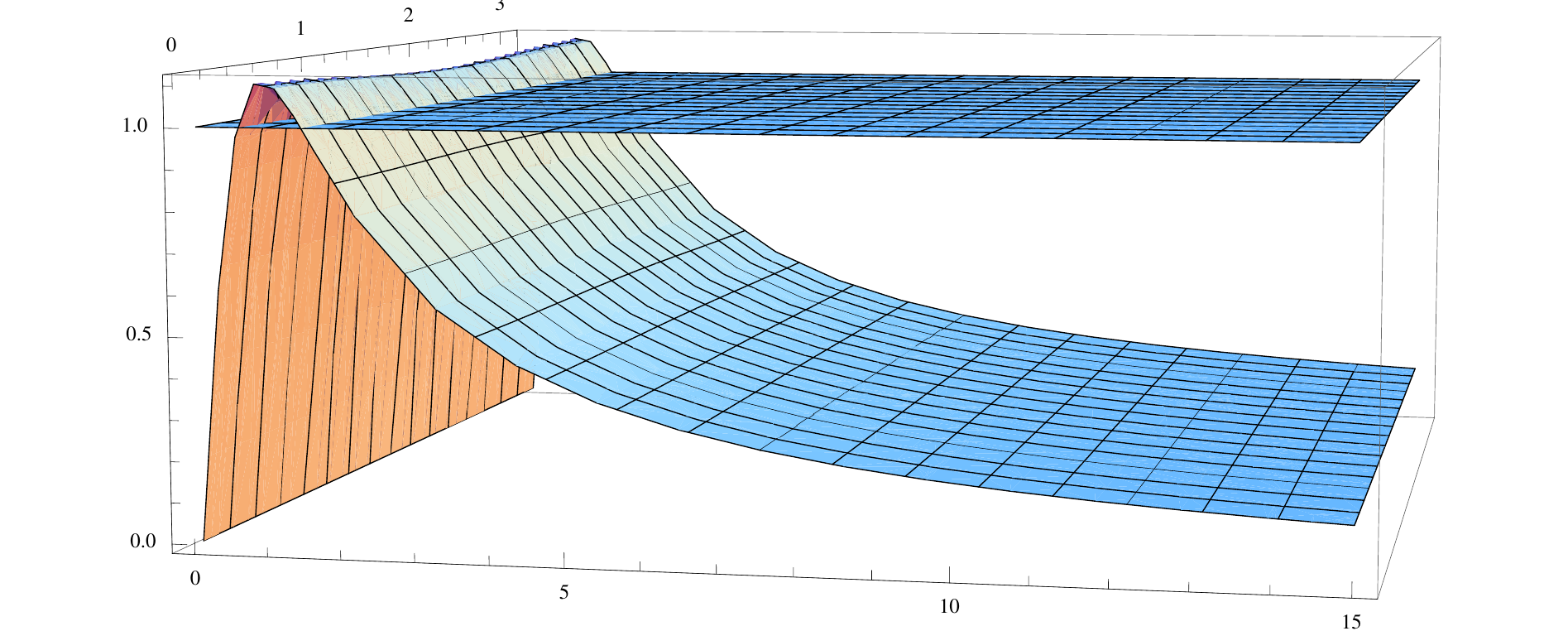}
\caption{The velocity parameter $v$ plotted against $r/M$ and $\theta$.  The plane is the surface $f=1$ and $a$ = $0.9$.}
\label{fig:KerrVelocityPlot}
\end{figure}

This term can be interpreted as the velocity of the 3-surface of constant R relative to the 3-surface of constant $\tau$..  Note that it is zero on the surface at spacelike infinity, as $r\rightarrow \infty$.  Since $g^{rt} = \frac{\beta^{r}}{\alpha^{2}} \rightarrow 0$ for asymptotically flat spacetimes, this result is only sensitive to asymptotic flatness, and not the details of the Kerr solution.  Furthemore, for the case of asymptotically flat spacetimes, one can use the fact that $v=0$ at infinity in order to interpret $v$ as the velocity of a point's reference frame with respect to conformal spacelike infinity.  In order to bolster this interpretation, note that $v$ takes the value of 1 for all points on either horizon  $r = M \pm \sqrt{M^{2}-a^2}$, is greater than one in the region between the two horizons, and less than 1 everywhere else\footnote{As another aside, note that $v$ takes the value 0 for $r=0$.  Observers inside the inner horizon cannot access the region outside the horizon, making it somewhat nonsensical to talk about ``local speeds of spacetime'' relative to conformal infinity.  One could, however, interpret $v$ as the velocity of the reference frames relative to the point at the center of the Kerr ring.}.  A \textit{Mathematica} plot of $v$ for $a=.9\,M$ in units of $r/M$ is included in figure \eqref {fig:KerrVelocityPlot}.

\nocite{*}


\bibliography{DissertationBib}                 

\begin{thesisauthorvita}
Jerry Schirmer was born in St. Charles, Missouri to Anne and Michael Schirmer.  He attended Marquette High School in Chesterfield Missouri, eventually attending college at Truman State University, where he graduated Magna Cum Laude and Phi Beta Kappa with degrees in Physics, Philosophy \& Religion, and Political Science.  He has a deep interest in the role that science and the scientist has in society.  

\end{thesisauthorvita}       

\end{document}